\documentclass[nofootinbib,eqsecnum,tightenlines,11pt]{revtex4}

\usepackage{graphicx}
\usepackage{amsmath,amssymb,amsfonts,amsthm,stmaryrd,mathtools,bm}
\usepackage{mathrsfs}
\usepackage{color}
\usepackage{multirow,bigdelim}
\usepackage{hyperref}
\usepackage{dsfont}
\usepackage{booktabs}
\allowdisplaybreaks[0]

\usepackage{pstricks}
\usepackage{tikz}

\usepackage[applemac]{inputenc}

\def\be#1\ee{\begin{align}#1\end{align}}

\def\ba{\begin{eqnarray}}
\def\ea{\end{eqnarray}}
\def\la{\langle}
\def\ra{\rangle}

\def\q{\qquad}

\def\i{\mathrm{i}}
\def\SU{{\rm{SU}}}

\def\R{\mathcal{R}}

\newcommand{\numlam}{\lambda_{*}}
\newcommand{\tristep}[1]{\Theta_{\sigma^{#1}}[\text{Tri}]}

\begin{document}

\title{Discrete gravity dynamics from effective spin foams  }

\author{Seth K. Asante}
\affiliation{Perimeter Institute, 31 Caroline Street North, Waterloo, ON, N2L 2Y5, CAN}
\author{Bianca Dittrich}
\affiliation{Perimeter Institute, 31 Caroline Street North, Waterloo, ON, N2L 2Y5, CAN}
\author{Hal M. Haggard}
\affiliation{Physics Program, Bard College, 30 Campus Road, Annandale-On-Hudson, NY 12504, USA}
\affiliation{Perimeter Institute, 31 Caroline Street North, Waterloo, ON, N2L 2Y5, CAN}

\begin{abstract}

The first computation of a spin foam dynamics that provides a test of the quantum equations of motions of  gravity is presented. Specifically, a triangulation that includes an inner edge is treated. The computation leverages the recently introduced effective spin foam models, which are particularly numerically efficient. Previous work has raised the concern of a flatness problem in spin foam dynamics, identifying the potential for the dynamics to lead to flat geometries in the small $\hbar$ semiclassical limit.  The numerical results presented here expose a rich semiclassical regime, but one that must be understood as an interplay between the various parameters of the spin foam model. In particular, the scale of the triangulation, fixed by the areas of its boundary triangles, the discreteness of the area spectrum, input from Loop Quantum Gravity, and the curvature scales around the bulk triangles, all enter the characterization of the semiclassical regime identified here. In addition to these results on the dynamics, we show that the subtle nature of the semiclassical regime is a generic feature of the path integral quantization of systems with second class constraints. 

\end{abstract}

\maketitle

\section{Introduction}

Einstein's theory of general relativity encodes the gravitational force into spacetime geometry.  A  promising strategy to quantize this theory of gravity, therefore,  is  to first construct a notion of quantum geometry and then to prescribe a gravitational dynamics for these quantum geometries. 

A rich notion of quantum geometry, in the form of  spatially diffeomorphism covariant Hilbert space representations of a suitable geometric operator algebra \cite{quantumgeometry,NewQG}, has been constructed in loop quantum gravity (LQG) \cite{LQG}.  A key result of this framework is  that areas, instead of lengths, arise as more fundamental variables and turn out to have discrete spectra \cite{DiscreteGeom}, with an area gap determined by the Barbero-Immirzi parameter \cite{Immirzi-Barbero}.

Discrete area spectra can also be motivated on a more heuristic level: the areas are conjugated to extrinsic curvature angles that take values in a compact circle.\footnote{ We consider spacetimes with Euclidean signature and leave the Lorentzian case for future work \cite{ADP}.}   Thus, their quantization leads to discrete area spectra with equidistant gaps in the limit of large areas. Several approaches to the black hole entropy counting also rely on a discrete area spectrum \cite{BHCounting}.

Discreteness, together with the fact that these areas arise as locally independent variables in loop quantum gravity, leads to an interesting challenge for the construction of a suitable quantum gravitational dynamics \cite{EffSFM}: area variables must be constrained in order to avoid suppression of curvature while their discreteness forbids strong imposition of these constraints.\footnote{A similar issue arises  in the constrained higher gauge theory formulation of gravity of \cite{MV}.}

In this paper we will discuss implementation of the dynamics via a path integral over geometries. We regularize the path integral by choosing a triangulation to which we associate geometrical data; these data will be summed over in the path integral. Specifically, we work with the areas of the triangles and allow only area values in the discrete spectrum, which enforces compatibility with the LQG Hilbert space. 

A free assignment of such areas to the triangles of a four-dimensional triangulation defines a much more general type of geometry than the piecewise flat geometries  defined through an assignment of lengths to the edges.  The latter type of length assignments have been shown to give a suitable space of metric geometries in the continuum limit \cite{SchraderEtAl, WilliamsEtAl}. It is also possible to define an action on the more general configuration space of areas, the so called area Regge action \cite{AreaRegge,ADHAreaR}. On the subspace of piecewise flat geometries the area Regge action agrees with the length Regge action \cite{Regge}. The length action is a discretization of the Einstein-Hilbert action and its variation leads to a discretized version of Einstein's equations. In contrast, free variation in the full space of area variables of the area Regge action does not lead to suitable gravitational dynamics---instead the resulting equations of motion force vanishing curvature. 

To obtain a gravitational dynamics, we can constrain the variations of the areas to those that correspond to varying only the lengths. To this end one can formulate so-called shape matching constraints \cite{AreaAngle,DittrichRyan,EffSFM}. These constraints vanish on the subspace of area configurations that arise from a consistent assignment of lengths to the edges of the triangulation and properly lead to gravitational dynamics.  These constraints do, however, suffer from a difficulty: for areas taking values in a discrete spectrum these constraints pose diophantine type equations and severely limit the solution space of discrete geometries.  Indeed, the constraints have  too few discrete area solutions to allow for a quantum dynamics with a suitable classical limit \cite{EffSFM}.

This problem also occurs for spin foams \cite{Perez}, but was first encountered in a more technical guise. Spin foams provide transition amplitudes for loop quantum gravity states. They are derived from a path integral based on the  Plebanski formulation of gravity  \cite{Plebanski}.  The variables of the Plebanski formulation are a gauge connection and conjugated bivector valued two-forms.\footnote{The same kind of phase space variables appear in the Hamiltonian formulation of Yang Mills theory. This lattice gauge theory also features holonomies and conjugated Lie algebra-valued fields, much like those described below.} This theory also features constraints, the so-called simplicity constraints, that ensure that the bivector valued two-forms are constructed from a wedge product of tetrad one-forms. 

In the classical, continuum theory the bivector valued two-forms commute, and, thus, the simplicity constraints commute too.\footnote{Here we only consider the so-called primary simplicity constraints.} In the discretized (or quantized) theory one replaces the Lie algebra-valued connection variable with Lie group-valued holonomy---this leads to a compactification of this variable and can be understood as the reason for the discreteness of the area spectrum. Replacement of the Lie algebra by the Lie group also introduces curvature into the holonomy configuration space. Thus, the bivector components, which are conjugated to the group valued holonomy variables, become non-commutative. This can be thought of as an anomaly in the algebra of bivector components.\footnote{See \cite{Corichi,FluxRepOfLQG,DittrichGeillerFlux,PerezCattaneo,CornerSimpl} for various explanations of how and why this non-commutativity arises.} For this reason, a certain part of the simplicity constraints, given by the so-called cross-simplicity constraints, become non-commutative, and more precisely, second class,  see e.g \cite{DittrichRyan,Perez}. 
The same process leads also to the non-commutativity of three-dimensional (3D) dihedral angles, which is the source for the second class shape matching constraints employed in our work.   

In short, both the discreteness of the areas and the second class nature of  the  constraint algebra arise from the same quantization step, and thus are deeply connected.  Both features are also characterized by the Barbero-Immirzi-parameter, which controls the area gap \cite{DiscreteGeom} and the anomaly  in the algebra of bivector components, or equivalently, 3D angles \cite{DittrichRyan,EffSFM,CornerSimpl}.

Second class constraints cannot be imposed sharply in the quantum theory.  The best one can achieve is to minimize the product of their uncertainties, e.g. by using coherent states to impose the constraints. This ``weak imposition" of the constraints has been utilized first in the Engle-Pereira-Rovelli-Livine (EPRL), Freidel-Krasnov (FK), and related spin foam models \cite{NSFM,Perez}, which are based on the Plebanski formulation of gravity. 

Study of the semiclassical regime \cite{SFLimit}, taken to be $\hbar\rightarrow  0$,\footnote{For spin foams this is equivalent to the so-called large $j$ limit, as the areas are quantized as $A\sim \gamma \hbar j$ with $j \in \mathbb{N}/2$. } revealed the possibility that these models impose vanishing curvature in this limit \cite{FlatProblem,ILQGS}, and that this was due to the constraints being imposed weakly. This became known as ``the flatness problem" in the spin foam literature.  An example of a suggested resolution was that to avoid suppressed curvature, one should, instead of just taking the $\hbar\rightarrow 0$ limit,  also scale the Barbero-Immirzi parameter to zero  \cite{CE,MHan}. 

A completely satisfying resolution of the flatness problem for the models \cite{NSFM} has  not been achieved so far for at least two reasons \cite{ILQGS}. Firstly, the structure of the amplitudes for these models is quite involved, making it hard to deduce how strongly the various constraints are imposed \cite{ILQGS}. This applies, in particular, to the shape matching constraints, which in the Plebanski formulation are part of the secondary simplicity constraints \cite{DittrichRyan} and, as such, are not imposed explicitly, but are essential for a correct gluing of the building blocks and therefore to achieve good dynamics.  Secondly, various features of the models make their numerical investigation hard, which has until now prevented an explicit numerical check of spin foam dynamics in even small complexes of several glued simplices, that said, see \cite{DonaNum} for recent progress in the EPRL-FK type models.

In summary,  the imposition of the constraints in quantum geometric models such as spin foams, is hindered by an anomaly in the algebra of spatial geometric observables. Thus, one must deal with second class constraints. Using a simple toy model, we will show that for path integrals with weakly imposed second class constraints, the semiclassical limit $\hbar \rightarrow 0$ does not, in general, reproduce the classical dynamics. The flatness problem in spin foams is just one example of this more general issue. 

A possible way to resolve this issue is to update the definition of the semiclassical limit. In a given physical system there is usually more than one parameter that can be tuned. Indeed, in the case of spin foams,  the anomaly is also parametrized by the Barbero-Immirzi parameter $\gamma$. It is not surprising that a suitable semiclassical regime might require a small Barbero-Immirzi parameter \cite{CE,MHan}. The dynamics also includes the curvature per triangle, another parameter whose behavior should be considered. 

The main aim of this paper will be to identify regimes for spin foam dynamics that feature a suitable semiclassical dynamics. These regimes will be characterized by a scale, given by the average value of the areas (evaluated on the classical solution), by the Barbero-Immirzi parameter $\gamma$, and by the average curvature $\epsilon$ per triangle. To find these regimes, we will compute expectation values of certain observables for different values of these parameters, and check when these match sufficiently well the corresponding classical values. To this end, we will need a model that is sufficiently amenable to numerical calculations. 

We have recently suggested such an `effective' spin foam model in \cite{EffSFM}. It is in our view the simplest possible model that implements a discrete area spectrum and defines transition amplitudes compatible with the loop quantum gravity Hilbert space. The model can be motivated as arising from a higher gauge formulation of gravity \cite{HigherGauge,MV}. A simpler starting point is the constrained area Regge action \cite{EffSFM}, where the constraints are the shape matching constraints mentioned above. These are---due to the  mechanism described above---second class. We have, therefore, implemented these constraints weakly, via coherent states. This effective spin foam model faces the same issue as the EPRL-FK type spin foam models \cite{NSFM}, which derive from the Plebanski formulation. In fact, as explained above, the need to implement the constraints weakly, can be understood as being due to the fact that the areas have discrete, approximately equidistant spectra.

We will see that the effective spin foam models indeed allow us to calculate the expectation values of observables for a triangulation that is sufficiently large to test the equations of motion. This allows us---for the first time---to test whether and how spin foam models impose the equations of motion of discrete gravity. Moreover, we have computed expectation values for quite a large range of scales and  for continuous values of $\gamma$, revealing a very rich structure.\footnote{ There is a (not often discussed)  restriction for the Euclidean signature EPRL model on the values of the Barbero-Immirzi parameter $\gamma$ and the spin representations \cite{Perez}.  If one wants to allow for all spins $j\in \mathbb{N}/2$, then only $\gamma \in \mathbb{N}$ is allowed.  A rational, non-integer, value for $\gamma$ leads to a restriction on the  spins summed over in the partition function, whereas non-rational values for $\gamma$ are not allowed at all. E.g. for $\gamma=1/2$, spins need to be even numbers, and for $\gamma=1/4$ spins need to be multiples of $4$.  This restriction does not arise for the effective spin foam models, nor does it arise for the Lorentzian signature version of the EPRL model.} 

These numerical results have allowed us to identify a semiclassical regime, that is, a region in parameter space for which the expectation values match well the classical values.  Although we will partially confirm  expectations, which derive from an  heuristic argument based on the stationary phase approximation \cite{MHan,EffSFM}, we will also find many unexpected features. These features support the hope that  spin foams admit a suitable continuum limit \cite{CG}. 

\bigskip

The outline of the paper is as follows: In section \ref{toymodel} we discuss a simple model with a weak implementation of constraints. We compute expectation values for this model and in this way study the consequences of the weak implementation for the semiclassical limit. We also show how expectation values with respect to  a partition function based on a (continuous) integration over an infinite range compares  with partition functions where one integrates over a finite range or sums over discrete variables. This introduces some features that also appear for the spin foam expectation values.

Section \ref{Sec-EffSFM}  provides a short introduction to the effective spin foam model. It is a quantization of constrained area Regge calculus, where the constraints are the shape matching constraints. These constraints are second class and we will implement them weakly.

In Section \ref{Sec-Triang} we introduce the triangulation in which we compute the expectation values.
The next section, Section \ref{SFExpV}, contains the main results: we compute a number of expectation values as functions of the Barbero-Immirzi parameter $\gamma$ and for a range of boundary data, inducing different scales and curvatures. We identify semiclassical regimes and discuss various new features, which became apparent in our results.

We close the main part of our paper with a discussion and an outlook in Section \ref{Disc}. The Appendices \ref{GTI} to \ref{speed}
contain information on the generalized triangle inequalities, on the roots for the area-length system of equations, and on the  implementation of the numerical computation.

\section{A toy model for the weak implementation of constraints}\label{toymodel}

Here we illustrate with a simple toy model how the weak imposition of a constraint leads to a restricted range of  parameters in which one can expect to regain a semiclassical regime. 

We will consider an integral with an oscillatory phase $\exp(\i \lambda S)$ over a set of constrained configurations. First we will impose the constraint $C$ strongly, that is, only integrate over configurations that satisfy the constraint exactly. In this case we expect that for large $\lambda$, we can apply the stationary phase approximation, in which the integral is replaced by the evaluation of the integrand at the critical point(s) of the constrained phase. 

We will compare this strong imposition of the constraint with a weak imposition. To do the latter we include a Gaussian factor  $\exp(- \mu C^2)$ into the integral.  Using these two procedures we  compute expectation values and can check for which parameters  $(\lambda, \mu)$ we reproduce (approximately) the critical value of the oscillatory constrained phase. 

The toy model allows us to consider infinite and finite integration ranges. We will see that introducing a finite integration range leads to peculiar effects. We can also replace the integral with a discrete sum.  These changes will effect the resulting expectation values and foreshadow similar effects for the spin foam model.

The unconstrained action $S$ in our toy model is a simple quadratic functional in two variables
\ba
S=x^2+y^2
\ea
with one critical point at $(x,y)=(0,0)$.  Imposing the constraint $C=y-(x-2)$, the constrained action 
\ba
S_C=x^2+(x-2)^2
\ea
has now a critical point at $(x,y)=(1,-1)$. We will call this point the critical point of the constrained theory or constrained critical point for short. 

We will compute the expectation value of an observable using  a  `path integral'  over $(x,y) \in \mathbb{R}^2$ with the constraint imposed strongly, as well as weakly.  We choose as observable ${\cal O}(x,y)=\exp(-x^2)$, 
which classically evaluates to $o(x,y)_{\text{class sol}}=e^{-1}\approx 0.368$.

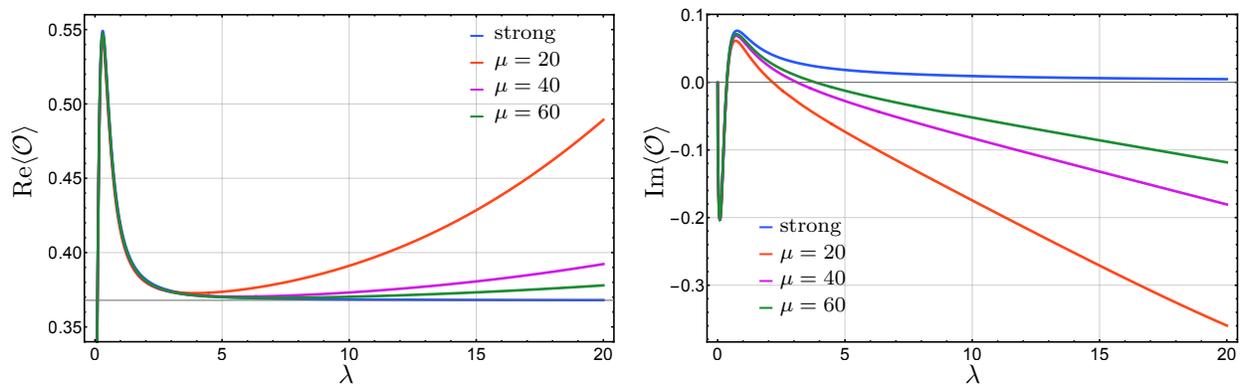
\begin{figure}[ht!]
\begin{picture}(500,150)
\put(15,7){ \includegraphics[scale=0.26]{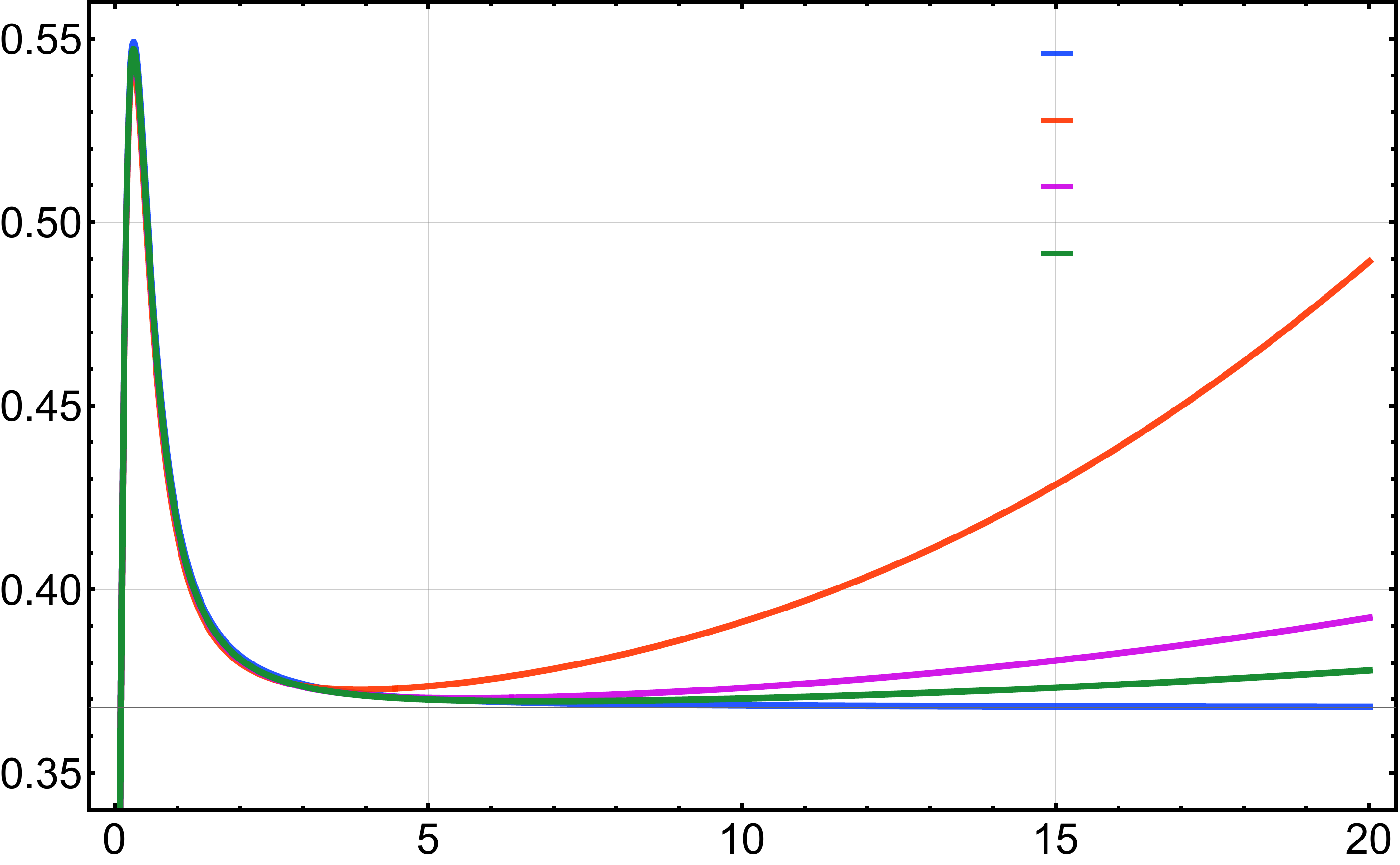} }
\put(250,7){ \includegraphics[scale=0.26]{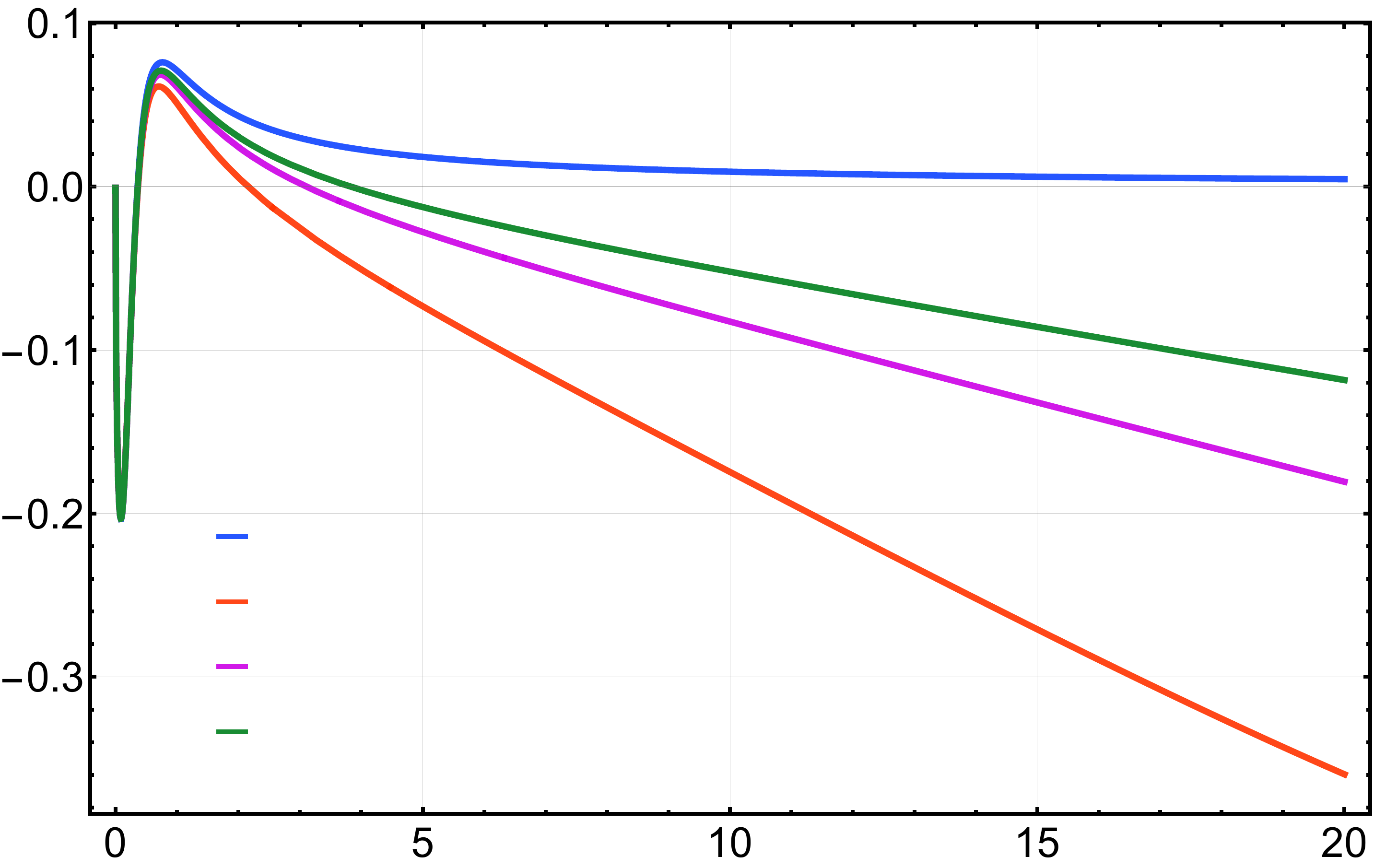} }

\put(1,70){ \rotatebox{90} {$\rm Re \langle  \cal O \rangle $ } }
\put(240,70){ \rotatebox{90} {$\rm Im \langle  \cal O \rangle $ } }

\put(187,130) {\scriptsize strong}
\put(187,120) {\scriptsize $\mu = 20$}
\put(187,110) {\scriptsize $\mu = 40$}
\put(187,100) {\scriptsize $\mu = 60$}

\put(295,57) {\scriptsize strong}
\put(295,47) {\scriptsize $\mu = 20$}
\put(295,37) {\scriptsize $\mu = 40$}
\put(295,27) {\scriptsize $\mu = 60$}



\put(128,0){$\lambda$}
\put(365,0){$\lambda$}
\end{picture}
\caption{Real and imaginary parts of the expectation values of ${\cal O}=\exp(-x^2)$ with strongly and weakly implemented constraints.  \label{Fig1}}
\end{figure}

To begin with we consider the strongly constrained path integral and compute\footnote{To obtain a convergent result we  assume here that $\lambda$ has a very small imaginary part.} the expectation value of ${\cal O}(x,y)$ 
\ba
\langle {\cal O}\rangle_{\rm strong}\, =\, 
\frac{\int_{-\infty}^\infty \exp \left(\i \lambda  (x^2 +(2-x)^2) \right) \exp\left(-x^2\right)d x }
{\int_{-\infty}^\infty \exp \left(\i \lambda  (x^2 +(2-x)^2) \right) d x  } \,=\,
\sqrt{\frac{2\i \lambda}{2\i\lambda-1}} \exp\left(  -\frac{2\lambda}{2\lambda+\i}  \right) \ . 
\ea
Figure \ref{Fig1} shows that the real part of the expectation value approximates well the constrained critical value for sufficiently large $\lambda$, e.g. for $\lambda=5$ one has ${\rm Re}(\langle {\cal O}\rangle_{\rm strong})\approx 0.370$, and the approximation gets better with growing $\lambda$. The expectation value has a non-vanishing imaginary part, which starting from a  maximum at $\lambda\approx 0.764$ decreases monotonically with growing $\lambda$. For $\lambda=5$ we have ${\rm Im}(\langle {\cal O}\rangle_{\rm strong})\approx 0.018$. Thus, in accordance with the stationary phase approximation, with larger and larger $\lambda$ the expectation value is better and better approximated by its classical value $o(x,y)_{\text{class sol}}$.
 
Next, we consider the expectation value as defined by a weakly constrained path integral
\ba
\langle {\cal O}\rangle_{\mu}\, =\, 
\frac{\int_{-\infty}^\infty \int_{-\infty}^\infty\exp \left(\i \lambda  (x^2 +y^2) \right) \exp\left(-\mu(y-x+2)^2\right) \exp\left(-x^2\right) dy d x }
{\int_{-\infty}^\infty \int_{-\infty}^\infty \exp \left(\i \lambda  (x^2 +y^2) \right)\exp\left(-\mu(y-x+2)^2\right)  d y d x  }  \ ,
\ea
where $\mu$ parametrizes the peakedness of the Gaussian implementing the constraint. (Note that for the computation of the expectation value the normalization of the Gaussians drops out.) The integrals in this example are convergent and the expectation values can be computed exactly. As we have an action and constraint that are quadratic in the variables, applying a saddle point approximation to the integrals in the numerator and denominator, leads to the same result.  The only subtle point is that the saddle points are not real, but shifted into the complex (hyper-)plane. 

As one can see from Figure \ref{Fig1}, 
the expectation values coming from the weakly constrained integrals approximate the one from the strongly constrained integral better and better for growing $\mu$. At the same time, the expectation values deviate more and more for growing $\lambda$. 

\begin{figure}[ht!]
\begin{picture}(500,140)
\put(120,7){ \includegraphics[scale=0.25]{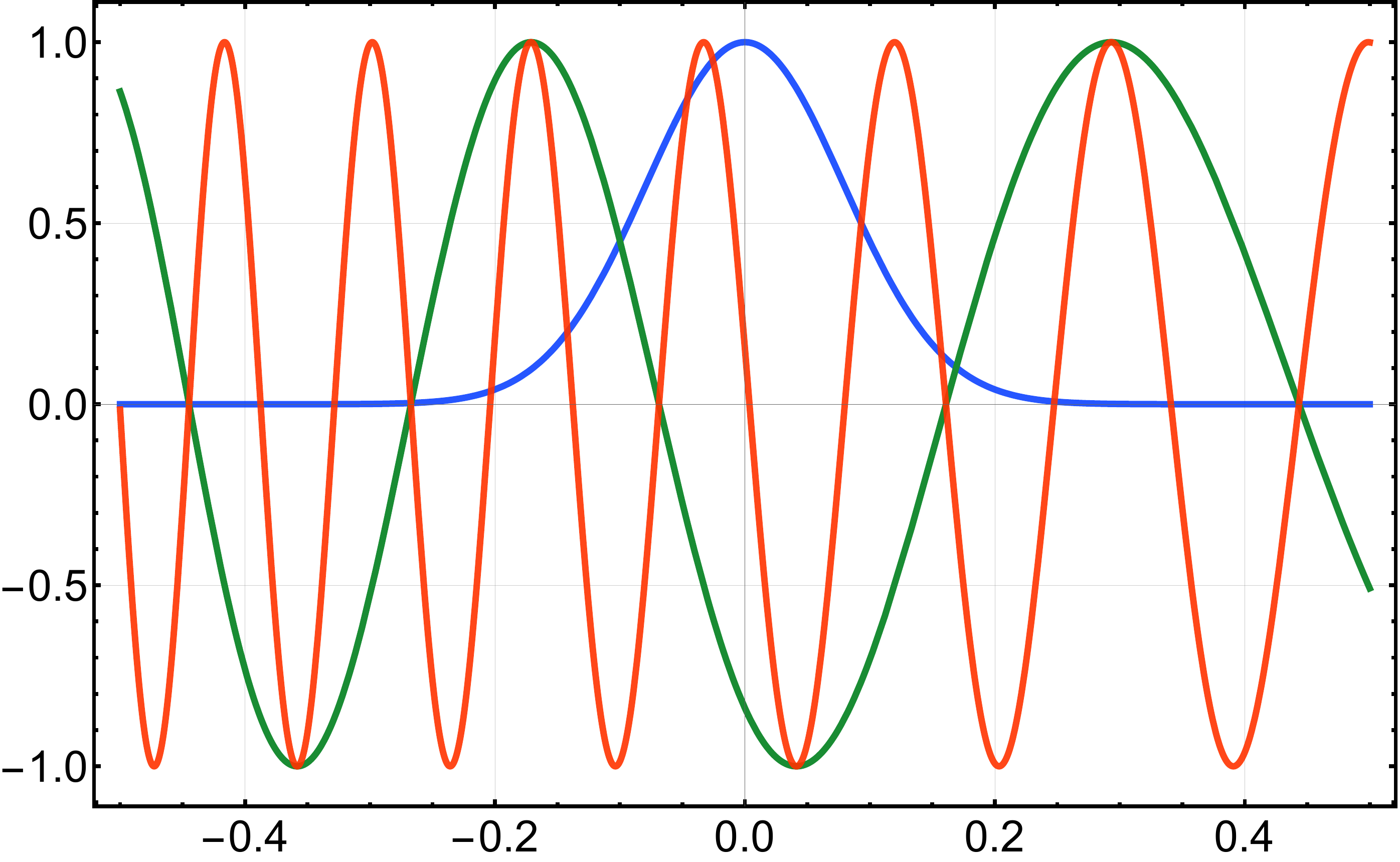} }

\put(280,0){ $ z $ }
\put(110,55){\rotatebox{90} {$ G\, ,\,\,{\rm Re} (\exp(\i \lambda S) )$ }}

\end{picture}
\caption{ The real part of the oscillatory factor $\exp \left(\i \lambda  (x^2 +y^2) \right)$  for $\lambda=5$ (green curve) and $\lambda=15$ (red curve), as well as the constraint implementing factor $G= \exp\left(-\mu(y-x+2)^2\right)$ (blue curve)  along the parametric curve $x=1+z,y=-1-z$ with parameter $z$.
 \label{Fig2}   }
\end{figure}

Figure \ref{Fig2} 
explains why this difference in the weakly and strongly constrained integrals increases with growing $\lambda$.  This figure shows the two contributing factors to the path integral, the Gaussian and the phase factors $\exp(\i\lambda S)$ for two different values of $\lambda$.  The plot shows these factors along a curve following the gradient of the constraints away from the critical point $(x,y)=(1,-1)$ of the constrained action. The Gaussian has a finite width, and over this width the phase factor oscillates. A strongly oscillating phase, which occurs for large $\lambda$, can average out the Gaussian and thus render the (weak) imposition of the constraints inoperative. With this insight we can derive a rule of thumb for when we might expect a reasonably good approximation to the expectation value of a weakly constrained (path) integral:  The number of oscillations of the phase factor occurring over the width of the Gaussian (around the critical point of the constrained action) should be less than a number of order 1.   Let us consider this condition around the constrained critical point and  in the direction of maximal change of the constraint,  that is along the gradient $\vec{n}=\vec{\nabla} C/|\vec{\nabla} C|$. We approximate the change in the phase with its derivative, and obtain the condition
\ba\label{SCcond}
\left. \lambda \times \left(|\vec{\nabla} S \cdot \vec{n}|\right)\right|_{ \text{constrained crit. pt.}} \times \sigma(\vec{n})  \lesssim \, {\cal O} (1)
\ea
with $\sigma^2(\vec{n})$ being the variance of the Gaussian along the $\vec{n}$-direction.\footnote{The form of the condition (\ref{SCcond}) also indicates a case, for which the weak imposition of second class constraints is  not  problematic: namely if the second class constraints gauge fix a symmetry, as is done in the Gupta-Bleuler quantization for electrodynamics. One would then have an action, and with it an oscillatory factor, which is constant along the gauge directions.} Absorbing  numerical factors into ${\cal O}(1)$ this gives for our toy model 
 the condition $\lambda/\sqrt{\mu} \lesssim {\cal O} (1)$.

In summary, we see that, keeping $\mu$ fixed, the classical value is well approximated by the expectation value only for a certain range of $\lambda$. A lower bound is needed for the stationary phase approximation, which requires a sufficiently oscillatory phase away from the critical points, and hence sufficiently large $\lambda$. An upper bound is needed for the weak implementation of the constraints to be effective and is due to the fact that a too oscillatory behavior of the phase factor (around the constrained critical points) renders the implementation of the constraints via Gaussians impotent.  Thus, given a fixed value of $\gamma$,  we have a semiclassical regime only for this bounded interval of $\lambda$. Alternatively, one can demand that with growing $\lambda$ the constraints are imposed more and more strongly, in other words, $\mu$ must also grow. 

For a quantum mechanical path integral, with a weak implementation of second class constraints, we have $\lambda \sim 1/\hbar$. The variance  $\sigma^2$ in the Gaussian implementing the constraints is determined by the commutator between the constraints, we therefore have $\mu\sim 1/\sigma^2\sim 1/\hbar$, and thus a scaling $\lambda/\sqrt{\mu} \sim 1/\sqrt{\hbar}$ in Planck's constant. The usual understanding of the semiclassical limit, as taking  $\hbar \rightarrow 0$ keeping all other parameters fixed, does not apply to the case of  second class constraints implemented weakly in a path integral. 

For the effective spin foam models considered here, as well as other standard models, we have additional parameters that can be tuned. Amongst these there is the average area of the boundary triangles,\footnote{We assume here that this leads to a solution where the bulk areas are of a similar size. Indeed, Regge gravity has a scaling symmetry:  for a given solution, multiplying all boundary lengths by a factor $c$ gives another solution, also scaled by a factor $c$.}  which are determined by $A\sim \gamma \ell_P^2 j$ , with $\gamma$ the Barbero-Immirzi parameter,  $j \in \mathbb{N}/2$ a discrete variable, often referred to as a spin, and $\ell_P^2=8\pi\hbar G_N/c^3$. The boundary data also determine the (average) curvature $\epsilon$ per triangle. The action has a peculiar property: it scales linearly in the areas and curvature $\tfrac{1}{\ell_P^2}S\sim\tfrac{1}{\ell_P^2} A \epsilon \sim \gamma j\epsilon$, whereas the commutator of the constraints leads to a variance $\sigma^2(j)\sim  \sqrt{A /(\ell_P^2 \gamma)} \sim \sqrt{ j}$. Thus, for spin foam models, the condition (\ref{SCcond}) can be expressed in terms of the spin $j$, the Barbero-Immirzi parameter $\gamma$, and the curvature per triangle $\epsilon$ \cite{EffSFM}:
\ba\label{SFcond1}
\gamma\, \sqrt{j}\, \epsilon  \lesssim {\cal O} (1) \ .
\ea
Alternatively one can replace the spin variable $j$ with the (average) area
\ba\label{SFcond2}
\frac{\sqrt{\gamma} }{\ell_P} \,\sqrt{A} \,\epsilon  \lesssim {\cal O} (1)\  ,
\ea
which reintroduces our previous scaling  with $\sqrt{\hbar} \sim \ell_P$.

\bigskip

As mentioned above the critical point for this system is not real, but rather lies in the complex plane.
Indeed we have to consider the complex phase
\ba
\Phi_{\lambda,\mu}=\lambda(x^2+y^2) +\i \mu(y-x+2)^2 \, .
\ea
For this quadratic phase we have one critical point 
\ba
x_c(\lambda,\mu)=\frac{2\i\mu}{\lambda + 2\i \mu}\; \stackrel{\mu \rightarrow \infty}{\longrightarrow}\;1\, , \q
y_c(\lambda,\mu)=-\frac{2\i\mu}{\lambda + 2\i \mu}\; \stackrel{\mu \rightarrow \infty}{\longrightarrow}\;-1\, ,
\ea
where we consider the limit $\mu \rightarrow \infty$ for fixed $\lambda$.  The $\mu \rightarrow \infty$ limit gives back the critical point of the strongly constrained system, whereas the limit $\lambda\rightarrow 0$, with fixed $\mu$, leads to the critical point  $\text{Re}(x_c)=0$ and $ \text{Im}(x_c)=0$ of the unconstrained system with action $S=x^2+y^2$. 
Evaluating our observable at the critical point, that is, $\exp(-x_c^2(\lambda,\mu))$, and using a sufficiently large $\lambda$, effectively approximates the exact expectation values of this observable, as computed for Figure \ref{Fig1}. For more complicated systems one should apply  Picard-Lefshetz theory in order to determine which critical points contribute \cite{Picard}.

Spin foam models  have two other features that can affect any approximation based on the theory of continuous, unbounded, integrals.
Firstly, the integration range for spin foams is finite or semi-infinite rather than infinite.\footnote{The integration is restricted to positive areas and furthermore restricted by the triangle inequalities} Secondly, integrals are replaced by sums. 


\begin{figure}[ht!]
\begin{picture}(400,160)
\put(10,7){ \includegraphics[scale=0.33]{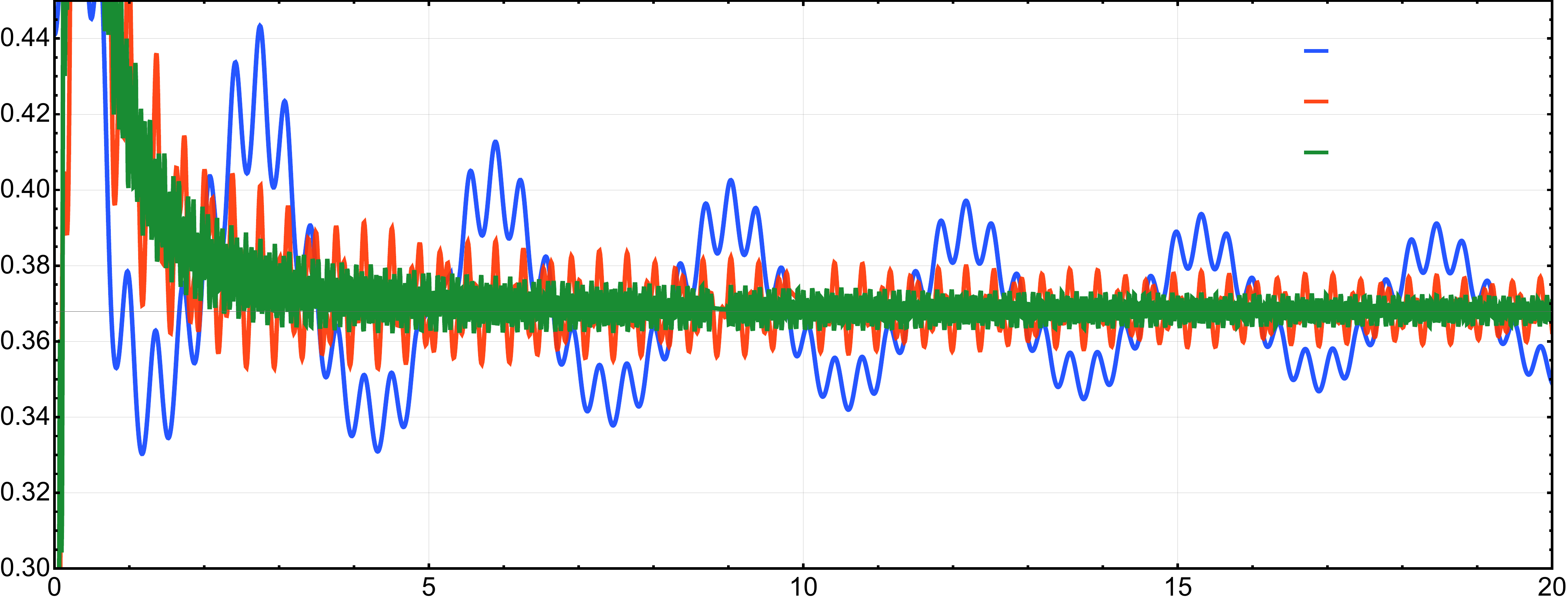} }

\put(230,0){ $ \lambda $ }
\put(-4,90){\rotatebox{90} {$ {\rm Re } \langle \exp(-x^2) \rangle $ }}

\put(355,146){\scriptsize  {$X=2 $ }}
\put(355,133){\scriptsize  {$X=4 $ }}
\put(355,120){\scriptsize  {$X=8 $ }}

\end{picture}
\caption{ Expectation values arising from an integral over finite intervals $(-X,X)$ and with a strong implementation of the constraints. 
 \label{Fig3}   }
\end{figure}
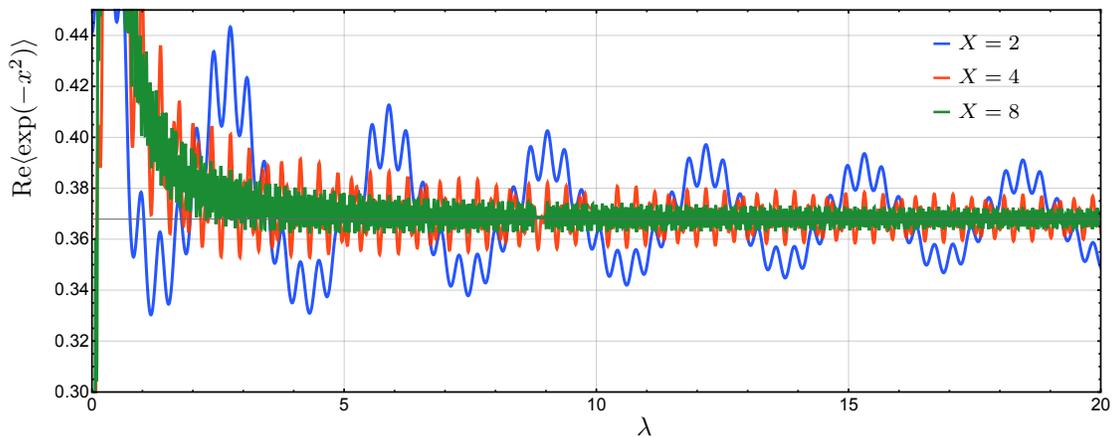

Figure \ref{Fig3} shows the expectation values for the strongly constrained path integral over the finite  interval $(-X,X)$ with $X\in \{2, 4,8\}$. This introduces, in the expectation value as a function of $\lambda$, two oscillations, one slow and the other much faster.  Increasing the size of the interval, the amplitude of the oscillations decreases (and the frequency of the fast oscillation increases), so that in the limit $X\rightarrow \infty$ one obtains the mostly monotonic function appearing in Fig. \ref{Fig1}.

We computed the  expectation values for the weakly constrained path integral numerically and this allowed us to also explore replacing the integral with a discrete sum.  By varying the density of the discretization we can see the effects of such a replacement. 


\begin{figure}[ht!]
\begin{picture}(500,250)
\put(80,7){ \includegraphics[scale=0.35]{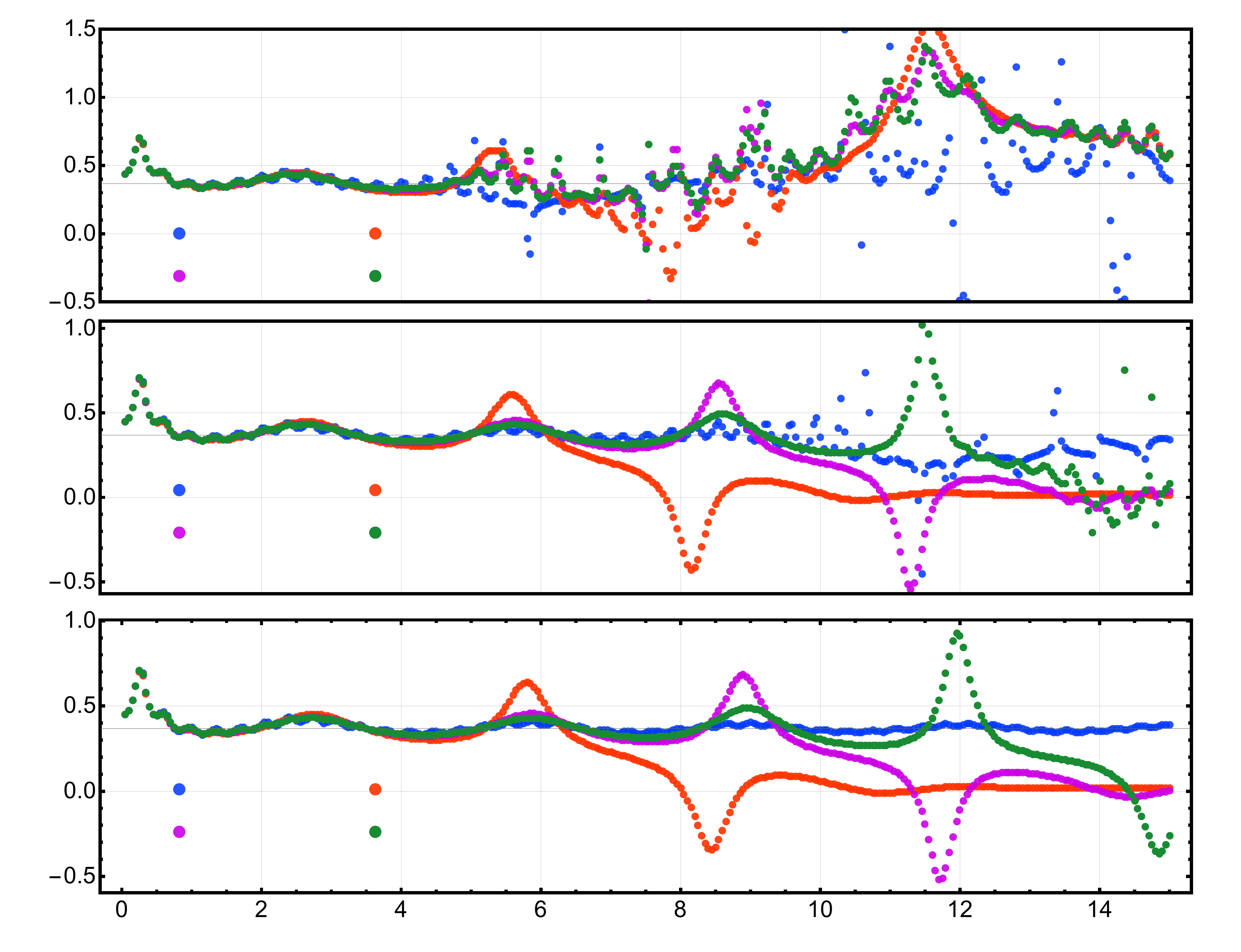} }
\put(145, 235){ $a=0.1$}
\put(145, 160){ $a=0.05$}
\put(145, 80){ $a=0.01$}

\put(250,0){ $ \lambda $ }
\put(80,115){\rotatebox{90} {$ {\rm Re } \langle \exp(-x^2) \rangle $ }}

\put(138,50){strong}
\put(191,50){$\mu = 20$}
\put(138,38){$\mu =40$}
\put(191,38){$\mu = 60$}

\put(138,130){strong}
\put(191,130){$\mu = 20$}
\put(138,120){$\mu =40$}
\put(191,120){$\mu = 60$}

\put(138,200){strong}
\put(191,200){$\mu = 20$}
\put(138,190){$\mu =40$}
\put(191,190){$\mu = 60$}

\end{picture}
\caption{ Expectation values arising from a summation over a finite range for different lattice constants, $a=0.1$ (top), $a=0.05$ (middle) and $a=0.01$ (bottom), and for different parameters $\mu$,  as well as with a strong implementation of the constraints. 
 \label{Fig4}   }
\end{figure}
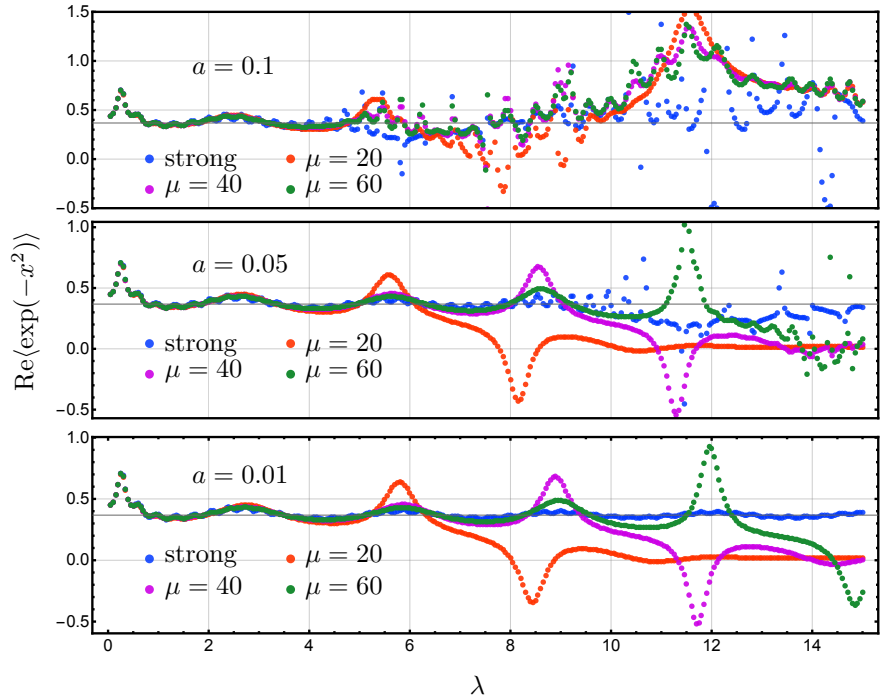

In Fig. \ref{Fig4} 
  we show the expectation values for the strongly and weakly constrained path integral, computed for a finite interval $x\in (-2,2)$ and $y\in(-6,2)$, and for three different discretizations with lattice constants $a \in \{0.01, 0.05, 0.1\}$ in the $x$- and $y$-coordinates. Interestingly, the fast oscillations of the strongly constrained path integral are somewhat washed out in the weakly constrained version. (These oscillations are visible for the $\mu=60$ version for $\lambda\lesssim 2$.) But, at a certain $\mu$-dependent threshold value for $\lambda$ the weakly constrained versions start to suddenly deviate from the expectation value of the strongly constrained version.

Effects due to the discretization occur in particular if the discretization scale is of the same order or larger than the wavelength defined by the oscillatory factor. Thus, these effects are  more pronounced for larger $\lambda$, and seem to start to affect the strongly constrained path integral for smaller values of $\lambda$ than the weakly constrained versions.

In the case of a continuous integration over an infinite range the deviation between the weakly implemented versions and the strongly implemented versions of the expectation value grow continuously with $\lambda$, see Fig. \ref{Fig1}.  Introducing a finite integration range we see more of a threshold behavior: the weakly implemented version   and the strongly implemented version are quite close up to a certain $\mu$-dependent value $\lambda_T(\mu)$, after which the deviation grows.

\vspace{0.3cm}

For the toy model discussed in this section we considered four different freely variable parameters: $\lambda$, which determines the frequency of the oscillatory factor in the amplitude, $\mu$, which determines the widths of the Gaussian implementing the constraint, as well as the integration range and the lattice constant for the discretization.   In the case of spin foams the Barbero-Immirzi parameter $\gamma$ determines the discretization. The frequency of the oscillatory part of the amplitude is determined by $\gamma$, the values of the boundary spins $j$, and the curvature $\epsilon$ they induce. The summation region is also determined by the boundary spins, as is the variance for the (approximately) Gaussian factors, which implement the constraints.  In section \ref{SFExpV}, we will consider examples where all three parameters, $j$, $\gamma$, and the curvature $\epsilon$, vary.

\section{ The effective spin foam model}\label{Sec-EffSFM}

The effective spin foam model was introduced, in  \cite{EffSFM}, by the authors. This model provides a dynamics for loop quantum gravity, in the form of transition amplitudes for states defined on the (kinematical) Hilbert space of loop quantum gravity. One aim of the model \cite{EffSFM} is to provide the simplest possible construction of amplitudes that implement a key feature of loop quantum gravity, namely, discrete area spectra \cite{DiscreteGeom}, and that still admits a semiclassical regime with the corresponding classical dynamics of (discrete) gravity.

More generally, the model provides a  path integral description for quantum gravity, by summing over geometric data associated to a (fixed) triangulation.  The geometric data are  given by areas, associated to the triangles, and 3D dihedral angles, associated to the tetrahedra, of the triangulation. These data need to satisfy certain constraints in order to define a piecewise flat geometry for the triangulation. These constraints, and the way they are implemented, are a crucial feature of spin foams.

The loop quantum gravity framework allows a rigorous construction of area operators. These turn out to have discrete spectra \cite{DiscreteGeom} given by
\ba
A(j) =\gamma \ell_{P}^2\sqrt{j(j+1)} \sim \gamma \ell_{P}^2 j \ , 
\ea
where $\gamma$ is the dimensionless Barbero-Immirzi parameter, $\ell_P=\sqrt{8\pi\hbar G/c^3}$ is the Planck length and $j$ are positive half-integers (if one uses $\SU(2)$ as internal symmetry group rotating the  triads defining the spatial metric) or positive integers (if one uses $\text{SO}(3)$ instead).  The $j$'s are often referred to as spins.\footnote{Other geometrical operators, such as the volume of a spatial region \cite{DiscreteGeom,BianchiHaggard1} and the length of edges \cite{BianchiLength} also have discrete spectra described in terms of the spins $j$ in LQG, but we will not be employing these operators here.}  In this work we are interested in the regime of large spin and we will therefore approximate $\sqrt{j(j+1)}$  by $j$ and we will also choose to work with integer $j$'s only.  Note, however, that one could just as well choose the $\sqrt{j(j+1)}$ form of the spectrum and half-integers for the effective spin foam models. Importantly, the area eigenvalues for each triangle can be chosen independently. 

To implement a discrete area spectrum, it is convenient to work with areas as basic variables and to sum, in the path integral, over spin values associated to the triangulation. In fact, the Regge action \cite{Regge} constitutes a discretization of the Einstein-Hilbert action (with the Gibbons-Hawking-York boundary term) for piecewise flat geometries, and can be expressed in terms of areas:
\ba \label{ARaction}
S_{AR}&=& \frac{c^3}{8\pi G} \left( \sum_t n_t \pi A_t - \sum_\sigma \sum_{t\subset \sigma} A_t  \Theta^\sigma_t(A)\right) \ ,
\ea
where $n_t=1$ for boundary triangles and $n_t=2$ for bulk triangles. The variables $\Theta^\sigma_t(A)=\Theta^\sigma_t(L^\sigma(A))$ are 4D (internal) dihedral angles in a 4-simplex $\sigma$, as determined by its area values.\footnote{See, e.g., the appendix of \cite{DFS07} for how to compute the dihedral angles as functions of the lengths of the simplex.}  These are computed using the fact that, for a given 4-simplex, one can locally invert the areas for the lengths, resulting in functions $L^\sigma(A)$. (There are global ambiguities, which we are suppressing here. These can be resolved by using the 3D dihedral angles that we introduce below.) 

Freely varying the area Regge action (\ref{ARaction}) with respect to the bulk areas one finds that the deficit angles,
\ba
\epsilon_t:=2\pi-\sum_{t\supset \sigma}  \Theta^\sigma_t(A) \stackrel{!}{=}0 \ , 
\ea
which are a measure of curvature, must vanish. (This follows from the Schl\"afli identity $\sum_{t\subset \sigma} A_t \delta \Theta_t^\sigma =0$, which ensures that the variations of the angles cancel each other \cite{Schlafli:1858, HaggardSch:2015}.)  Instead, the Regge action is usually formulated in terms of length variables. There are (typically) more triangles than edges in a 4D triangulation and by treating the areas as independent variables we get more restrictive equations of motion,  which, in the case of the area Regge action, demand flatness. 

To obtain the equations of motion of length Regge calculus, we have to constrain the areas such that they describe configurations with a consistent assignment  of lengths to the edges of the triangulation. Indeed, starting with a free assignment of areas to the triangles, and inverting, within each 4-simplex, the areas for the lengths, one will typically find that the length values for a given edge, as calculated from the different neighboring 4-simplices, do not coincide. 
To specify the constraints on the area variables, which ensure a consistent length assignment, we consider 
the gluing of two 4-simplices along a shared tetrahedron. The tetrahedron's geometry is uniquely specified by its six edge lengths. Working with area variables specifies only four quantities for this tetrahedron. 
 We thus need two more variables. Choosing two 3D dihedral angles $\phi^\tau_{i},i=1,2$ at non-opposite edges will fix the lengths in the tetrahedron  $\tau$ uniquely.\footnote{One can also work with all six 3D dihedral angles as variables, but then has to add so-called closure constraints (resulting from the internal rotation symmetry) that reduce these six angles to two variables.} (A proof can be found in \cite{AreaAngle,ADHAreaR}.) 
 
 Adding further variables seems to make the problem of having too many variables even worse.  But, we can now easily express the constraints that ensure that the 4-simplices glue properly to each other, that is, that the edge lengths in the shared tetrahedron are consistently determined by the areas of the 4-simplex. For this we just need to impose that the two (and therefore all) 3D angles in the shared tetrahedron are fixed  to the same values as those induced by the areas of the neighbouring 4-simplices:
\ba\label{Constraint1}
{\cal  C}^{\tau,\sigma}_i = \phi^\tau_{i}-\Phi^{\tau,\sigma}_i(A) \stackrel{!}{=}0 \ .
\ea
Here the $\Phi^{\tau,\sigma}_i(A)=\Phi^{\tau}_i(L^\sigma(A))$ are the 3D dihedral angles in the tetrahedron $\tau$ as determined by the areas of the 4-simplex $\sigma$.  Demanding (\ref{Constraint1}) implies that, for a bulk tetrahedron shared by two 4-simplices $\sigma$ and $\sigma'$, we have
\ba\label{Constraint2}
{\frak  C}^\tau_i =\Phi^{\tau,\sigma}_i(A) -\Phi^{\tau,\sigma'}_i(A)\stackrel{!}{=}0 \ ,
\ea
that is, that the two pairs of tetrahedral dihedral angles, as determined by the areas of the simplices $\sigma$ and $\sigma'$, respectively, agree. 

The difference between version (\ref{Constraint1}) and (\ref{Constraint2}), is that (\ref{Constraint1}) involves only the data of one 4-simplex $\sigma$, whereas (\ref{Constraint2}) involves the data of the two simplices sharing $\tau$. That is, introducing the 3D angles as additional variables, we can express the constraints more locally. 

As the 3D angles appear only in the constraints (\ref{Constraint1})---and not in the action (\ref{ARaction})---they can be integrated out and lead to the constraints (\ref{Constraint2}). This can only be achieved, however, for the angles in the bulk tetrahedra, as the 3D angles in the boundary are part of the boundary data. 

There is an alternative---and more local---version of the constraints (\ref{Constraint1}), known as gluing or shape matching constraints \cite{AreaAngle,DittrichRyan,Twisted}. Here one demands that pairs of tetrahedra glue properly along their shared triangle. That is, given the areas $A_t$ and dihedral angles $\phi^\tau_i,\phi^{\tau'}_i$ of two neighbouring tetrdrahedra $\tau,\tau'$, we can determine the 2D angles $\alpha^\tau_{t,v}$ and $\alpha^{\tau'}_{t,v}$ in the shared triangle \cite{ADHAreaR}. The shape matching constraints demand that the angles, as computed from tetrahedron $\tau$ and $\tau'$, agree
\ba\label{Constraint3}
\alpha^\tau_{t,v}(A_t,\phi^\tau_i) \stackrel{!}{=}\alpha^{\tau'}_{t,v}(A_t,\phi^{\tau'}_i) \ .
\ea
This imposes that  the geometry and, in particular, the shape of the triangles, as computed from $\tau$ and $\tau'$, respectively, match. In a given 4D triangulation, it is sufficient to demand the constraints  (\ref{Constraint3}) for any pair of tetrahedra $\{\tau,\tau'\} \subset \sigma$ in a given 4-simplex $\sigma$, see \cite{AreaAngle}. The constraints (\ref{Constraint3}), for a given simplex $\sigma$, imply the constraints (\ref{Constraint1}), and vice versa. In this sense, the constraints (\ref{Constraint1}) and (\ref{Constraint3}) are equivalent.  The constraints (\ref{Constraint1}) can also be understood more directly as shape matching constraints, as they guarantee that the shapes of a given tetrahedron, as determined by the areas of the two neighbouring simplices, agree. 

The constraints (\ref{Constraint1}) are linear in the 3D angles, which we use as basic variables, and, therefore, are much easier to implement than the version (\ref{Constraint3}). On the other hand, (\ref{Constraint3}) provides the more local formulation: a given constraint only involves a pair of tetrahedra, and can, thus, be applied to the 3D boundary of a 4D triangulation. Restricted to such a 3D boundary these constraints   specify the boundary data---given by the areas and the two 3D angles per boundary tetrahedron---that define a piecewise flat geometry for the boundary triangulation.

The 3D dihedral angles are also part of the loop quantum gravity phase space (gauge reduced by the internal rotation group) adapted to a triangulation \cite{DittrichRyan}. There are two independent 3D dihedral angles $\phi^\tau_{i},i=1,2$ (associated to non-opposite edges) per tetrahedron. These Poisson commute with the areas but are conjugated to each other (see e.g. \cite{Barbieri,DittrichRyan,EffSFM}):
\ba\label{PB1}
\hbar \{ \phi^\tau_{1}, \phi^\tau_{2}\}=\pm \ell_{P}^2 \gamma \frac{\sin \alpha^{t,\tau}_v}{A_t}\,=\, \pm \ell_{P}^2 \gamma \frac{2}{ l_1 l_2} \ ,
\ea
where $ \alpha^{t,\tau}_v$ is the 2D angle between the two edges carrying the 3D dihedral angles, $A_t$ is the area spanned by these two edges, and $l_1$ and $l_2$ are the lengths of these two edges.

This means that the two constraints associated to a given tetrahedron (\ref{Constraint1})  also do not commute
\ba\label{PB2}
\hbar \{ {\cal C}^\tau_{1}, {\cal C}^\tau_{2}\}=\pm \ell_{P}^2 \gamma \frac{\sin \alpha^{t,\tau}_v}{A_t} 
\ea
and, are, more precisely, second class. This means that they cannot be imposed strongly on the loop quantum gravity Hilbert space: there do not exist sufficiently many states satisfying $\hat {\cal C}^\tau_{1} \psi =\hat {\cal C}^\tau_{2}\psi=0$. (There might be such states, but for those states the commutator (\ref{PB2}) must also vanish.)

This issue can also be understood from a less technical viewpoint \cite{EffSFM}: the constraints (\ref{Constraint2}), if expressed in the discretized areas $A_t\sim \gamma \ell_{P}^2 j_t$, constitute diophantine equations for the (half-) integer-valued spins $j_t$. There will be only very few exact solutions, typically for configurations that have special symmetries. This leads to a drastic reduction in the density of states \cite{EffSFM}, which prevents a suitable semiclassical limit.

Spin foam models deal with this issue by implementing the (second class part of the) constraints weakly \cite{NSFM}.  We follow the same route  and, to this end, employ so-called intertwiner coherent states  \cite{Coherent}. This allows us to implement the constraints as strongly as possible. The intertwiner coherent states are diagonal in the areas and  are peaked on the conjugated pairs of 3D angle values $\Phi^\tau_i$. For a given tetrahedron $\tau$ we will denote such states by ${\cal K}_\tau(\phi^\tau_i; \Phi^\tau_i)$. Here $\phi^\tau_i,\, i=1,2$ denote the arguments of the wave function ${\cal K}_\tau$ and $\Phi^\tau_i,\, i=1,2$ the angular values on which the coherent state is peaked.\footnote{As in the Segal-Bargmann representation we have two arguments encoding one quantum degree of freedom.}  The coherent states come  with a measure $d\mu^\tau_{\cal K}$, which we will use for the integration over the angles in the path integral. 

To construct the gravitational path integral, we will associate a coherent state ${\cal K}_\tau(\phi^\tau_i; \Phi^{\tau,\sigma}_i(A))$ (or, depending on the  orientation of the tetrahedron, its complex conjugate) to each pair $\tau \subset \sigma$ of a 4-simplex and sub-tetrahedron. Here $\Phi^{\tau,\sigma}_i(A)$ are the 3D dihedral angles in $\tau$ as determined by the areas of the 4-simplex $\sigma$. Each bulk tetrahedron $\tau$ will, therefore, carry two such coherent states, which arise from the two simplices $\sigma, \sigma'$ that share it.  The angle variables $\phi^\tau_i$ only appear in those two coherent states. Integrating  out these angles gives a factor
\ba\label{Gfactor}
G^{\sigma,\sigma'}_\tau(A)=\langle {\cal K}_\tau(\cdot, \Phi^{\tau,\sigma}_i(A)) \,|\, {\cal K}_\tau(\cdot, \Phi^{\tau,\sigma'}_i(A))\rangle\sim \exp\left( -\frac{ ({\frak C}^\tau_1)^2+({\frak C}^\tau_2)^2 }{ 4 \Sigma^2}\right) \ ,
\ea
 which depends on the areas in the simplex $\sigma$ and the simplex $\sigma'$. These $G$-factors are, as inner products of coherent states, approximately Gaussian in the constraints ${\frak  C}^\tau_i$ with squared deviation $\Sigma^2$ determined by the Poisson brackets (\ref{PB2}), that is,\footnote{Note that $\Sigma^2$ defines the variance in the 3D angle variables. To obtain the variance for the spin variables one has to first transform the Gaussian to these variables.}
\ba\label{Gfactor2}
\Sigma^2 = \ell_{P}^2 \gamma \frac{2}{l_1l_2}\,=\,  \ell_{P}^2 \gamma \frac{\sin \alpha^{t,\tau}_v}{A_t} \,=\, \frac{\sin \alpha^{t,\tau}_v}{j_t}\ .
\ea
This quantity can be defined from either the geometric data of the simplex $\sigma$ or the geometric data of the simplex $\sigma'$. The constraints ensure that these geometries are approximately equal. In order to obtain a more symmetric expression we take the average  $\Sigma^2=\tfrac{1}{2}(\Sigma^2(\sigma)+\Sigma^2(\sigma'))$.

For the calculation of expectation values in section \ref{SFExpV} we are interested in the semiclassical regime. Hence, it is safe to approximate the inner product (\ref{Gfactor}) between the coherent states with its Gaussian form, as defined on the right hand side of (\ref{Gfactor}). 

Having understood how to implement the constraints weakly in the path integral, we can now define the path integral itself.  We define it as 
\ba\label{DefEffSM}
{\cal Z}=\sum_{j_t} \mu(j) \int \prod_\tau d\mu^\tau_{\cal K}(\phi) \,\, \exp\left( \frac{\i}{\hbar} S_{AR}(j) \right) 
\prod_\sigma \Theta_{\sigma}[\text{Tri}(j)]
\prod_{(\tau \subset \sigma)}  {\cal K}^\pm_\tau(\phi^\tau, \Phi^{\tau,\sigma}_i(j) ) \ , 
\ea
where $\Theta_{\sigma}[\text{Tri}]=1$ if the generalized triangle inequalities (see Appendix \ref{GTI}) are satisfied, and is vanishing otherwise; this is like a step function for the generalized triangle allowed region.  The $\pm$ super-index for the coherent state ${\cal K}^\pm_\tau$ indicates whether to take the complex conjugate or not, depending on the orientation of $\tau$. We included a general measure term $\mu(j)$, which can be fixed by, e.g., studying the coarse graining flow of this model \cite{PImeasure}. Here we will set this measure term to $\mu(j)=1$.\footnote{Assuming a saddle point approximation can be applied, and there is only one critical point, the measure term drops out for expectation values.}   We sum in the partition function (\ref{DefEffSM}) only over positively oriented 4-simplices. We can easily include a sum over both orientations, by replacing $\exp(\i/\hbar S_{AR})$ with $\cos(\i/\hbar S_{AR})$ in (\ref{DefEffSM}). Indeed, effective spin foams can be used to study the effect of including---or not---a sum over orientations into a quantum gravity partition function \cite{Engle,3DHol}.

Due to the additive structure of the action (\ref{ARaction}), the amplitude factorizes over simplices and triangles, with factors
\ba
{\cal A}_t=  \exp\left( \i \gamma n_t \pi j_t\right) \, ,\ \ \text{and} \ \  {\cal A}_\sigma =\exp\left( -\i \gamma   \sum_{t\subset \sigma}  j_t \Theta^\sigma_t(j) \right) \Theta_\sigma(\text{Tri})\prod_{\tau \subset \sigma} {\cal K}^\pm_\tau(\phi^\tau, \Phi^{\tau,\sigma}_i(j)) \ ,
\ea 
as is typical for spin foams.  Here $n_t=1$ for boundary triangles and $n_t=2$ for bulk triangles.

This factorization no longer holds after integrating out the dihedral angles from the bulk tetrahedra: the factors $G^{\sigma,\sigma'}_\tau(A)$, defined in (\ref{Gfactor}), depend on the areas of the two neighbouring simplices $\sigma, \sigma'$ and do not factorize.

Integrating out the angles from the bulk tetrahedra still leaves us with the coherent state factors ${\cal K}_\tau$ for the boundary tetrahedra. Here, we will assume boundary states that are diagonal in the areas and include, for each tetrahedron, a coherent state ${\cal K}^\pm_\tau(\cdot, \Phi^\tau_i)$ peaked on some  boundary values  for the pair of 3D dihedral angles.  In this way we can also integrate out the 3D dihedral angles for the boundary tetrahedra, which results in boundary  $G$-factors peaked on $\Phi^{\tau,\sigma}_i(j)=\Phi^\tau_i$.  We will assume  boundary data $A_t$ and $\Phi^\tau_i$, which satisfy the (3D) shape matching constraints (\ref{Constraint3}). If this were not the case, we would obtain an  exponentially suppressed amplitude (as measured by the square of the constraint).

\section{Triangulation with a bulk edge}\label{Sec-Triang}

Here we will describe in more detail  the triangulation, for which we will calculate various expectation values.

The triangulation consist of six simplices glued around a bulk edge $e(12)$ bounded by vertices $v_1$ and $v_2$. The triangulation has seven vertices $v_1,\ldots, v_7$, which are all in the boundary. The simplices of the triangulation  are given by
\ba\label{triang1}
\{(12346),(12356),(12456),(12347),(12357),(12457)\}\ .
\ea
To simplify the calculations, we will set certain edge lengths equal, that is, we apply a symmetry reduction. To this end,  it is convenient to introduce the indices $m\in\{1,2\}$ and $i,j\in\{3,4,5\}$ as well as $k\in \{6,7\}$. The six simplices (\ref{triang1}) are then of the form $(12ijk)$, and with our symmetry reduction we will have the same geometry for the three simplices $(12ij6)$ (referred to as simplices of type I) and the three simplices $(12ij7)$ (referred to as simplices of type II), respectively.

  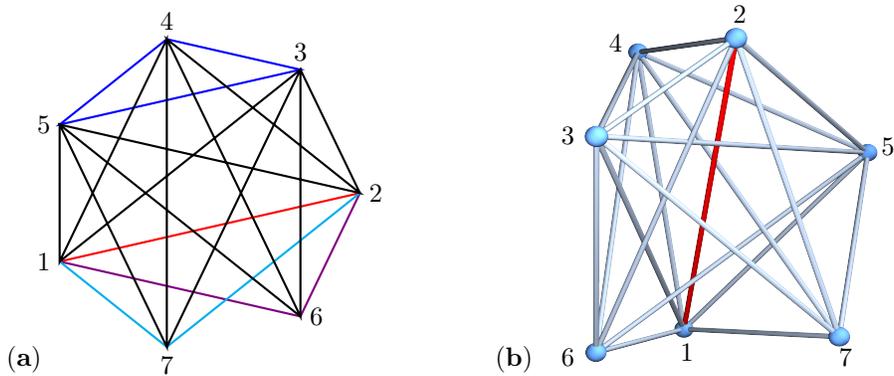
\begin{figure}[ht!]
\begin{picture}(350,140)
\newdimen\R
\R=2.1cm

\put(10,0){ \begin{tikzpicture}[scale=1.]
       \path (51.4286:\R) coordinate (p3) node[right, above] {3} ; 
       \path (2*51.4286:\R) coordinate (p4)  node[right, above] {4} ; 
       \path (3*51.4286:\R) coordinate (p5)  node[left] {5} ;  
       \path (4*51.4286:\R) coordinate (p1)  node[left] {1} ; 
       \path (5*51.4286:\R) coordinate (p7)  node[below] {7} ; 
       \path (6*51.4286:\R) coordinate (p6)  node[right] {6} ;
       \path (7*51.4286:\R) coordinate (p2)  node[right] {2} ;     
       
       \draw[thick,red]  (p1)--(p2) ;
       \draw[thick,blue]  (p3)--(p4)--(p5)--cycle ;
       \draw[thick,violet] (p1)--(p6)--(p2) ;
       \draw[thick,cyan] (p1)--(p7)--(p2) ;
       \draw[thick] (p1)--(p3)--(p6)--(p4)--(p2)--(p5)--(p7)--(p4)--(p1)--(p5)--(p6) (p2)--(p3)--(p7) ;
       
       \end{tikzpicture} }

\put(220,8){ \includegraphics[scale=0.21]{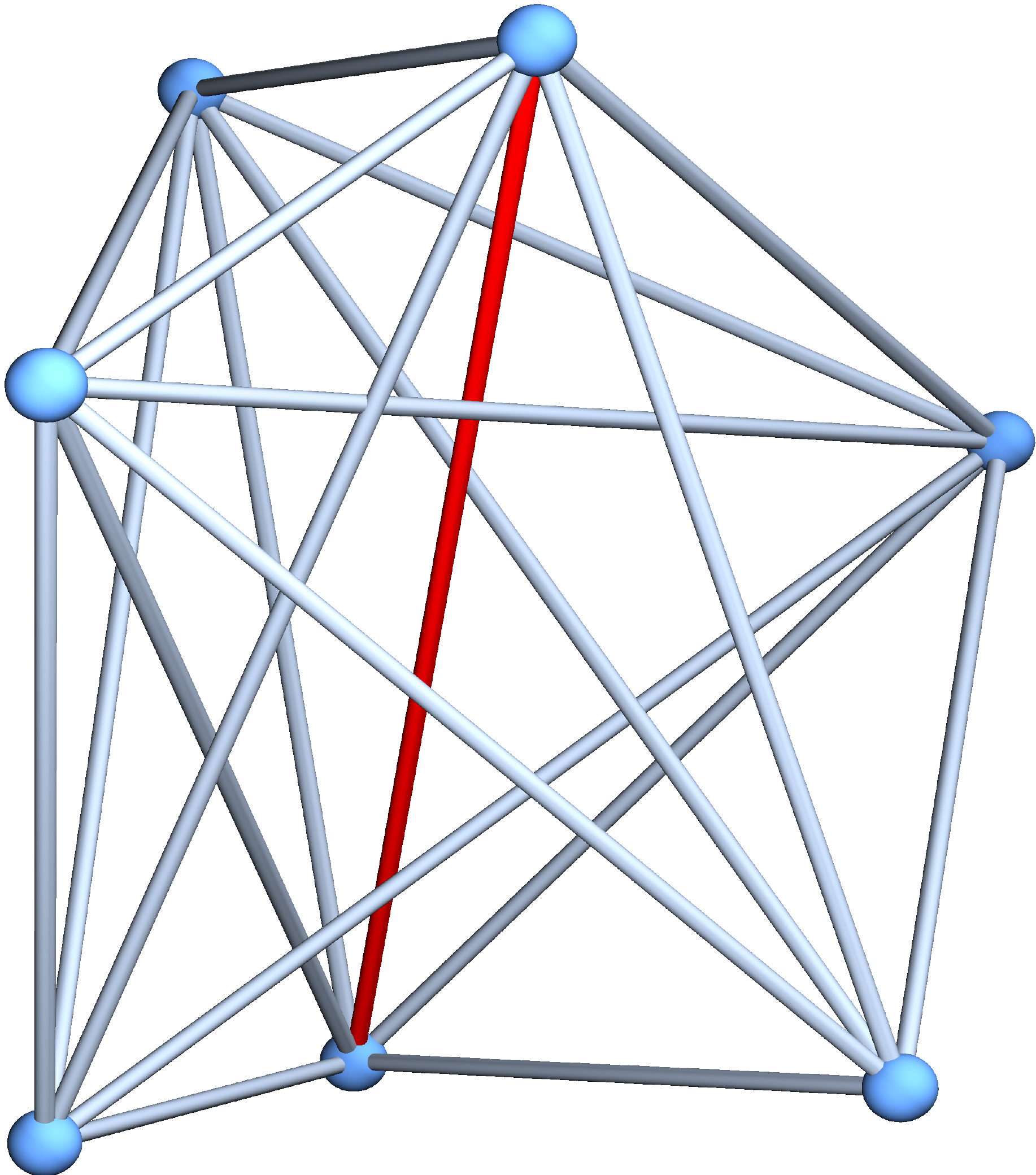} }

\put(5,6){({\bf a})}
\put(190,6){({\bf b})}

\put(260,8){1}
\put(280,136){2}
\put(215,6){6}
\put(320,6){7}

\put(215,90){3}
\put(233,125){4}
\put(336,86){5}

\end{picture}
\caption{\label{Triangulation} A complex made of six simplices  sharing the bulk edge $e(12)$ with length $t$ (the red line in both panels). In panel ({\bf a}) the boundary edges are colored according to their lengths. The edges in black, blue, violet and cyan  have lengths $a,b,c$, and $d$, respectively. Panel ({\bf b}) is a 3D projection of the complex. }
 \end{figure}

The reduced edge lengths and  resulting area parameters  are summarized in Table \ref{SymmReduc}. The triangulation has five length parameters and six area parameters. Amongst the length parameters only one, $l(12)=t$, is a bulk variable, whereas there are three bulk area variables $A_1,A_2$, and $A_3$. 

 \begin{table}[h]
\caption{\label{SymmReduc} The five length and six area parameters used in the symmetry reduced triangulation described in this section and depicted in Fig. \ref{Triangulation}. The symbol $\stackrel{\cdot}{=}$ is used to indicate an equation that holds if the geometricity constraints between the areas are satisfied and $A(x,y,z)$  denotes the area of a triangle with edge lengths $x,y,z$.}
\begin{ruledtabular}
\begin{tabular}{ll}
lengths & areas\phantom{$\Big|$}  \\
\hline
    $l(12) = t$  & $A(mij)=A(kij) = B_1\stackrel{\cdot}{=}A(a,a,b)$\\
    $l(mi)=l(ik)=a$ & $A(im6)=B_2\stackrel{\cdot}{=}A(a,a,c)$ \\
$l(ij)=b$ & $A(im7)=B_3\stackrel{\cdot}{=}A(a,a,d)$ \\
$l(m6)=c$ &  $A(12i)=A_1\stackrel{\cdot}{=}A(a,a,t)$\\
$l(m7)=d$ & $A(126)=A_2\stackrel{\cdot}{=}A(c,c,t)$\\
	& $A(127)=A_3\stackrel{\cdot}{=}A(d,d,t)$
\end{tabular}
\end{ruledtabular}
\end{table}

%
%

Note, that for a given simplex we have four length parameters  and four area parameters and thus the same number of lengths and areas.  For simplices of type I these are $\{t,a,b,c\}$ and $\{B_1,B_2,A_1,A_2\}$, and for simplices of type II they are $\{t,a,b,d\}$ and $\{B_1,B_3,A_1,A_3\}$.   Were we not to have implemented a symmetry reduction, we would still have had the same number of area and length parameters per simplex, namely, 10 areas and 10 lengths. Working with the symmetry reduced case simplifies the process of solving for the lengths in terms of the areas on a given simplex. In fact, we are able to solve these equations analytically, but the resulting expressions are involved enough to not be so useful. The system leads to eight  roots, four of which lead to positive length solutions. We explain the structure of the roots in more detail in Appendix \ref{sec-roots}. 

The triangulation has one bulk edge with length $t$. Varying the (length) Regge action with respect to this length we obtain one equation of motion
\ba\label{EOMt}
3\tfrac{\partial A_1}{\partial t} \,\epsilon_{1} + \tfrac{\partial A_2}{\partial t} \,\epsilon_{2} + \tfrac{\partial A_3}{\partial t}\, \epsilon_{3} \stackrel{!}{=}0 \ ,
\ea
where with $\epsilon_{i}$ we denote the deficit angle at the triangle with area $A_i$. 

\bigskip

We also need to discuss the boundary and bulk tetrahedra and the associated $G$-factors. 

The boundary consists of 12 tetrahedra. Six of these are of type I, those with $(mij6)$, and six are of type II, with $(mij7)$.  The type I tetrahedra have edge lengths $(a,a,a,a,b,c)$ with the $b$- and $c$- edges opposite.  For our 3D angle variables we choose the dihedral angles at two of the non-opposite edges with length $a$. The dihedral angles at the $a$-edges all agree as functions of the lengths $(a,a,a,a,b,c)$ of the tetrahedron and are given by
\ba\label{phi1}
\Phi^{\text{I}}\stackrel{\cdot}{=}\arccos\left(    \frac{ c^2 b^2 }{\sqrt{(4 a^2 b^2-b^4)( 4 a^2 c^2-c^4) }} \right) \ .
\ea
To obtain the dihedral angle as function of the areas of a given simplex, we use the solutions for the lengths as function of the areas for this simplex.  In this manner we get one Gaussian peaked on the area configurations where this 3D angle is equal to the angle as determined by the boundary data, e.g., the boundary lengths,  $\Phi^{\text{I}}_{\rm bdry}(a,b,c)$. This Gaussian is
\ba\label{GI}
G^{\text{I}}=\exp\left( - \frac{6}{2 \Sigma^2}  ( \Phi^{\text{I}}(A)-\Phi^{\text{I}}_{\rm bdry}(a,b,c) )^2 \right) \ .
\ea
Here $\Sigma^2= \ell^2_{\rm P} \gamma(1/(a^\text{I}(A))^2+1/a^2)$, where $a^\text{I}(A)$ is the length of the $a$-edges as computed from the areas of a simplex of type I. In (\ref{GI}) we have a factor of $6/2$ appearing (instead of the factor $1/4$ in (\ref{Gfactor})). This is because there are six boundary tetrahedra of type I and because the constraints $\frak{C}^\tau_1$ and $\frak{C}^\tau_2$ appearing in (\ref{Gfactor}) coincide in our symmetry reduction.

We get the $G$-factor for the type II tetrahedra in the same way, by replacing the edge length $c$ with $d$.

There are nine additional bulk tetrahedra.  Three of these have $(12i6)$, and are shared between simplices of type I. As the simplices of type I have all the same geometry, the associated shape matching constraints will be automatically satisfied. The same holds for the three tetrahedra with $(12i7)$, which are shared between simplices of type II. 

The remaining three bulk tetrahedra have vertices $(12ij)$ and are shared between one simplex of type I and one simplex of type II. These are subject to an area constraint, which is not automatically satisfied. The tetrahedra have edge lengths $(a,a,a,a,b,t)$ with the $t$- and $b$- edges opposite. We can again choose as angle variables the  (coinciding) 3D dihedral angles at two non-opposite $a$-edges. The 3D dihedral angles $\Phi^{\text{B,I}}$  and $\Phi^{\text{B,II}}$  at the $a$-edges are, as functions of the tetrahedral edge lengths,  given as in (\ref{phi1}), just that $c$ needs to be replaced with $t$.  
The Gaussian factor coming from the bulk tetrahedra is given by
\ba
G^{\rm B}=\exp\left( - \frac{3}{2 \Sigma^2}  ( \Phi^{\text{B,I}}-\Phi^{\text{B,II}} )^2 \right) \ .
\ea
with $\Sigma^2 = \ell_P^2\gamma (1/(a^\text{I}(A))^2 + 1/(a^\text{II}(A))^2)$. 

Overall we have three area constraints, which are implemented weakly by the Gaussian factors.  Classically, the constraints reduce the three bulk area variables to only one independent variable, which can be parametrized by the bulk length $t$.  There is furthermore one constraint on the boundary data---here given by a set of three area parameters $B_1,B_2,B_3$ and  two dihedral angles $\Phi_{\rm bdry}^{\text{I}}$ and $\Phi_{\rm bdry}^{\text{II}}$. If these data  satisfy the (boundary) shape matching constraint (\ref{Constraint3}), then they uniquely define the boundary lengths $(a,b,c,d)$.

\section{Computation of expectation values}\label{SFExpV}

In section \ref{toymodel}  we discussed the circumstances in which a stationary phase approximation does not reliably identify a semiclassical regime for spin foams. This is, on the one hand, due to the weak implementation of the constraints, but also impacted by having a finite range of summation and discrete summations rather than continuous integrations.

In this section we will test more directly whether the expectation values of certain observables reproduce the classical solutions.  To this end we will consider boundary values for which there is only one classical solution.\footnote{This is generally the case for our triangulation. There are, however, special boundary values that have, e.g., two solutions.} 
 We will compute the expectation values for different values of three parameters: (i) for different values of the Barbero-Immirzi parameter $\gamma$; (ii) for different scales, that is, for different multiples of a given set of boundary areas;  and (iii) for boundary data leading to different classical values for the bulk curvature angles.  This will allow us to identify a regime for these three parameters in which the expectation values approximate well the classical values, and which one can, therefore, consider to be semiclassical. 
 
 In the next subsection \ref{DefObs} we will define the expectation values we will be computing for the effective spin foams. In subsection \ref{S-Curv} to \ref{L-Curv} we present in detail the results of the computations for three different choices of boundary data,  leading to a low curvature, an intermediate curvature, and a large curvature solution, respectively.  We summarize the main features of these results in subsection \ref{SummExs}. 
 
 \subsection{Definition of the observables}\label{DefObs}
We note that the observables considered here are not diffeomorphism invariant or Dirac observables in the sense of, e.g., \cite{Ditt05}. Indeed, diffeomorphism symmetry is generically broken in 4D discrete gravity, including Regge gravity, due to the presence of curvature \cite{BahrDittrich09a}. In addition, violations of the shape matching constraints can also lead to a breaking of diffeomorphism symmetry, even for flat configurations \cite{ADHAreaR}.  

Here we aim to compare expectation values of the effective spin foam model with classical solutions of Regge calculus. We can therefore  use functions of the configuration variables, which can be defined in the Regge calculus setup, as observables. This includes areas and deficit angles of specific triangles in the triangulation. This will be sufficient for our purposes; a more sophisticated approach will be needed, however, if one aims to compare spin foam observables with those in continuum general relativity.

For the definition of the expectation values we must also specify boundary states. We will choose boundary states that are eigenstates of the area operators and are coherent with respect to the 3D angle degrees of freedom, in other words, the intertwiners associated to the boundary tetrahedra \cite{Coherent}.  The coherent states are peaked on angles satisfying the shape matching constraints. This allows us to match our boundary data with the data of (length) Regge calculus.

 Choosing the same coherent states as for the definition of the effective spin foam model in (\ref{DefEffSM}), we can formally integrate out the 3D angles in the bulk and the boundary tetrahedra. This results, as in (\ref{Gfactor}), in inner products between the coherent states, which we approximate here by  Gaussian factors. We specified these Gaussian factors in section \ref{Sec-Triang}.

The expectation value of an observable ${\cal O}$ is then given by 
\ba\label{ExpV1}
\langle {\cal O}\rangle =
\frac{
\sum_{j_{A_1},j_{A_2},j_{A_3}} \exp\left( \frac{\i}{\hbar} S_{AR} \right)\,\, \tristep{\text{I}} \, \tristep{\text{II}}  \,\, G^{\text{I}}\, G^{\text{II}}  G^\text{B} \,\, {\cal O}
}
{
\sum_{j_{A_1},j_{A_2},j_{A_3}} \exp\left( \frac{\i}{\hbar} S_{AR} \right)\,\, \tristep{\text{I}} \, \tristep{\text{II}} \,\, G^{\text{I}}\, G^{\text{II}}  G^\text{B} \,\,
} \ ,
\ea
where the action $S_{AR}$, the factors implementing the triangle inequalities $\tristep{}$, and the Gaussian factors $G^{\text{I}},G^{\text{II}}$ and $G^\text{B}$, have been introduced in section \ref{Sec-EffSFM} and \ref{Sec-Triang}.  As we discussed in section \ref{Sec-EffSFM} we sum over integer spins $j_{A_i}>0$.

The summations of Eq. \eqref{ExpV1} are, at the outset, over an unbounded range of spins $j_{A_1},j_{A_2},j_{A_3}>0$. The triangle inequalities reduce this to a finite range, and a further, quite drastic reduction for large spins, appears through the Gaussian factors, which implement the constraints weakly. Indeed, classically, the constraints reduce the three bulk area parameters to just one bulk length parameter.
The summation over several variables is the most time consuming step in the numerics, and so it is advantageous to first identify the range of spins where $(a)$ the triangle inequalities are satisfied, and $(b)$ the product of the Gaussian factors is above a certain threshold value. Appendix \ref{speed} gives more details on this and other aspects of the numerical implementation.


\subsection{Examples with small curvature angles}\label{S-Curv}

In this section we consider boundary values for which the solution has very small deficit angles. The deficit angles, boundary lengths and areas, bulk areas, and scales are summarized in table \ref{LowCurv}.

 \begin{table}[h]
\caption{\label{LowCurv} The lengths, areas, and deficit angles for the small curvature examples. The parameters are defined in Sec.~\ref{Sec-Triang}, Table~\ref{SymmReduc}. Here we will consider scales $\lambda=\numlam \gamma \ell_P^2$ with $\numlam \in \{1,10,20,40,80\}$.}
\begin{ruledtabular}
\begin{tabular}{lll}
lengths & areas\phantom{$\Big|$} & deficit angles \\
\hline
       $a= 5.590\sqrt{\lambda}$ & $B_1= 15\lambda$ &$\epsilon^{\rm sol}_1=0$      \\
        $b= 8.944\sqrt{\lambda}$     &  $B_2=B_3= 8\lambda$ & $ \epsilon^{\rm sol}_2= 0.034$      \\
       $c=2.969\sqrt{\lambda}$  & $A_1^{\rm sol}=10.87\lambda$ & $\epsilon^{\rm sol}_3  =0.034$      \\
       							 $d=2.969\sqrt{\lambda}$&  $A_2^{\rm sol}=4.407\lambda$ & \\
              $t_{\rm sol}=4.198\sqrt{\lambda}$  & $A_3^{\rm sol}= 4.407\lambda$ & \\
\end{tabular}
\end{ruledtabular}
\end{table}

For these solutions we have the special situation that $c=d$. Thus, the simplices of type I and type II have the same geometry, and the expectation values of $A_2$ and $A_3$, as well as those for $\epsilon_2$ and $\epsilon_3$, agree. We have $\epsilon^{\rm sol}_2=\epsilon^{\rm sol}_3$ and $\epsilon^{\rm sol}_1=0$. Looking at the equation of motion (\ref{EOMt}), one might wonder whether it can be satisfied. It can be, and this is achieved with $\partial A(c,c,t)/\partial t= \partial A(d,d,t)/\partial t=0$ for $t=t_{\rm sol}$; geometrically, these conditions mean  that the triangles are right-angled at the solution. 

This solution is a special member of a family of solutions, where $\epsilon_{1}(t_{\rm sol})=0$ and $\epsilon_{2}(t_{\rm sol})=\epsilon_{3}(t_{\rm sol})$, but with, in general, $c\neq d$. In these cases one has $\partial A(c,c,t)/\partial t= -\partial A(d,d,t)/\partial t \neq 0$, so the equation of motion (\ref{EOMt}) is satisfied, and the triangles $(c,c,t)$ and $(d,d,t)$ are not right-angled. The examples studied in subsection \ref{M-Curv} are of this type.

We will see that, within a certain $\numlam$-dependent range for $\gamma$, the expectation values match well for the area $A_1^{\rm sol}$ and the deficit angle $\epsilon_1^{\rm sol}$. However, the expectation values do tend to under-estimate $A_2^{\rm sol}$ and  $\epsilon_2^{\rm sol}$. The matching for these expectation values improves with growing $\numlam$.

\begin{figure}[ht!]
\begin{picture}(500,150)
\put(160,7){\includegraphics[scale=0.4]{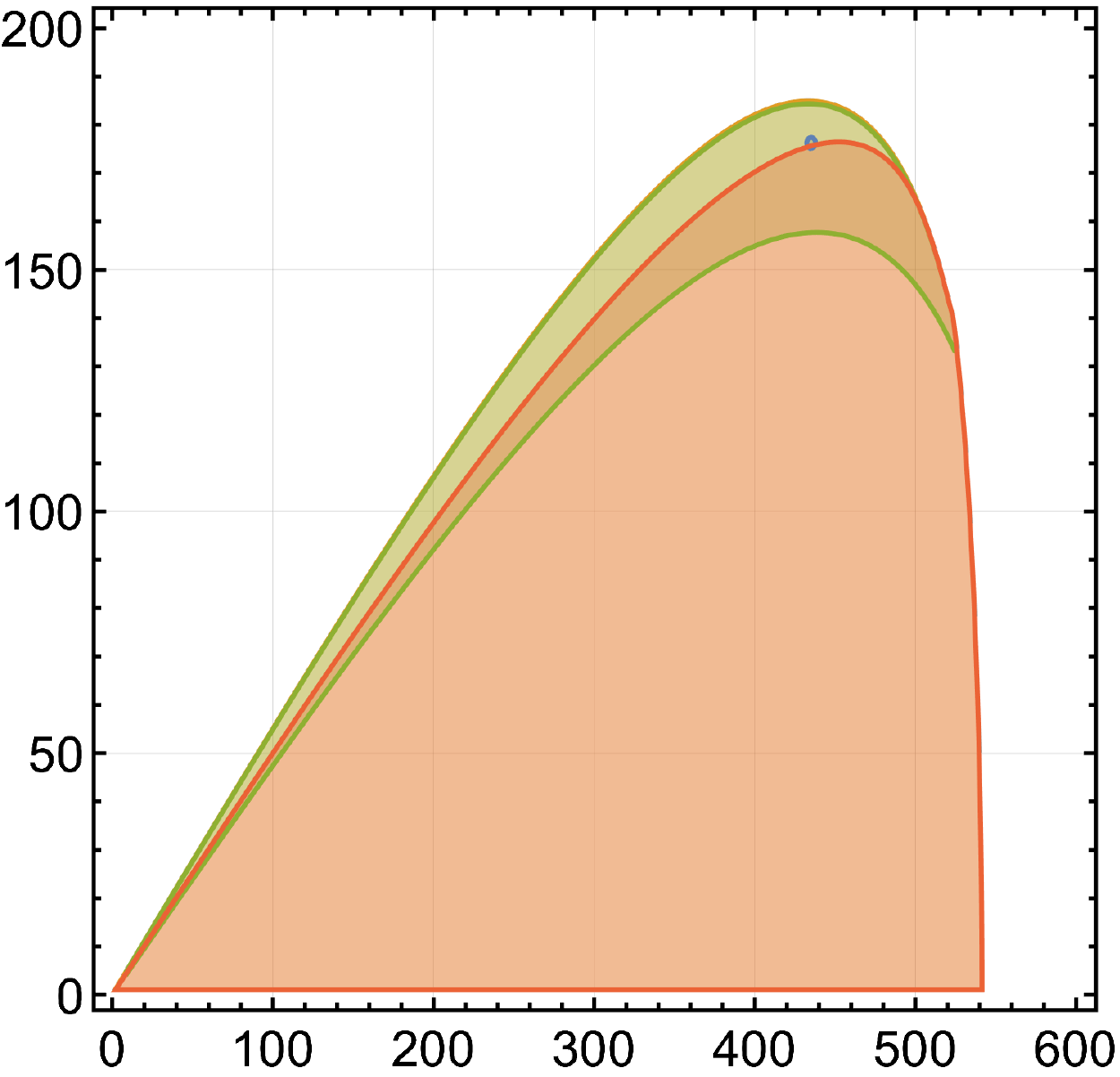}}
\put(290,0){\footnotesize $A_1$}
\put(148,130){\rotatebox{90}{\footnotesize $A_2$}}
\end{picture}
\caption{The $(A_1,A_2)$ plane of the configuration space for scaling $\numlam=40$. The union of the green (upper) and red (lower) shaded regions indicates the range of the bulk areas, $A_1$ and $A_2$, for simplices of type I, where the generalized triangle inequalities are satisfied. The areas are given in units of $\gamma \ell_P^2$. The green shaded region is picked out as those values for which  $G^\text{I}$  is larger than $10^{-3}$. The deficit angle $\epsilon_2$ evaluates to a negative number for values in the red shaded region. The (very small) blue circle encloses the classical solution. 
 \label{Plot001}}
\end{figure}

The reason for this is that the classical solution, $(A_1^{\rm sol},A_2^{\rm sol})$, lies quite near the boundary of the region where the triangle inequalities are satisfied, see Fig. \ref{Plot001}. 
In particular $A_2^{\rm sol}$ gives almost the maximally allowed value for $A_2$. In fact, for $\numlam=1$, we have $A_2^{\rm sol}$ being larger than the maximum of the allowed values for $A_2=j \times\gamma \ell^2_p$ (with $j$ integer). 

Similarly, we have that $\epsilon_2<0$ for almost the entire set of bulk areas $A_1$ and $A_2$, allowed by the triangle inequalities, see Fig. \ref{Plot001}. 
Thus, for small $\numlam$ there are only very few (or no) discrete area values that lead to positive $\epsilon_2$ and are allowed by the triangle inequalities. Increasing $\numlam$ more and more allowed values are possible and a better matching of the expectation value with the area values of the classical solution results.

At the same time, the $G$-factors suppress a growing portion of the region allowed by the triangle inequalities for growing $\numlam$, and here, in particular, the portion where $\epsilon_2$ (and $\epsilon_3$) are negative, see Fig \ref{Plot001}.  This is because the triangle allowed region grows with $\numlam$, whereas the deviation $\Sigma(j)$ for the $G$-factors grows only with $\sqrt{\numlam}$, see (\ref{Gfactor2}). The mismatch of the expectation values with the classical solution essentially disappears for sufficiently large  $\numlam$. The presence of this mismatch for smaller $\numlam$  is a result---at least in this example---of the weak imposition of the constraints.

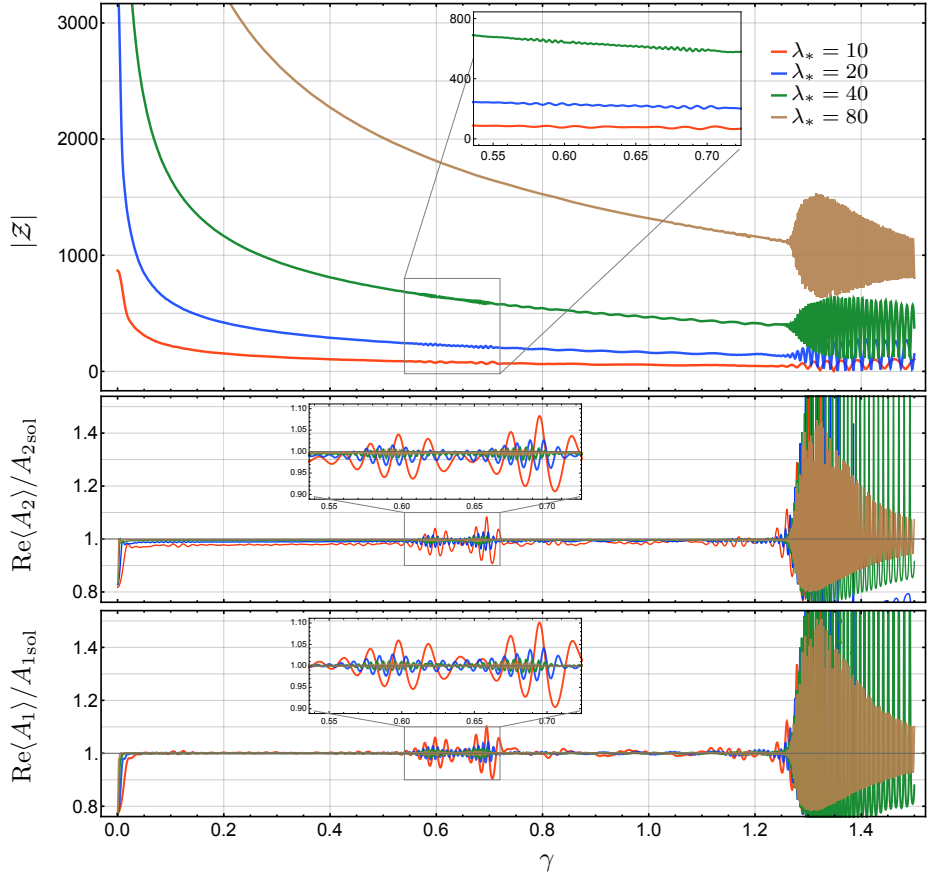
\begin{figure}[h!]
\begin{picture}(500,330)
\put(60,7){ \includegraphics[scale=0.317]{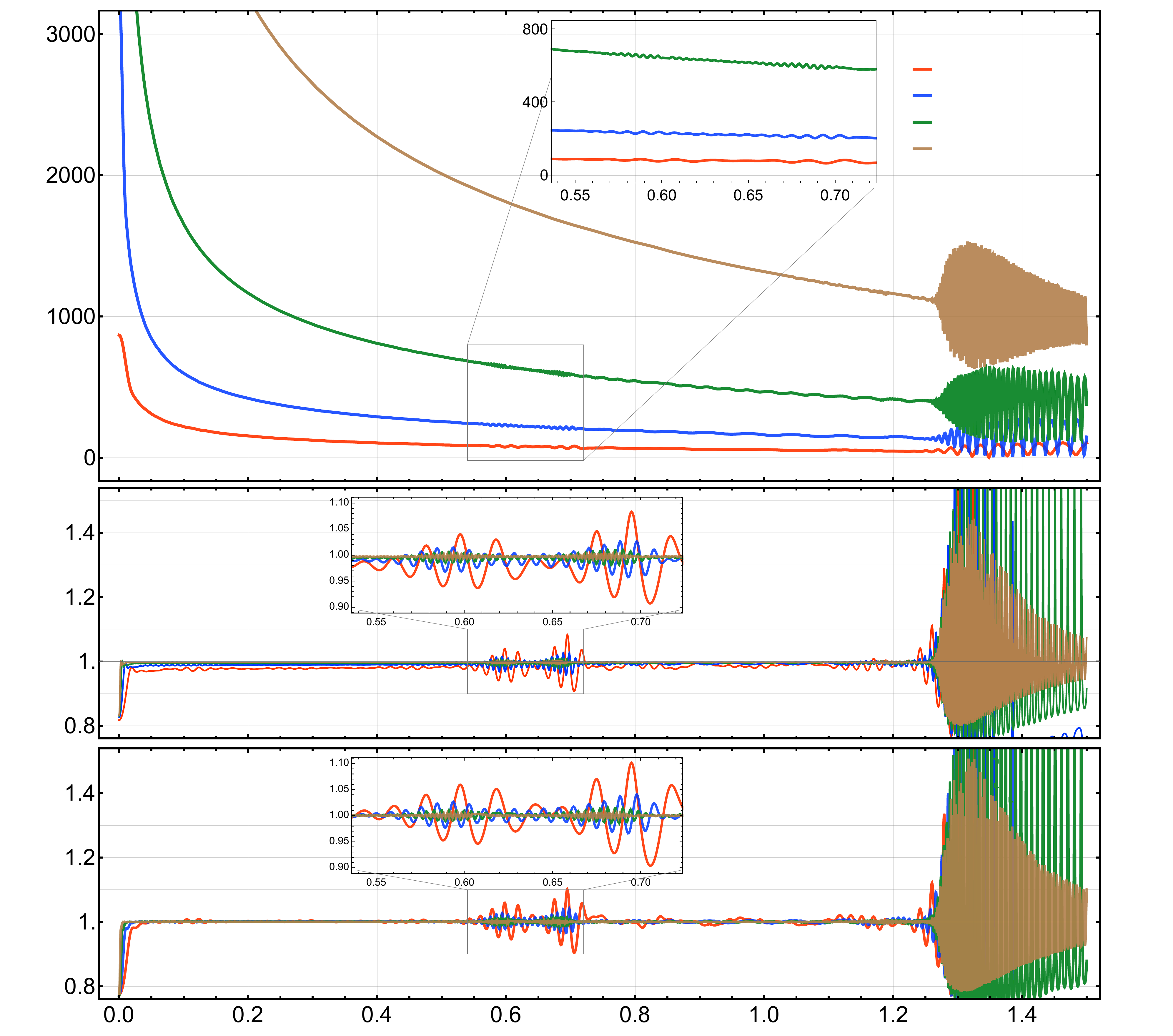} }

\put(60, 230){\rotatebox{90}{  $| {\cal Z} | $ }}
\put(60, 108){\rotatebox{90}{  $ {\rm Re} \langle A_2 \rangle / { {A_2}_{\rm sol} }  $ }}
\put(60, 28){\rotatebox{90}{  $ {\rm Re} \langle A_1 \rangle / { {A_1}_{\rm sol} }  $ }}

\put(260,0){$\gamma$}

\put(352, 306){ \scriptsize   $\numlam = 10 $ }
\put(352, 298){ \scriptsize   $\numlam = 20 $ }
\put(352, 290){ \scriptsize   $\numlam = 40 $ }
\put(352, 281){ \scriptsize   $\numlam = 80 $ }

\end{picture}
\caption{Absolute value of the partition function and normalized expectation values for the areas $A_1$ and $A_2$ for boundary values leading to a very small curvature solution. 
\label{Fig6} }
\end{figure}

To explain some features of the expectation values it is helpful to consider first the behavior of the partition function as a function of the Barbero-Immirzi parameter $\gamma$. The sums appearing in (\ref{ExpV1}) were computed keeping the $\gamma$-parameter general, and consequently, the absolute value of the partition function and the expectation values can be plotted as continuous functions of $\gamma$. Fig. \ref{Fig6} shows these functions for $\numlam\in\{10,20,40,80\}$. For a certain region of $\gamma$ these functions oscillate rapidly and the curves can appear to be thickened into a region.  (The phase shows extremely rapid oscillations over the entire range $0 < \gamma<1.5$.)\\

The absolute values for the partition functions appear to decrease monotonically  up through $\gamma \sim 0.55$. There is then an oscillatory behavior with relatively high frequency and small amplitude up through $\gamma \sim 0.72$.  The value for the onset of these oscillations seems to change very little with scale $\numlam$, whereas the frequency of these oscillations (as a function of $\gamma$) increases and the amplitude decreases with $\numlam$. In fact, for $\numlam=80$ the oscillations are no longer noticeable.  For $\gamma\sim0.72$ to $\gamma\sim 1.25$  oscillations of a much lower frequency and smaller amplitude than in the previous regime appear.

Starting near $\gamma\sim 1.25$ we have fast oscillations, which grow rapidly in amplitude until reaching a somewhat stable value. Again, the onset of these oscillations does not seem to depend on the scale $\numlam$, but the frequency and amplitude increase with $\numlam$. Due to these oscillations the absolute values of the partition function reach relatively small values repeatedly, and the smaller $\numlam$, the closer the values come to zero.

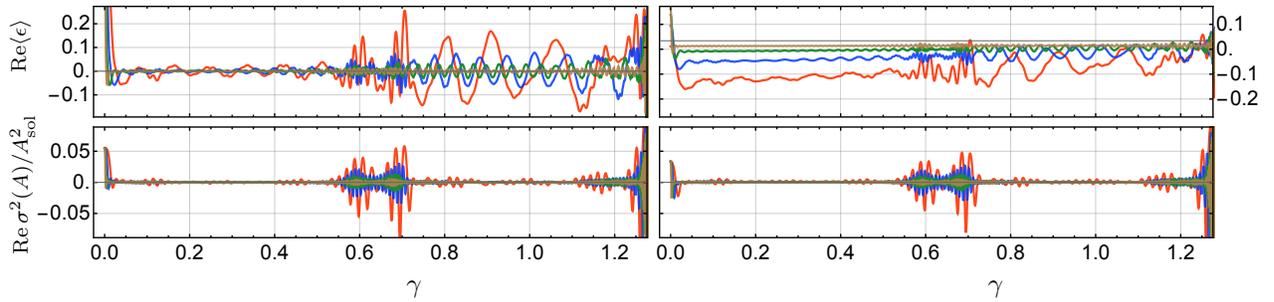
\begin{figure}[ht!]
\centering
\begin{picture}(550,120)
\put(10,7){\includegraphics[scale = 0.43]{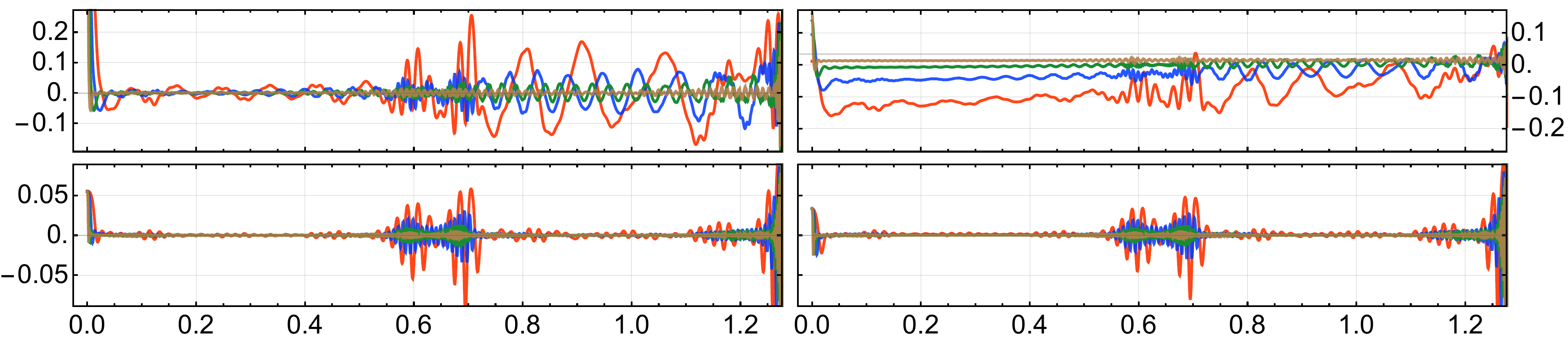}}

\put(150,0){$\gamma$}
\put(370,0){$\gamma$}

\put(0,77){\rotatebox{90}{ \scriptsize { Re $\!\! \langle \epsilon \rangle$}}}
\put(0,11){\rotatebox{90}{ \scriptsize  Re $\!\sigma^2(A)/A_{\rm sol}^2$ }}

\end{picture}
\caption{At left, for the triangle with area $A_1$, and at right, for the triangle with area $A_2$,  the  real parts of the deficit angle expectation values (top) and the normalized variances of the areas (bottom) for $\numlam=10$ (red), $\numlam=20$ (blue), $\numlam=40$ (green) and $\numlam=80$ (brown).  
\label{Fig7}}
\end{figure}

We can see in Fig.~\ref{Fig6}  and Fig.~\ref{Fig7}  that these oscillations with $\gamma$ in the absolute value of the partition function  are reflected in the oscillatory behavior of the expectation values. 
In particular, for the region where the partition function reaches relatively small values, the expectation values show oscillations with relatively large amplitude. The amplitudes of the oscillations are particularly large when the absolute values of the partition function nearly reaches zero, which tends to happen for the smaller scale examples.
 Thus,  for $\gamma > 1.25$ we have the onset of a regime where close matching with the classical solution is not reliable.  Contrary to expectations coming from the discussion in section \ref{toymodel}, the $\gamma$-value for this onset does not seem to depend  on the scale $\numlam$. We will see that this behavior is specific to the small curvature case.

Fig.~\ref{Fig7}  shows the  real parts of the expectation values for  the deficit angles $\epsilon_1$ and $\epsilon_2$, as well as the $A_1$ and $A_2$ variances for $\numlam \in \{10,20,40,80\}$.  Note that  the area expectation values  have been normalized by their classical values, and the area-variances by $(A_1^{\rm sol})^2$ and $(A_2^{\rm sol})^2$, respectively. We computed the expectation values for  general values   $0\leq \gamma \leq 2$, but  the plots show only $0\leq \gamma\leq 1.3$. 

In general, we can distinguish various regimes for the behavior of these expectation values:
\begin{itemize}
\item Regime I  ($\gamma \lesssim 0.05$): Here $\gamma$ is too small (and correspondingly the oscillations of the amplitude in the partition function too slow) to lead to a reliable semiclassical regime.  The onset of a reliable regime moves to smaller  $\gamma$ for growing $\numlam$, and for $\numlam=10$ it is around $\gamma\sim 0.05$.
\item Regime II ($0.05\lesssim\gamma \lesssim 0.55$):  For this interval of $\gamma$ the expectation values show the most stable behavior and the best matching to the classical values. We have oscillations appearing throughout this regime but with very small amplitude, which decreases with increasing $\numlam$; e.g., for $\numlam=10$ the $A_1$ expectation value 
remains within an error interval of $\pm 0.01 \times A_1^{\rm sol}$, whereas for $\numlam=40$ the $A_1$ expectation value  remains within an error interval of $\pm 0.002 \times A_1^{\rm sol}$. For $A_2$ and $\epsilon_2$ we have, resulting from the mechanism described above,  a `systematic' mismatch of expectation values and classical solutions. The (relative) error of this mismatch decreases with  increasing $\numlam$: for $\numlam=10$, the $A_2$ expectation value oscillates around a value of about $0.97 A_2^{\rm sol}$, whereas for $\numlam=80$ we have $\langle A_2\rangle$ between $0.995 \times A_2^{\rm sol}$ and $1\times A_2^{\rm sol}$.
\item Regime III ($0.55\lesssim\gamma \lesssim 0.72$): Here we have a regime where the oscillations come with a larger amplitude as compared to Regime II. E.g., for $A_1$ and $\numlam=10$ the maximal deviation of the expectation value is around $\pm 0.1  \times A_1^{\rm sol}$ whereas for $\numlam =40$ it is around $\pm 0.02  \times A_1^{\rm sol}$. 
\item Regime IV ($0.72\lesssim\gamma \lesssim 1.25$): The expectation values for the areas show oscillations with an amplitude smaller than in Regime III, but larger than in Regime II. However, the expectation values for the deficit angles oscillate with almost the same amplitude as in Regime III. 

\item Regime V ($1.25\lesssim\gamma \leq 2$): (The upper value is the maximal value of $\gamma$ for which we computed the expectation values.)  In this regime we have an onset of oscillations of very large amplitude. The amplitude decreases with $\numlam$, but is much larger than in the preceding regimes: for $A_1$ and $\numlam=40$ the maximal deviation of the expectation value from the classical value is between $2\times  A_1^{\rm sol}$ and $3\times  A_1^{\rm sol}$
 and for $\numlam=80$ it is around $1\times A_1^{\rm sol}$. These  oscillations mirror those in the absolute value of the partition function, and one reason for their large amplitude is that these absolute values almost reach zero (and the minima for these absolute values are smaller for smaller $\numlam$). 
\end{itemize}

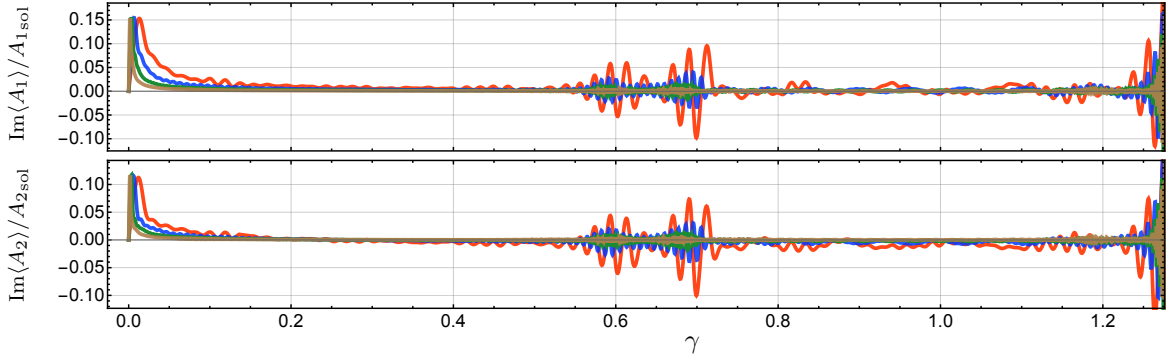
\begin{figure}[ht!]
\centering
\begin{picture}(550,120)
\put(12,2){\includegraphics[scale = 0.38]{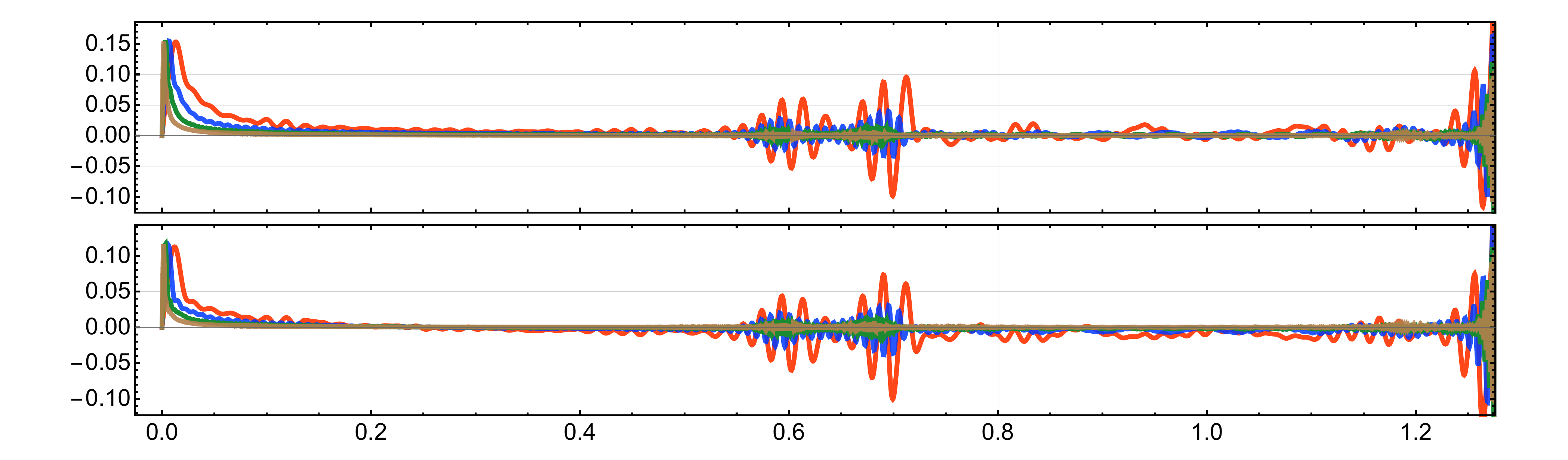}}

\put(270,0){$\gamma$}

\put(14,75){\rotatebox{90}{ \scriptsize  Im $\!\! \langle A_1 \rangle/{A_1}_{\rm sol}$ }}
\put(14,14){\rotatebox{90}{ \scriptsize  Im $\!\! \langle A_2 \rangle/{A_2}_{\rm sol}$ }}
\end{picture}
\caption{Imaginary parts of expectation values for the  areas. \label{Fig8}}
\end{figure}

The expectation values all have a relatively small imaginary part, see Fig. \ref{Fig8} for the imaginary parts of the normalized expectation values for the areas. The oscillatory behavior of the imaginary parts is similar to those of the real parts. Thus, as for the real parts, the (relative) size of the oscillations in the imaginary parts decreases with growing $\numlam$.

\vspace{0.5cm}

In summary, Regime II can be considered a reliable semiclassical regime for $\numlam\geq10$. Whether one accepts Regime III and IV as semiclassical depends on what one considers as an acceptable mismatch from the classical values, and on the scale $\numlam$. For example, for $\numlam=10$ the mismatch for $A_1$ in Regime III is around 10\% whereas for $\numlam=80$ it is within one percent. 

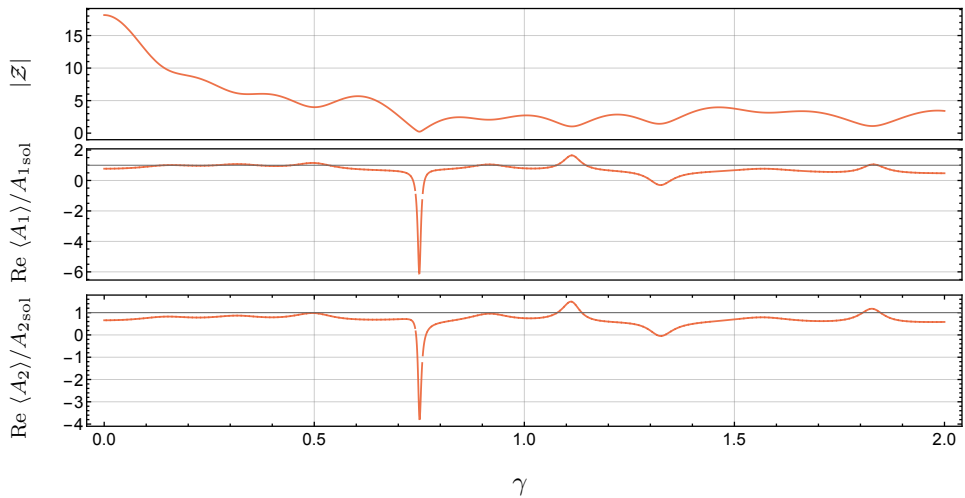
\begin{figure}[ht!]
\centering
\begin{picture}(350,185)
\put(10,7){\includegraphics[scale = 0.425]{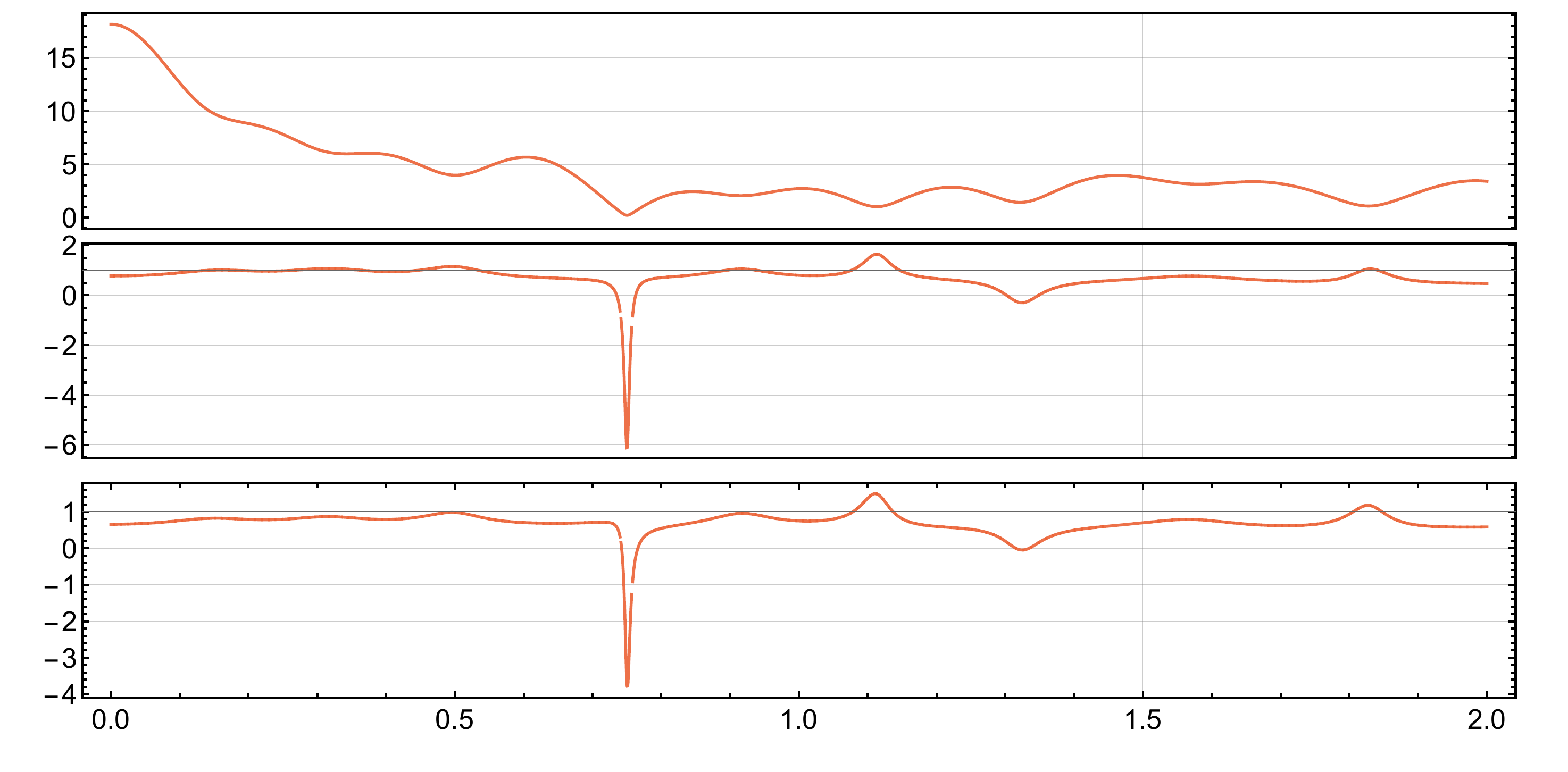}}

\put(190,0){$\gamma$}

\put(0,146){\rotatebox{90}{ \scriptsize {  $| \cal Z| $}}}
\put(0,72){\rotatebox{90}{ \scriptsize { Re $\langle A_1 \rangle/{A_1}_{\rm sol} $}}}
\put(0,13){\rotatebox{90}{ \scriptsize { Re $\langle A_2 \rangle/{A_2}_{\rm sol} $}}}

\end{picture}
\caption{The panels show the absolute value of the partition function and expectation values for the areas $A_1$ and $A_2$ for the scale $\numlam=1$. 
\label{FigL1} }
\end{figure}

We have also calculated expectation values for $\numlam=1$, see Fig. \ref{FigL1}. Here we do not see the same regimes as described above. Nonetheless, for this example we see that the absolute value of the partition function is a good indicator for the behavior of the expectation values. The absolute value of the partition function goes almost to zero for $\gamma\approx 0.75$ and for these value we see large mismatches. Given that this example is at a very small scale one might consider the expectation values for the interval between $\gamma\approx 0.15$ and $\gamma \approx 0.5$ as acceptable approximations---at least for $A_1$, which does not suffer from the mechanism described at the start of this section. 

In this $\numlam=1$ example, the number of area configurations is basically determined by the triangle inequalities, as the Gaussian $G$-factors are quite spread over the entire region of bulk areas allowed by the triangle inequalities.  For our computations we excluded configurations from the partition function for which the product of the $G$-functions gave a value smaller than $10^{-10}$, see Appendix \ref{speed} for more details.   For $\numlam=1$ this does not exclude any value allowed by the triangle inequalities and the partition function is a sum over only 117 area triplets  $(A_1,A_2,A_3)$. By contrast for $\numlam=20$ we have a sum over $191,475$ terms.


\subsection{Examples with intermediate curvature angles}\label{M-Curv}

Next we consider a family of examples  with intermediate curvature angles. The deficit angles, boundary lengths and areas, bulk areas, and scales are summarized in table \ref{MedCurv}.

 \begin{table}[h]
\caption{\label{MedCurv}   The lengths, areas, and deficit angles for the intermediate curvature examples. The parameters are defined in Sec.~\ref{Sec-Triang}, Table~\ref{SymmReduc}. Here we will consider scales $\lambda=\numlam \gamma \ell_P^2$ with $\numlam \in \{10,20,40\}$.}
\begin{ruledtabular}
\begin{tabular}{lll}
lengths & areas\phantom{$\Big|$} & deficit angles \\
\hline
       $a= 5.590\sqrt{\lambda}$ & $B_1= 15\lambda$ &$\epsilon^{\rm sol}_1=0$      \\
        $b= 8.944\sqrt{\lambda}$     &  $B_2= 11\lambda$ & $ \epsilon^{\rm sol}_2= -1.0405$      \\
       $c=4.256\sqrt{\lambda}$  & $B_3=10.5\lambda$ & $\epsilon^{\rm sol}_3  =-1.0405$      \\
       							 $d=4.027\sqrt{\lambda}$&  $A_1^{\rm sol}=13.934\lambda$ & \\
              $t_{\rm sol}=1.850\sqrt{\lambda}$  & $A_2^{\rm sol}= 9.042\lambda$, \quad $A_3^{\rm sol} = 8.095 \lambda$ & \\
\end{tabular}
\end{ruledtabular}
\end{table}

%
%

\begin{figure}[ht!]
\begin{picture}(550,260)
\put(60,7) {\includegraphics[scale=0.31]{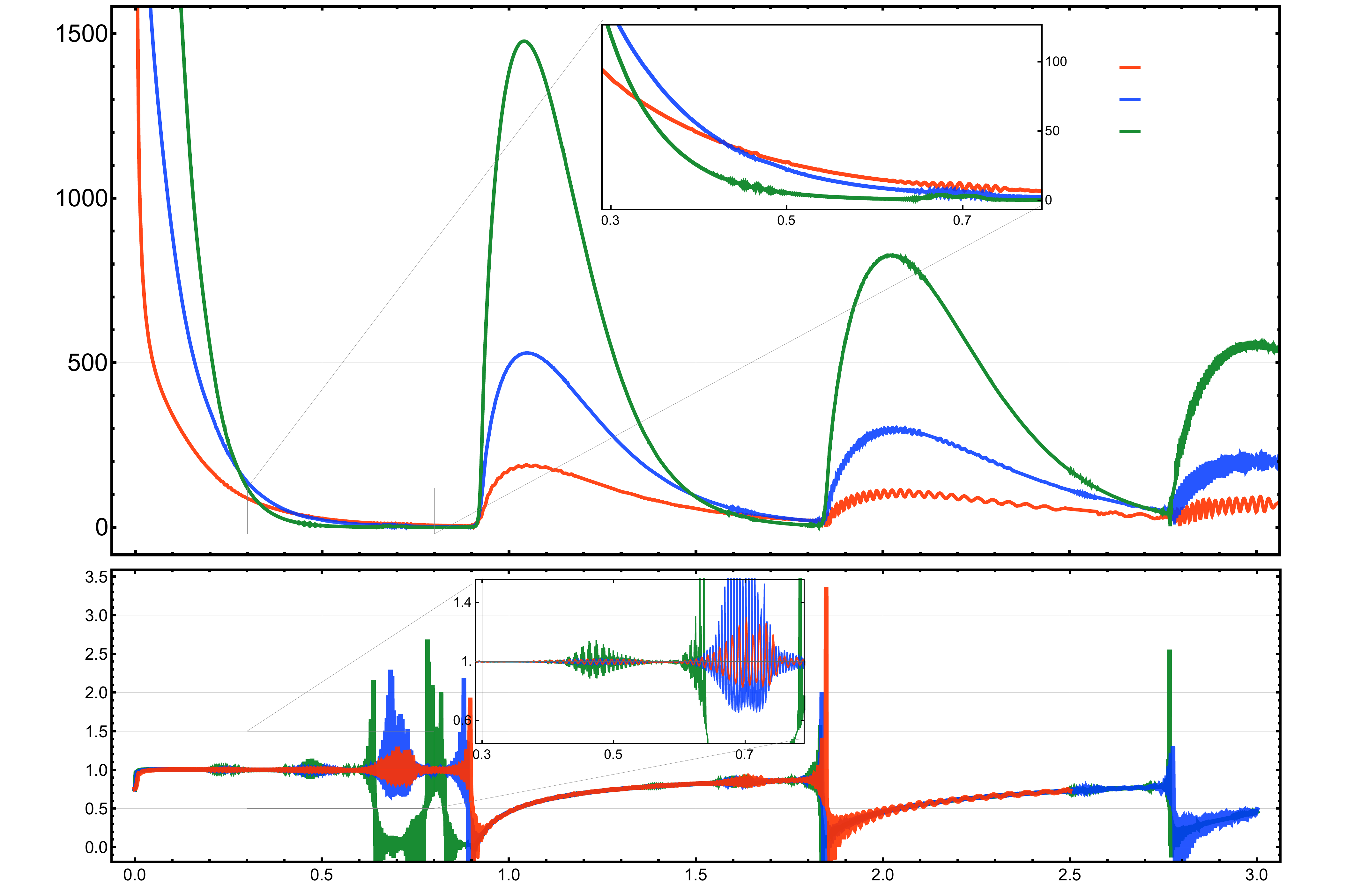} }
\put(240,0) {$\gamma$} 

\put(55,170){ \rotatebox{90}{ $| { \cal Z} |$} }
\put(55,30){ \rotatebox{90}{ Re$\langle A_1\rangle / {A_1}_{\rm sol} $ } }

\put(382,239) {\scriptsize $\numlam = 10$}
\put(382,229) {\scriptsize $\numlam = 20$}
\put(382,219) {\scriptsize $\numlam = 40$}

\end{picture}
\caption{Absolute value of the partition function and real part of the expectation value for $A_1$ for the example with medium size curvature angles.  \label{Fig9}
}
\end{figure}
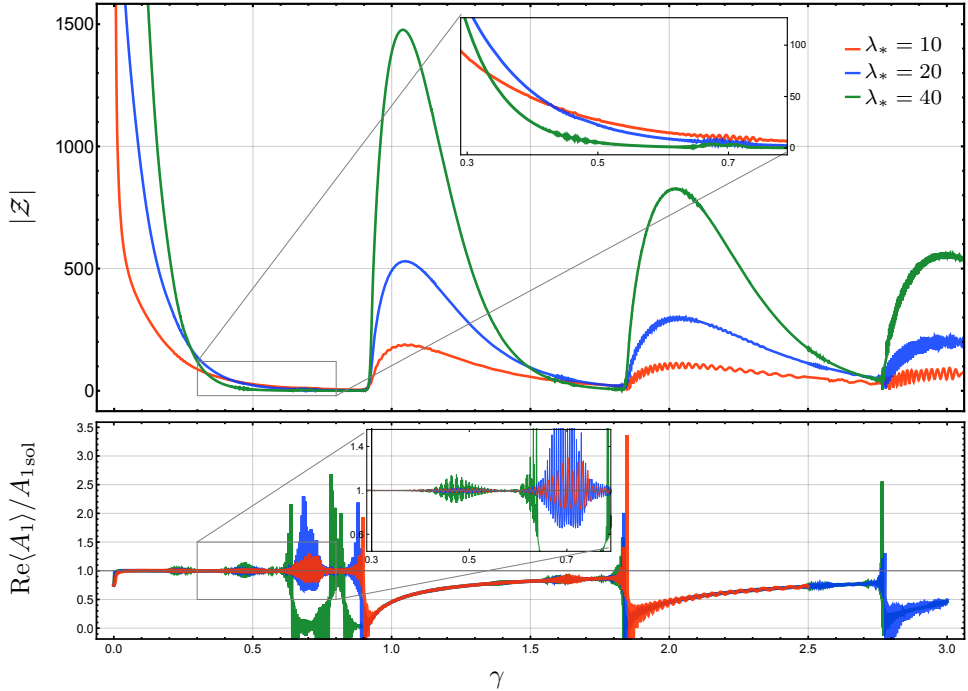

The behavior of the expectation values as a function of the Barbero-Immirzi parameter $\gamma$ can again be deduced from the behavior of the absolute value of the partition function. Fig. \ref{Fig9} shows the absolute value of the partition functions, as well as the expectation value for $A_1$ for a range $0<\gamma<3.0$. The other expectation values show a very similar behavior.

We see that the relatively small oscillations in the absolute value of the partition functions are reflected in, at some places, quite large oscillations in the expectation value. The amplitudes of the oscillations in the expectation values are particularly large if the absolute values of the partition function are near zero. 

The smallest value for which this happens is around $\gamma \sim0.6$, and that is where the expectation values start to oscillate with a relatively large amplitude. Beyond this value of $\gamma$ the expectation values mismatch the classical solutions by a considerable amount, even for regions where the absolute values of the partition functions are quite large.
 In the following we will therefore only discuss the range $0<\gamma<0.9$.

 Figure \ref{Fig9} shows also a zoom in of the range of $0.3<\gamma<0.8$.    Here the behavior is roughly similar to that of the small curvature example over the range $0<\gamma<2$ in subsection \ref{S-Curv}. But there is an important difference: In the small curvature example the norms of the partition functions were generally larger for larger scale $\numlam$. Thus, in the regimes III and IV,  the norms tended to approach nearer to zero for smaller values of $\numlam$. Accordingly, the oscillations in the expectation values tended to be larger for smaller $\numlam$. 
By contrast, in the example here, the relative size of the norms changes with $\gamma$: for $\gamma<0.2$ the ordering is as in the small curvature case, but at around $\gamma\sim 0.44$ the order is inverted and the oscillations in the expectation values tend to be larger for larger values of $\numlam$,   starting with $\gamma \approx 0.44$. This means that, here, the argumentation of Eq. (\ref{SCcond}) applies, and having a large scale reduces the range of $\gamma$ for which the expectation values approximate the classical solution well.

Roughly we have the following regimes for the current example:
\begin{itemize}
\item Regime I  ($\gamma \lesssim 0.05$): The $\gamma$ is too small (and correspondingly the oscillations of the amplitude in the path integral too slow) to lead to a reliable semiclassical regime.  The onset of a reliable regime moves to smaller values of $\gamma$ for growing $\numlam$, and for $\numlam=10$ it is around $\gamma\sim 0.05$.
\item Regime II ($0.05\lesssim\gamma \lesssim 0.18$):  For this interval of $\gamma$, the expectation values show the most stable behavior and the best matching to the classical values. We have oscillations appearing throughout this regime but with very small amplitude.
\item Regime III ($0.18\lesssim\gamma \lesssim 0.3$): Here we have an interval with slightly larger oscillations---for the expectation value of $A_1$ these remain below 2\%.  The amplitudes of these oscillations are almost the same size for the different scales.

\item Regime IV ($0.3\lesssim\gamma \lesssim 0.38$): The expectation values show oscillations with an amplitude  much smaller than in Regime III. 

\item Regime V ($0.38\lesssim\gamma \leq 0.6$): The amplitudes of the oscillations are larger than in Regime III, and their size depends on $\numlam$. The amplitudes for $\numlam=10$ and $\numlam=20$ are roughly of the same size---reflecting the fact that the norm of the partition functions for this interval of $\gamma$ are approximately the same. Compared to these, the amplitudes for $\numlam=40$ are much larger, as the norm for the $\numlam=40$ partition function is much closer to zero. The maximal mismatch of the expectation value for $A_1$ from the classical value is approximately 15\%. 

\item Regime VI ($0.6\lesssim\gamma \leq 0.9$): The expectation values oscillate with  large to very large amplitude, and the maximal mismatches grow with $\numlam$.

\item Although one can also identify  different regimes for $\gamma>0.9$, we generically have rather large differences between expectation values and classical solutions.

\end{itemize}

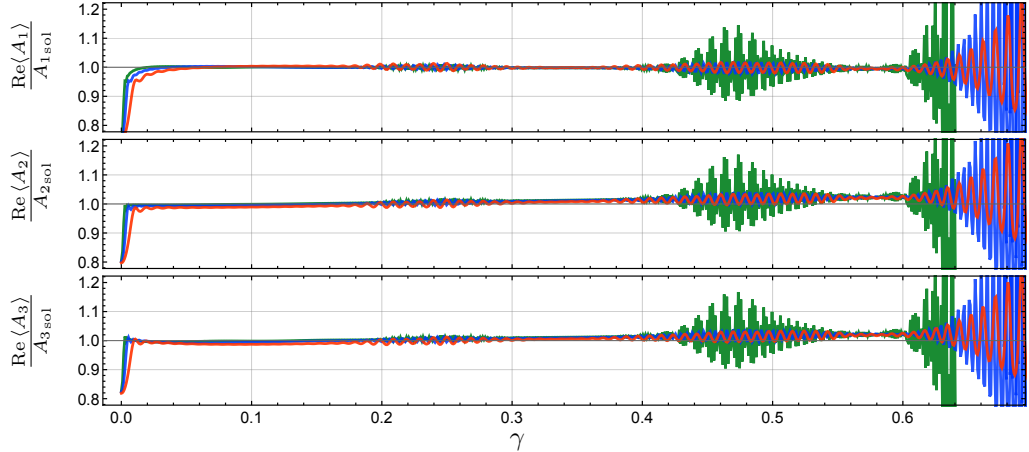
\begin{figure}[ht!]
\begin{picture}(500,165)
\put(80,5) {\includegraphics[scale=0.4]{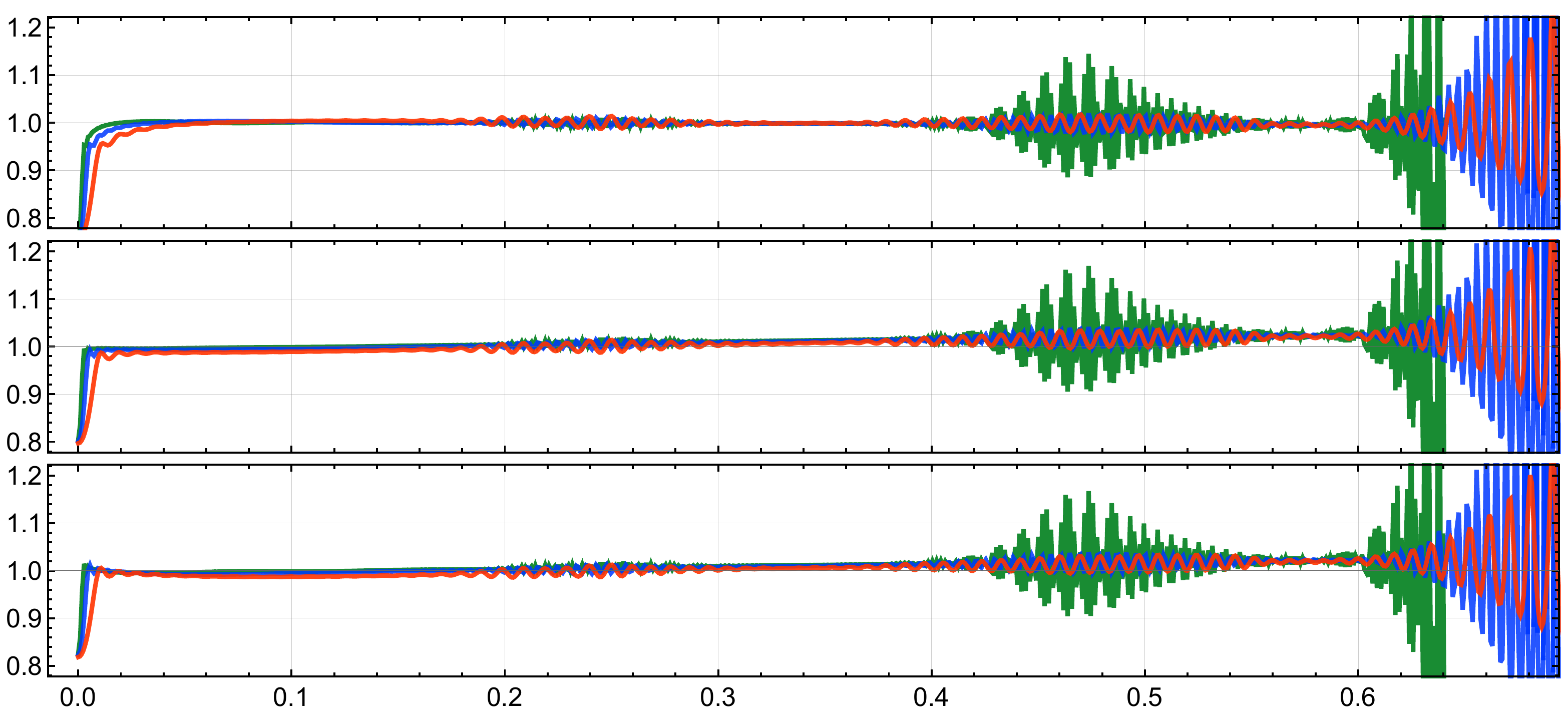} }
\put(245,0) {$\gamma$} 

\put(53,130){ \rotatebox{90}{ $\frac{  {\rm Re } \langle A_1 \rangle }{ {A_1}_{\rm sol} }$} }
\put(53,80){ \rotatebox{90}{ $\frac{  {\rm Re } \,\langle A_2 \rangle }{ {A_2}_{\rm sol} }$} }
\put(53,25){ \rotatebox{90}{ $\frac{  {\rm Re } \,\langle A_3 \rangle }{ {A_3}_{\rm sol} }$ } }
%

\end{picture}
\caption{ Real part of the expectation values for the bulk areas in the medium size curvature example normalized with corresponding classical values for $\numlam=10$ (red), $\numlam=20$ (blue) and $\numlam=40$ (green). \label{FigAmc}
}
\end{figure}

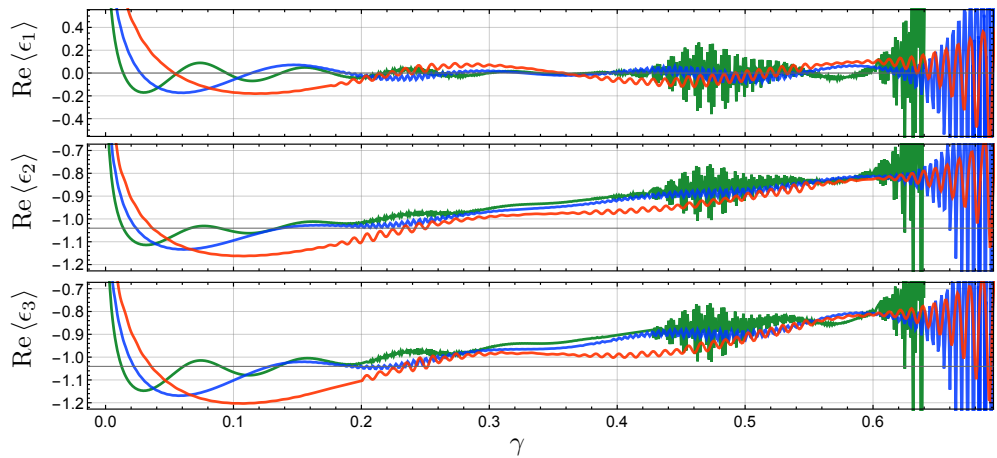
\begin{figure}[ht!]
\begin{picture}(550,165)
\put(70,5) {\includegraphics[scale=0.4]{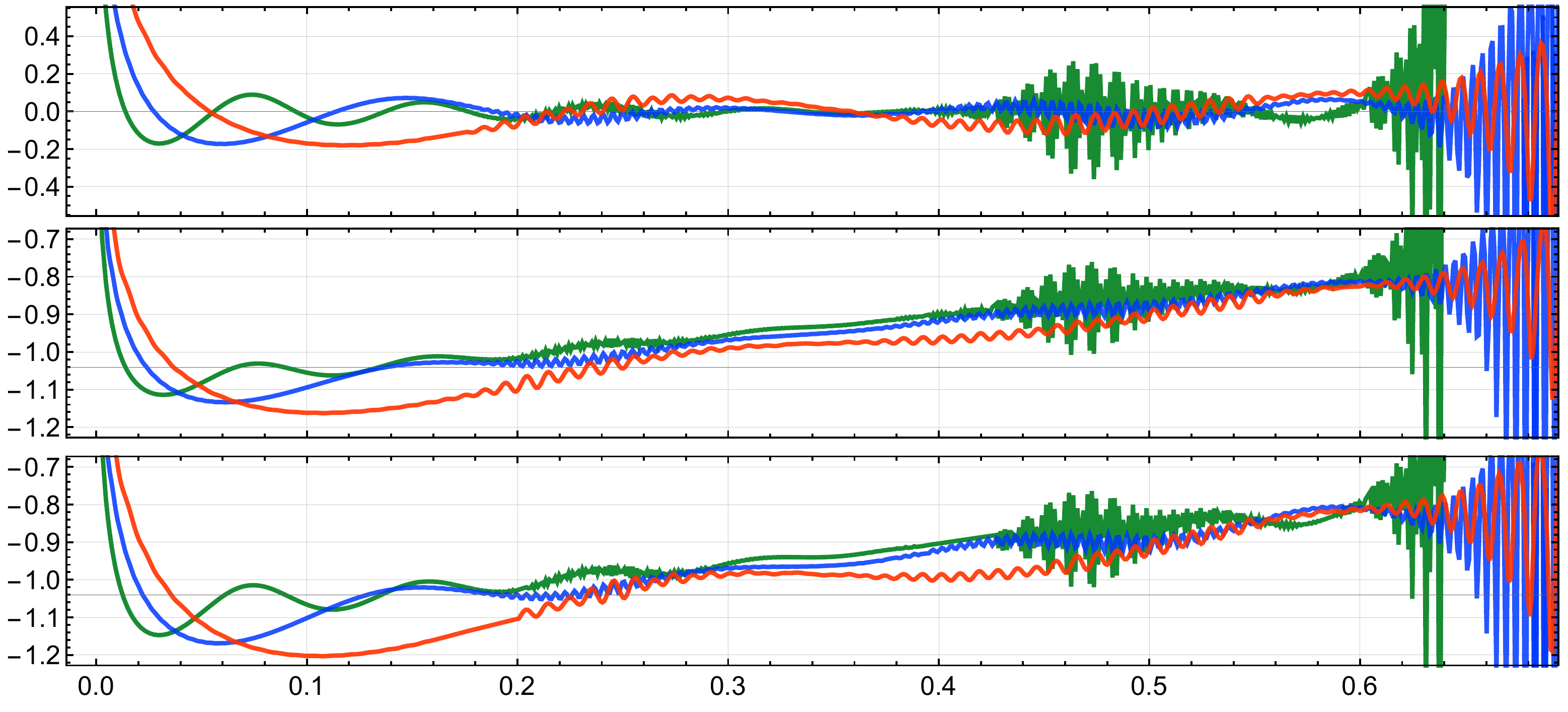} }
\put(245,0) {$\gamma$} 

\put(53,130){ \rotatebox{90}{ $ {\rm Re } \,\langle \epsilon_1 \rangle$} }
\put(53,80){ \rotatebox{90}{ $ {\rm Re } \, \langle \epsilon_2 \rangle$} }
\put(53,28){ \rotatebox{90}{ $  {\rm Re }\,  \langle \epsilon_3 \rangle$ } }
%

\end{picture}
\caption{ Real part of the expectation values for the bulk deficit angles in the medium size curvature example, for $\numlam=10$ (red), $\numlam=20$ (blue) and $\numlam=40$ (green).  The classical values for the deficit angle are 
$\epsilon_1^{\rm sol}=0\, , \epsilon_2^{\rm sol}= \epsilon_3^{\rm sol}=-1.0405$ (indicated by  dark gray  horizontal  lines in the plots).
\label{Fig11}
}
\end{figure}

Fig. \ref{FigAmc}  and Fig. \ref{Fig11} show the real part of the expectation values for the areas and for the deficit angles.  While the Regimes II and IV still give the best agreement for the $A_2$ and $A_3$ expectation values, there is a small, but consistent mismatch in these regimes; this might be related to a similar issue as in the small curvature case. The mismatch in Regime II is smaller for larger $\numlam$ and in Regime IV smaller for smaller $\numlam$.

A similar behavior can be observed for $\epsilon_2$ and $\epsilon_3$. For all three angles the matching is somewhat less perfect, but in Regime II the mismatch is smaller for larger $\numlam$.

In summary, depending on the amount of mismatch one deems acceptable, one can declare Regime II-V as semiclassical, with Regime II showing minimal mismatches for sufficiently large scale. However, in regime V the (maximal) size of the mismatch is $\numlam$-dependent and grows with $\numlam$, for $\numlam$ larger than a certain threshold value of around $\numlam=20$. Thus, one might not accept Regime V as semiclassical anymore for, say, $\numlam=40$.

\subsection{Examples with large curvature angles}\label{L-Curv}

The last family of examples we will discuss has  relatively large deficit angles. The deficit angles, boundary lengths and areas, bulk areas, and scales are summarized in table \ref{LargeCurv}.

 \begin{table}[h]
\caption{\label{LargeCurv}  The lengths, areas, and deficit angles for the intermediate curvature examples. The parameters are defined in Sec.~\ref{Sec-Triang}, Table~\ref{SymmReduc}. Here we will consider scales $\lambda=\numlam \gamma \ell_P^2$ with $\numlam \in \{1,2,4\}$. Note that these scales  correspond approximately to the scales $\numlam=10,20,40$ in the previous examples.}
\begin{ruledtabular}
\begin{tabular}{lll}
lengths & areas\phantom{$\Big|$} & deficit angles \\
\hline
       $a= 15.280\sqrt{\lambda}$ & $B_1= 114\lambda$ &$\epsilon^{\rm sol}_1=4.193$      \\
        $b= 19.140\sqrt{\lambda}$     &  $B_2= 116\lambda$ & $ \epsilon^{\rm sol}_2= -1.790$      \\
       $c=20.356\sqrt{\lambda}$  & $B_3=115\lambda$ & $\epsilon^{\rm sol}_3  =-1.1432$      \\
       $d=19.661\sqrt{\lambda}$&  $A_1^{\rm sol}=115.7\lambda$ & \\
              $t_{\rm sol}=20.125\sqrt{\lambda}$  & $A_2^{\rm sol}= 178.1\lambda$, \quad $A_3^{\rm sol} = 170.0 \lambda$ & \\
\end{tabular}
\end{ruledtabular}
\end{table}

%
%

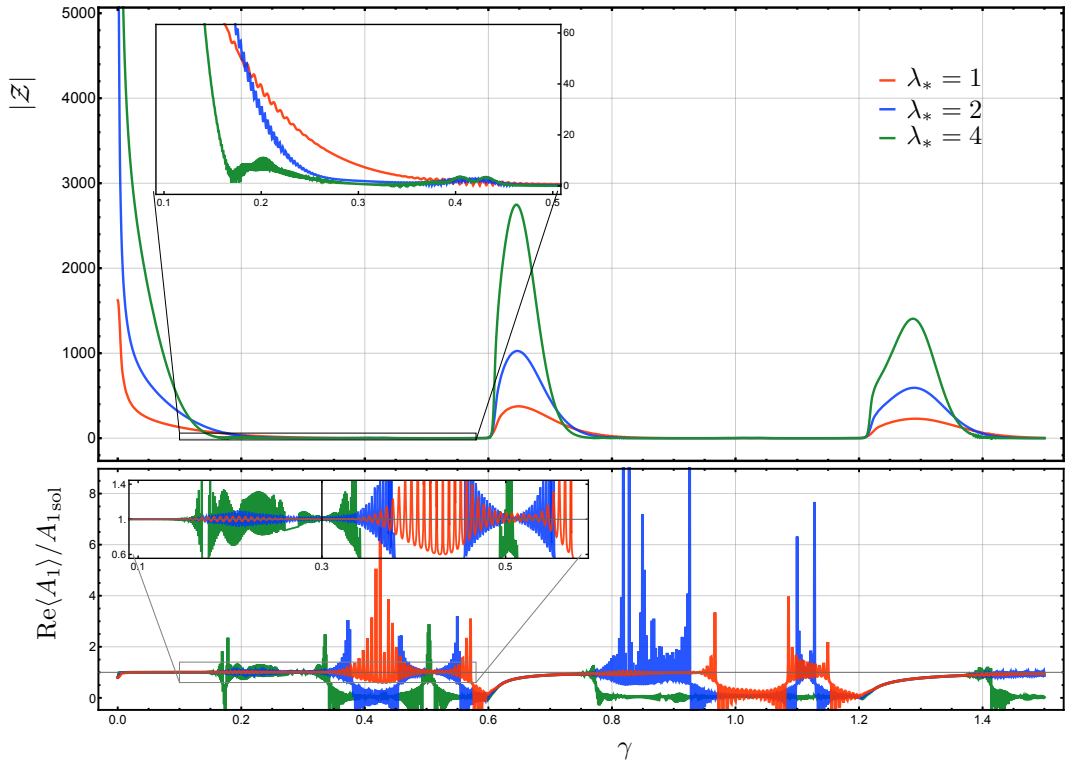
\begin{figure}[ht!]
\begin{picture}(500,290)
\put(30,7){ \includegraphics[scale=0.32]{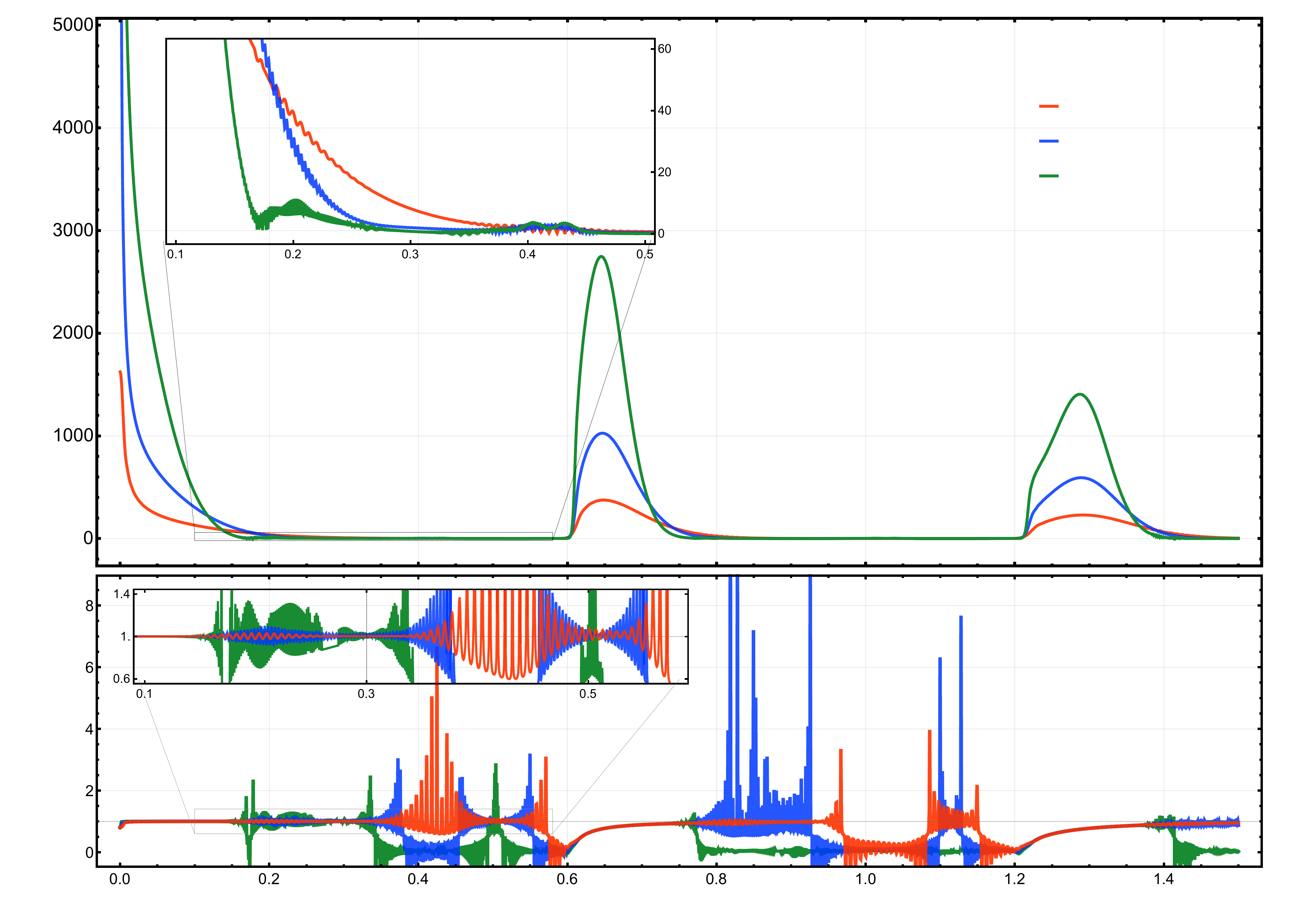} }
\put(30, 240){\rotatebox{90}{ $| \cal Z | $ }}
\put(40, 40){\rotatebox{90}{ $ {\rm Re} \langle A_1 \rangle /{A_1}_{\rm sol} $ }}

\put(260,0){$\gamma$}

\put(366,253){ $\numlam = 1$ }
\put(366,241){ $\numlam = 2$ }
\put(366,231){ $\numlam = 4$ }
\end{picture}
\caption{ Absolute value of the partition function and real part of the expectation value for $A_1$ for an example with large curvature angles.
\label{Fig12}}
\end{figure}

Here we see again that the behavior in the norm of the partition function is reflected in the behavior of the expectation values, see Fig. \ref{Fig12}.  Beyond $\gamma>0.4$  we do not have a semiclassical regime and so we focus on $\gamma\lesssim 0.4$.

For $\gamma<0.1$ the norm of the partition function is larger for larger $\numlam$. This order is however inverted for $\gamma>0.2$ and the $\numlam=4$ norm shows strongly oscillatory behavior, starting at approximately $\gamma \sim 0.15$.  Thus, the oscillations of the expectation values in $\gamma$ will have larger amplitude for larger scale $\numlam$.

We can differentiate between the following regimes:
\begin{itemize}
\item Regime I ($\gamma \lesssim 0.01$): The $\gamma$ is too small  to lead to a reliable semiclassical regime. The onset of a reliable regime moves to smaller values of $\gamma$ for growing $\numlam$.
\item Regime II ($0.01 \lesssim \gamma \lesssim 0.14$): The (area) expectation values match well the classical values.
\item Regime III ($0.14 \lesssim \gamma \lesssim 0.34$): The expectation values show oscillations, with an amplitude depending quite strongly on the scale. For $\numlam=1$ the maximal mismatch for $A_1$ is under 10\%, and for $A_2$ it is between 10\% and 20\%. For $\numlam=4$ the oscillations are too large for this regime to be considered semiclassical.

\item Regime IV ($0.34 \lesssim \gamma \lesssim 0.6$): The expectation values for $\numlam=1$ and $\numlam=2$ are strongly oscillating, and this regime cannot be considered semiclassical.

\end{itemize}

\begin{figure}[ht!]
\begin{picture}(400,160)
\put(30,7){ \includegraphics[scale=0.38]{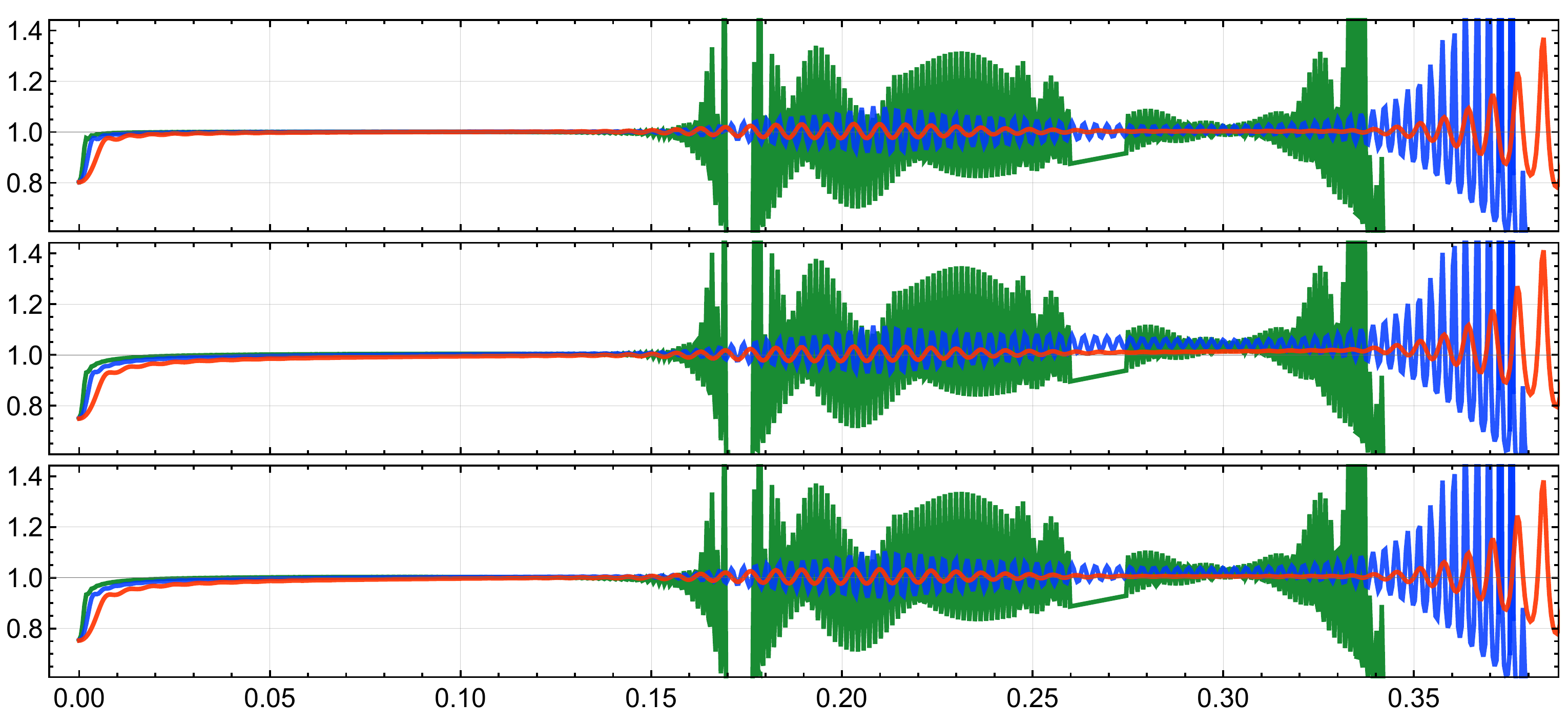} }
\put(10,120){ \rotatebox{90}{ $\frac{  {\rm Re } \langle A_1 \rangle }{ {A_1}_{\rm sol} }$} }
\put(10,70){ \rotatebox{90}{ $\frac{  {\rm Re } \langle A_1 \rangle }{ {A_1}_{\rm sol} }$} }
\put(10,25){ \rotatebox{90}{ $\frac{  {\rm Re } \langle A_1 \rangle }{ {A_1}_{\rm sol} }$} }

\put(260,0){$\gamma$}

\end{picture}
\caption{ Real part of the expectation value for bulk areas normalized with corresponding classical values for an example with large curvature angles and for scales $\numlam=1$ (red), $\numlam=2$ (blue) and $\numlam=4$ (green).
\label{FigAhc}}
\end{figure}
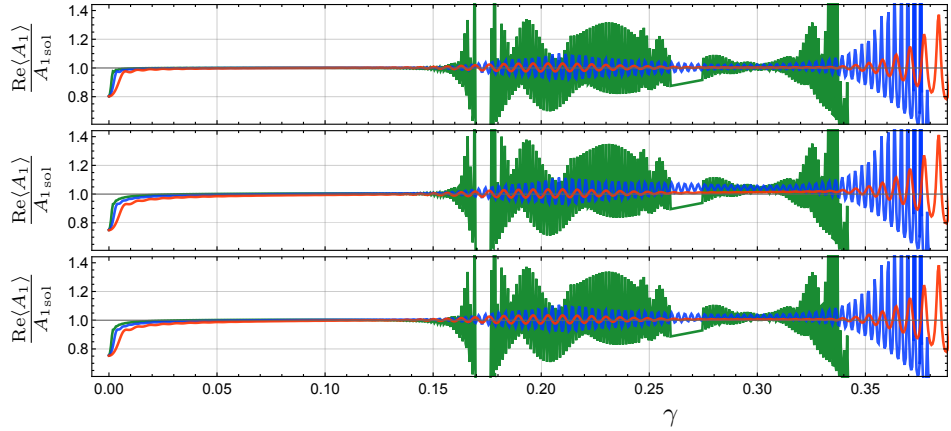

\begin{figure}[ht!]
\begin{picture}(400,160)
\put(30,7){ \includegraphics[scale=0.38]{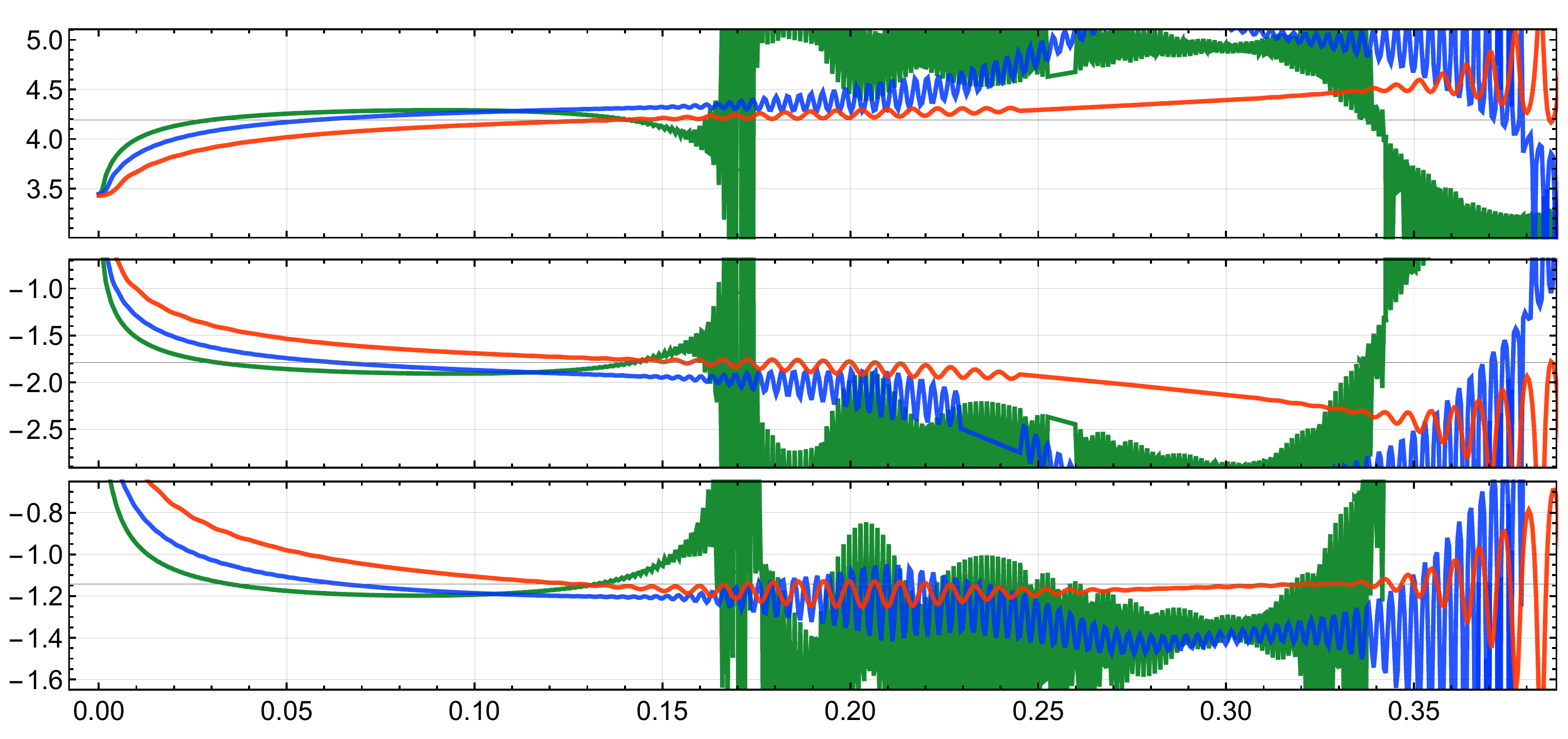} }

\put(15,120){ \rotatebox{90}{ $ {\rm Re } \,\langle \epsilon_1 \rangle$} }
\put(15,70){ \rotatebox{90}{ $ {\rm Re } \, \langle \epsilon_2 \rangle$} }
\put(15,25){ \rotatebox{90}{ $  {\rm Re }\,  \langle \epsilon_3 \rangle$ } }

\put(260,0){$\gamma$}

\end{picture}
\caption{ Real part of the expectation value for deficit angles $\epsilon_1,\epsilon_2$ and $ \epsilon_3$ for an example with large curvature angles and for scales $\numlam=1$ (red), $\numlam=2$ (blue) and $\numlam=4$ (green).
\label{FigEpshc}}
\end{figure}
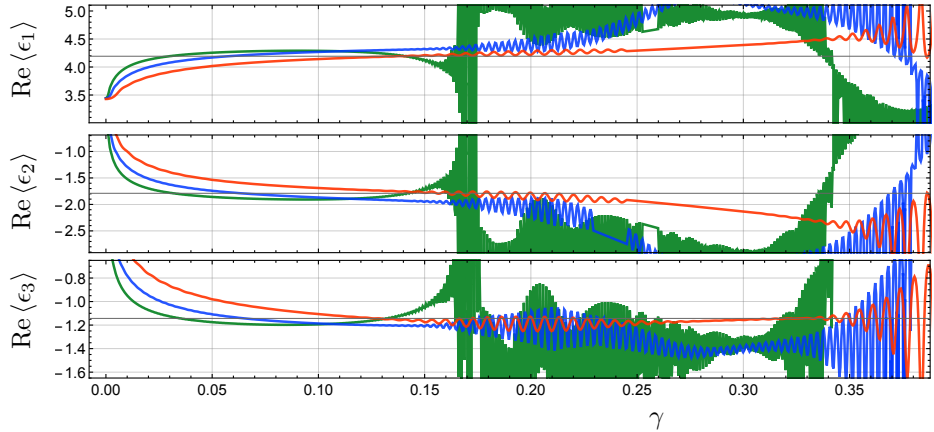

The areas are quite well matched in Regime II, see Fig. \ref{FigAhc},  but this is less the case for the deficit angles, see Fig. \ref{FigEpshc}. There is also more of a systematic mismatch for the deficit angles in Regime III for $\numlam=2$ and, less so, for $\numlam=1$. 

To summarize, in contrast to the low curvature example, the identification of a semiclassical regime is, for this example, quite strongly scale dependent. Regime II can be considered as semiclassical for $\numlam=1,2$ and $\numlam=4$. Regime III is not semiclassical anymore for $\numlam=4$, and depending on how much mismatch one wants to accept for the deficit angles, one may only declare a part of Regime III semiclassical for $\numlam=2$. 

We could therefore argue that the heuristic argument leading to equation (\ref{SFcond1}), which suggested a $\gamma$-dependent bound on the scale for the semiclassical regime, applies in this example. However, we see also here, that an important factor determining a reliable semiclassical regime is the occurrence of oscillations in the norm of the partition function as a function of $\gamma$.  

\subsection{Summary of behavior found in the examples}\label{SummExs}

We have uncovered a surprisingly rich behavior of the expectation values with varying Barbero-Immirzi parameter $\gamma$, different scales, and different classical curvature values. 

The arguments in \cite{EffSFM} and in section \ref{toymodel} suggested a bound $\gamma \sqrt{j} \epsilon \lesssim {\cal O}(1)$ for the identification of a semiclassical regime. Our results suggest some nuance in this bound. We have seen, in particular, that the identification of a semiclassical regime depends a lot on the appearance of oscillations at certain threshold values for $\gamma$:

\begin{figure}[ht!]
\begin{picture}(450,320)
\put(50,5){\includegraphics[scale=0.32]{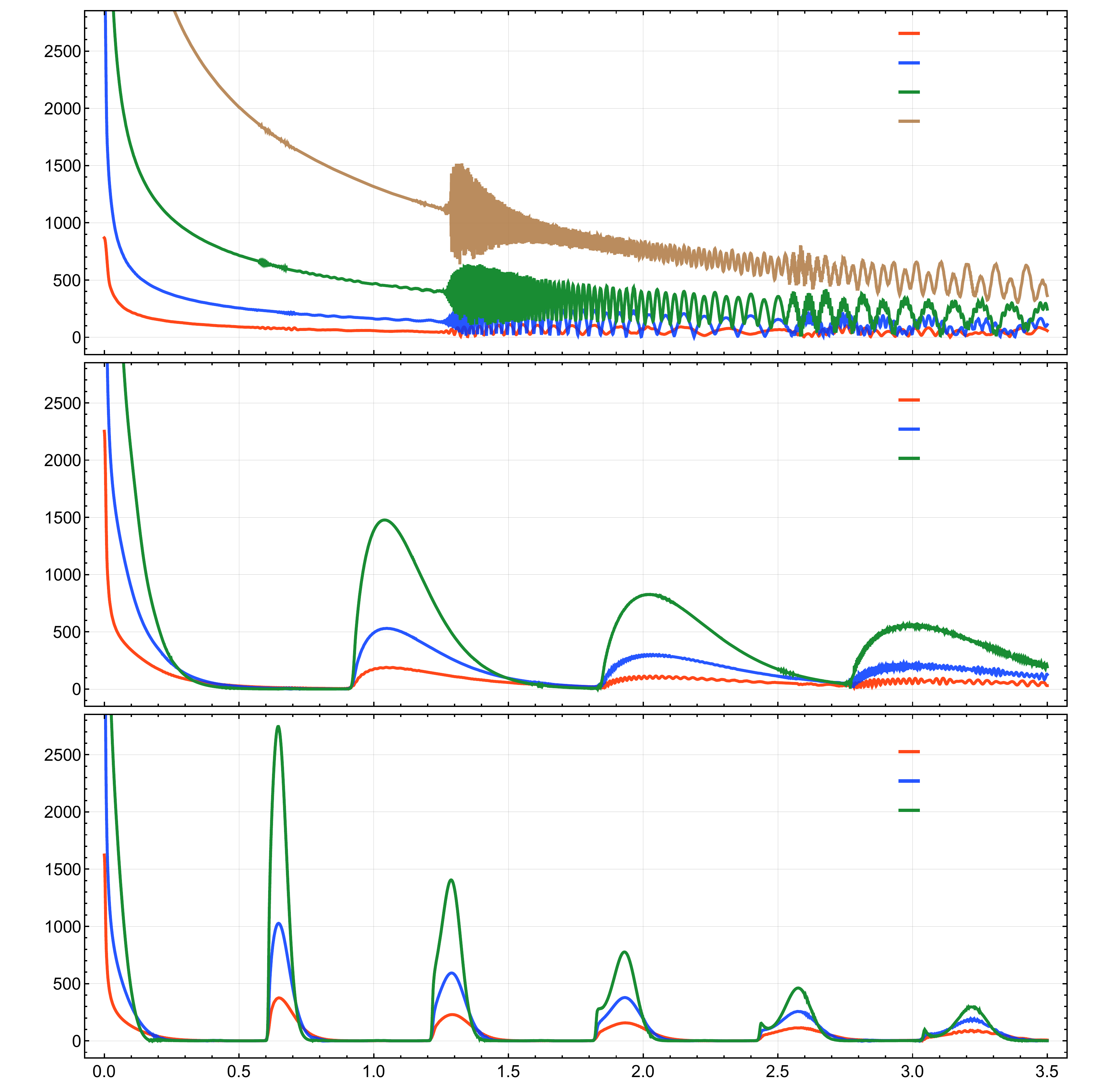}}

\put(40, 160){\rotatebox{90}{ $| \cal Z | $ }}
\put(210,2){ $\gamma$ }
\put(320,100){\footnotesize $\numlam = 1$ }
\put(320,91){\footnotesize $\numlam = 2$ }
\put(320,82){\footnotesize $\numlam = 4$ }
\put(320,200){\footnotesize $\numlam = 10$ }
\put(320,191){\footnotesize $\numlam = 20$ }
\put(320,182){\footnotesize $\numlam = 40$ }
\put(320,305){\footnotesize $\numlam = 10$ }
\put(320,296){\footnotesize $\numlam = 20$ }
\put(320,287){\footnotesize $\numlam = 40$ }
\put(320,279){\footnotesize $\numlam = 80$ }

\end{picture}
\caption{The panels show the absolute value of the partition function as a function of the Barbero-Immirzi parameter $\gamma$, for the very small curvature example (top panel), the medium size curvature example (middle panel), and the large curvature example (top panel). \label{FigAbsZ}
}
\end{figure}
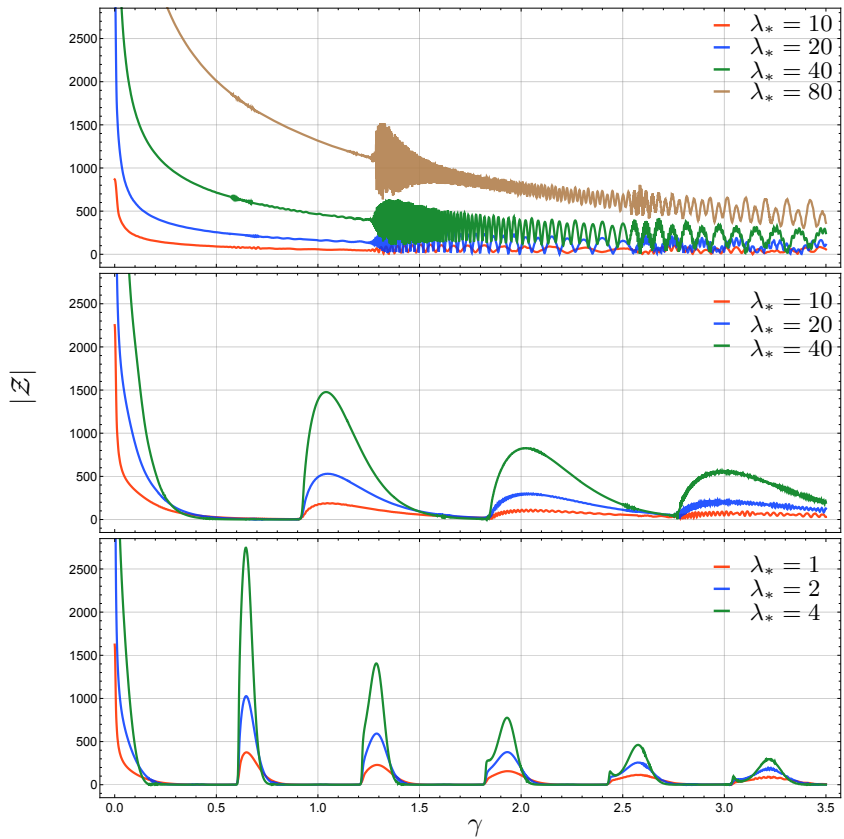

\begin{itemize}
\item We found a semiclassical regime for all examples. The upper bounds on $(\gamma, \numlam)$ for this regime depend heavily on the bulk curvature.

\item One needs a minimal amount of oscillations in the amplitude, and therefore a minimal value for the Barbero-Immirzi parameter, for a semiclassical regime to emerge. We find that this minimal value is quite small: for the low curvature example the areas are well approximated  for values  starting at $\gamma \approx 0.05$, and for the large curvature case for values starting at $\gamma \approx 0.02$. For context, one can compare to the regimes of $\gamma$ preferred in the LQG literature on black hole entropy. There is some spread in results reported in the literature that has to do with the specifics of the model considered, e.g. the gauge group and the number of spins taken to contribute to the entropy \cite{AshtekarBaez:1998, DomagalaLewandowski:2004, PigozzoBacelarCarneiro:2021}. Nonetheless, the most frequently used values range from [0.12, 0.35]. While in the largest curvature example that we presented the semiclassical regime begins to degrade at the large end of this interval, for the most part this interval sits squarely amongst the values where we found semiclassical behavior.

\item Whether a certain regime is acceptable or not as semiclassical seems to depend mostly on the appearance of oscillations in the expectation value as a function of $\gamma$. Surprisingly, these oscillatory regimes appear  at approximately the same values of $\gamma$ for different scales. But the smallest $\gamma$-value for which they appear depends on the bulk curvatures---for the lowest curvature example these regimes appear for larger $\gamma$'s as compared to the larger curvature examples.
\item A regime with oscillatory behavior might still be acceptable as semiclassical if the oscillations come with a relatively small amplitude.  The amplitude of the oscillations in a given regime is scale dependent in a way that depends on the bulk curvature, and---in the intermediate curvature example---also on the $\gamma$-values. In particular, for the example with small curvature the larger scale examples show smaller scale oscillations, whereas for the large curvature example larger scale leads (in the relevant regimes) to larger oscillations. For the intermediate curvature example, we have the small curvature behavior for smaller values of $\gamma$ and the large curvature behavior for larger values of $\gamma$. This behavior seems to be  determined by the relative size of the absolute values of the partition functions for the different scales $\numlam$. 
\item The oscillatory regimes for the expectation values can be predicted from the oscillatory regimes appearing for the absolute value of the partition function $|{\cal Z}|$, see Fig.\ref{FigAbsZ}. Oscillations with large amplitude appear for the expectation values if  $|{\cal Z}|$ oscillates around a relatively small average. 
\item We have also seen that systematic mismatches between expectation values and classical values can appear, e.g., in the very small curvature examples for $A_2$ and $\epsilon_2$. These mismatches can be explained by considering the region of bulk areas allowed by the triangle inequalities and the $G$-factors, and the position of the classical solution within this range.  Mismatches tend to appear if the classical solution is very near the boundary of  this region. For the low curvature example, where such mismatches appear, their size does get smaller (in the appropriate semiclassical regimes)
for larger scales $\numlam$.  

\end{itemize}

The fact that, for the  small curvature regime, we find better matching for larger scales is quite encouraging for the prospects of finding a suitable continuum limit.  In a continuum limit we would expect small curvature per triangle. The results here indicate that, for almost flat continuum configurations, we can use a discretization with a relatively large lattice constant, and still expect a semiclassical regime.

We have computed the expectation values for a number of other examples. These other examples could always be classified as having small or large curvature, or as a cross-over between these two cases, like the medium curvature example presented above. According to their classification they showed very similar behavior as for the examples illustrated here.

The behavior uncovered here opens up many questions for future research. It would be good to understand why the oscillatory regimes appear---we conjecture that it is due to certain resonances, e.g., between the frequency of the phase and the  the discrete spectra of the areas.  Such an understanding would also help us to predict whether this behavior might be stable if we change certain features in the definition of the partition function, such as the area spectrum or the measure in the path integral.  In any event, our computations show that the effective spin foam models are very amenable to numerical computations, and such changes can also be studied more directly through numerical explorations.

\section{Discussion}\label{Disc}

Loop quantum gravity \cite{LQG} suggests that, for quantum gravity in four spacetime dimensions, areas, instead of lengths, are the fundamental variables. In this approach the areas have a discrete, asymptotically equidistant, spectrum. This leads to interesting proposals for the origin of black hole entropy \cite{BHCounting} and resonates with more recent developments on deriving geometry from entanglement \cite{RT}, e.g. via holographic tensor network states \cite{QGTNW}. 

Spin foam models, which implement a regularized dynamics for loop quantum gravity,  are computationally challenging \cite{DonaNum}. An explicit check of whether, in the semiclassical limit, one regains the classical equations of discretized gravity, has so far been outstanding. Indeed, concerns have been voiced \cite{FlatProblem,ILQGS}, that the semiclassical limit imposes a flat dynamics in the $\hbar \rightarrow 0$ limit. 

Using the recently introduced effective spin foam model \cite{EffSFM}---which was designed, in part, to be particularly numerically efficient---we have performed the first explicit test of how the equations of motion are implemented in effective spin foams. To this end we simulated a triangulation with an inner edge, which provides a non-trivial equation of motion, but still  allows us to  perform the calculations on laptop computers for a rather wide range of parameters.

This illustrates the efficiency of the effective spin foam models; they are the  simplest and numerically fastest  spin foam model available to date. The quantum simulations performed here are in excellent agreement with expectations from the classical theory. They do, however, exhibit the surprising feature predicted in \cite{EffSFM}: their semiclassical regime has a non-trivial dependence on the model's parameters.  The simple limit $\hbar \rightarrow 0$ (or  $j\rightarrow \infty$), with fixed Barbero-Immirzi parameter $\gamma$ and curvature, will not, in general, recover the classical equations of motion.\footnote{Instead of imposing flatness, this limit leads to a complicated behavior for the expectation values, illustrated by a variety of oscillatory regimes, see section \ref{SFExpV}. We conjecture that this is due to the discreteness of the areas, which, in interplay with the oscillatory phase of the amplitude, can lead to resonance effects. In contrast, the analysis that leads to the so-called flatness problem \cite{FlatProblem} replaces the sum over area values with an integral over continuous area variables, and misses these resonances.}

This is due to  the weak implementation of certain constraints, which ensure a proper gluing between the simplicies.  For spin foams, as well as other discrete models of quantum gravity, simplices serve as elementary building blocks of the triangulated spacetime.  An exact gluing of simplices would ensure an unambiguous association of length variables to the edges of the triangulation. However, the discreteness of the area spectra prevents an exact gluing.  This also shows up in the algebra of the constraints, which do not commute, rendering these constraints second class. The noncommutativity of the constraints is proportional to $\hbar$ and can be viewed as a quantum anomaly.  These second class constraints cannot be imposed exactly, they are imposed weakly---but as strongly as allowed by the constraint algebra---in the effective spin foam models. 

In order to better understand the generality of this feature of the effective spin foam models, we have introduced a simple toy model where constraints on the path integral can be imposed both strongly and weakly.  This toy model shows that a non-trivial dependence of the semiclassical regime on model parameters should be understood as a general feature of the quantization of systems with second class constraints. It illustrates, in particular, that the semiclassical regime of the theory cannot be understood simply as the $\hbar \rightarrow 0$ limit, but rather requires the tuning of additional parameters.

This sheds light on one aspect of the consternation surrounding the flatness problem of loop quantum gravity.  While the $\hbar \rightarrow 0$ limit is sufficient for understanding the asymptotic behavior of a single 4-simplex of a spin foam theory, it will not suffice for understanding the semiclassical regime of a full complex, which requires the weak imposition of constraints discussed above. The toy model also helps in understanding other features that are apparent in the numerical results found for the effective spin foams. A weak imposition of constraints  leads to a complex action, and such an action will, generically, lead to complex expectation values. 

Certainly, a prerequisite for a semiclassical regime is that the imaginary parts of expectation values (of hermitian observables) should be small.  This issue can be studied in more detail using, e.g., tools from Picard-Lefschetz theory \cite{Picard}. However, numerical treatment of the toy model allows us to go beyond this continuum analysis and to replace the path integral with a discrete sum and to study finite summation ranges. We find that strong deviations are to be expected if the discretization scale becomes comparable  to  or even larger than the wavelength of the oscillatory part of the amplitude. This happens, in particular, in the $\hbar \rightarrow 0$ limit.

While we continue to develop our analytic understanding, effective spin foams also open the door to explicit numerical explorations of the semiclassical regime. Our simulations have revealed a multitude of interesting and unexpected behavior, both for the partition functions and the expectation values of the effective spin foam models, and their dependencies on the Barbero-Immirzi parameter, the scale (or spin $j$), and the curvature.  We have studied three different curvature regimes: low, intermediate and large curvatures. In each case we identified a semiclassical regime where an, in general, interdependent range for the scale $\numlam$ and the Barbero-Immirzi parameter $\gamma$ gave expectation values that were reasonably good approximations to the classical values. 

Surprisingly, we found that the  acceptable range for $\gamma$ only depends  for larger curvature on the scale $\numlam$ (at least in the range of scales we considered). This is due to an onset of oscillatory behavior in the expectation values, which seem to happen almost always for the same values of $\gamma$, independently of $\numlam$. 

Moreover we find that, for the low curvature case, the semiclassical regime is not bounded by a maximal value for the scale. In fact, the naive semiclassical limit $j\rightarrow \infty$ rather seems to `work' here. This is good news for the prospects of regaining diffeomorphism symmetry via a continuum limit \cite{PImeasure,Improved}; in such a limit one distributes the curvature over many building blocks leading to small curvature values per building block. Discrete diffeomorphisms act by changing locally the scale \cite{BahrDittrich09a,DittrichHoehn1}, so it is fortuitous that a semiclassical regime does not seem to imply a bound on the scale.

We have tested a particular triangulation with a non-trivial equation of motion. But we believe that our conclusions will hold for more general triangulations, where the scale and curvature of the bulk can be controlled through the boundary data. 

A crucial question will be whether and how the semiclassical regime survives a continuum and refinement limit \cite{CG}. To this end it will be crucial to understand how the weak implementation of the second class constraints interacts with a coarse graining and renormalization flow. This question can also be  studied directly in the continuum, e.g. using the family of models described in \cite{Krasnov}.

The numerical explorations in this paper have opened up interesting questions that can be investigated in future work. These include a better understanding of the onset of oscillatory behavior of the expectation values  as functions of the Barbero-Immirzi parameter $\gamma$, which we have found to be foreshadowed by the oscillatory behavior of the absolute value of the partition function. We are also interested by certain features of the absolute value of the partition function, for example, its oscillatory regimes and its maxima, which seem to be independent of scale.  Knowing how this behavior can be explained will help us to estimate whether such behavior is stable when one changes certain details of the model, and whether a similar behavior can be expected also from other spin foam models.First steps to generalize the results of this paper to the Lorentzian case have been made in \cite{ADP}.  One should furthermore allow for triangulations constructed from constant curvature simplices, which are connected to the cosmological constant, infrared finiteness, and course graining and renormalization flow \cite{NewRegge,TuraevViro:1992,Crane:94,Major:96,Barrett:03,Dupuis:14,Bonzom:14,Haggard:15,Haggard:16,Haggard:162,Haggard:18,Dittrich:17,Dittrich:172}.

\appendix

\section{Generalized triangle inequalities}\label{GTI}

An Euclidean flat $n$-simplex is the smallest convex set that contains $n+1$ points (vertices) $p_0,p_1,\cdots , p_n \in \mathbb R^n$. 
 The edge lengths $l_{ij} = |p_i - p_j|$ of the $n$-simplex are given by the distances between all pairs of vertices. With these edge lengths we form an $n\times n$ matrix $M$ with entries
\ba
m_{ij}  = l_{0i}^2 + l_{0j}^2 -l_{ij}^2 \equiv 2\la p_i ,p_j\ra.
\ea
Then, the edge lengths $l_{ij}$ determine an $n$-simplex, if and only if the matrix $M$ has real and positive eigenvalues, that is, if $M$ is positive definite \cite{HaggardLittlejohn,Dekster-Wilker}.

The matrix $M$ is symmetric and hence the conditions on the eigenvalues can be translated to determinants, i.e, the principal minors of $M$ should have positive determinants. This gives the generalized triangle inequalities for an $n$-simplex. For a triangle $t$ with edge lengths $(a,b,c)$, the determinant of the corresponding matrix $M$ is given by
\ba
\det(M)= (a+b+c)(a+b-c)(c+a-b)(b+c-a),
\ea
and hence $\det(M)>0$ leads to the triangle inequalities $a+b>c, a+c>b, b+c>a$.

This expresses the triangle inequalities in terms of lengths. To obtain the inequalities for a 4-simplex in terms of areas, one has to use one of the roots expressing the lengths of this 4-simplex as a function of its areas. That is the triangle inequalities in term of the areas will depend on the choice of root.

\section{On the roots of the area-length equations}\label{sec-roots}

One advantage of the symmetry reduction, which we employ for the triangulation described in section \ref{Sec-Triang}, is that the system of equations  determining the lengths from the areas of a given simplex simplifies considerably. In fact, one can still solve this system analytically. 

Let us consider the simplices of type I---the discussion for type II proceeds in the same way.  For a simplex of type I we have four length parameters $(t,a,b,c)$ and four area parameters $(B_1,B_2,A_1,A_2)$. These are connected by the following system of equations
\ba
B_1&=&A(a,a,b)\, , \label{B1} \\
B_2&=& A(a,a,c)\, ,\label{B2}\\
A_1&=& A(a,a,t)\, ,\label{A1}\\
A_2&=& A(c,c,t)\, , \label{A2} \, 
\ea
where $A(x,y,z)=\tfrac{1}{4}\sqrt{2x^2y^2+2x^2z^2+2y^2z^2-x^4-y^4-z^4}$ is the area of a triangle with edge lengths $x,y,$ and $z$.  All equations are of the form $A(x,x,z)=A$ and one can solve one of these equations for either $x$ or $z$:
\ba
x^2=\frac{16A^2+z^4}{4z^2} \, , \q\q z^2_{\pm}=2(x^2\pm \sqrt{x^4-4A^2}) \ .
\ea
Thus, we get a unique (positive length) solution if we solve for the edge length appearing twice in the triangle, and two (positive length) solutions if we solve for the edge length appearing only once.  For the latter case, the two solutions agree, if $2x^2=z^2=4A$, that is, if the triangle has a right angle between the two $x$-edges. 

\vspace{0.3cm}

To solve the system of equations (\ref{B1}--\ref{A2}) one can proceed as follows: 
\begin{itemize}
\item Solve (\ref{A2}) for $c$, resulting in a unique solution ${\cal C}^{(1)}(A_2,t)$.
\item Solve (\ref{A1}) for $t$, which gives two solutions ${\cal T}^{(1)}_{1,2}(A_1,a)$. 
\item Setting $t={\cal T}^{(1)}_{1,2}(A_1,a)$ in ${\cal C}^{(1)}(A_2,t)$ gives two solutions ${\cal C}^{(2)}_{1,2}(A_1,A_2,a)={\cal C}^{(1)}(A_2,{\cal T}^{(1)}_{1,2}(A_1,a))$.
\item Setting $c={\cal C}^{(2)}_{1,2}(A_1,A_2,a)$  in Eq. (\ref{B2})  gives  an equation (for each choice of $t$-solution) involving the length $a$ and the areas $A_1,A_2,B_2$. 
We can solve each of these equations for $a$, which gives two positive length solutions for each equation. We seem to have reached four different solutions for $a$. However, there are two pairs of solutions that agree with each other, and we are still left with only two solutions ${\cal A}_{1,2}(B_2,A_1,A_2)$.  Inserting these solutions ${\cal A}_{1,2}$ into ${\cal C}^{(2)}_{1,2}$ and ${\cal T}^{(2)}_{1,2}$ (matching the sub-indices) we get two solutions for $t,a$ and $c$ in terms of $A_1,A_2,B_2$.
\item Finally, we solve Eq. (\ref{B1}) for $b$, leading to two solutions ${\cal B}^{(1)}_{1,2}(B_1,a)$.
Replacing $a$ with ${\cal A}_{1,2}(B_2,A_1,A_2)$ in ${\cal B}^{(1)}_{1,2}$ gives four different solutions ${\cal B}^{(2)}_{ij}(B_1,B_2,A_1,A_2)={\cal B}^{(1)}_{i}(B_1,{\cal A}_j(B_2,A_1,A_2))$.

\end{itemize}

We can furthermore consider the Jacobian $\partial A_t/\partial l^2_t$ of this system. The determinant of this Jacobian vanishes for configurations satisfying
\ba
(i)\,\, t=0, \q\text{or} \q(ii)\,\, \, b^2=2a^2, \q\text{or} \q (iii)\,\, \, t^2=\frac{2c^2(4a^2-c^2)}{2a^2+c^2} \ .
\ea
Around these length configurations the areas cannot be inverted for the lengths.  Indeed, if $(i)$ holds we are left with three non-vanishing length parameters but only two non-vanishing area parameters. 

Mapping the length configurations satisfying $(ii)$ to the area configuration space gives points where the ${\cal B}^{(2)}$ roots coincide pairwise, that is ${\cal B}^{(2)}_{1j}={\cal B}^{(2)}_{2j}$. The configurations $(iii)$ describe areas for which ${\cal A}_1={\cal A}_2$ and ${\cal T}^{(2)}_1={\cal T}^{(2)}_2$ and ${\cal C}^{(2)}_1={\cal C}^{(2)}_2$, and therefore ${\cal B}^{(2)}_{i1}={\cal B}_{i2}^{(2)}$. 

The condition $(ii)$ is satisfied if the triangle  with edge lengths $(a,a,b)$ has a right angle. Condition $(iii)$ connects the lengths of more than one triangle. 
It is satisfied, if both triangles $(a,a,c)$ and $(a,a,t)$ are right-angled, or if both triangles $(c,c,t)$ and $(a,a,t)$ are right angled.  (However, these configurations do not exhaust the points in $(iii)$.) Generally this condition is not satisfied if only one of the triangles is right-angled.

We see that a given set of area values may correspond to several sets of length values. In our case the maximal number of length configurations corresponding to a given set of areas is four, but this number can be smaller as the triangle inequalities might forbid some of the area configurations.  

Adding the 3D dihedral angles to the area variables one can distinguish between the different length solutions. The 3D angles can, however, be expressed as (four different) functions of the areas, that is, we only need a discrete amount of information from the 3D angles to distinguish between the roots.  

\begin{figure}[ht!]
\begin{picture}(500,160)
\put(130,10){\includegraphics[scale=0.5]{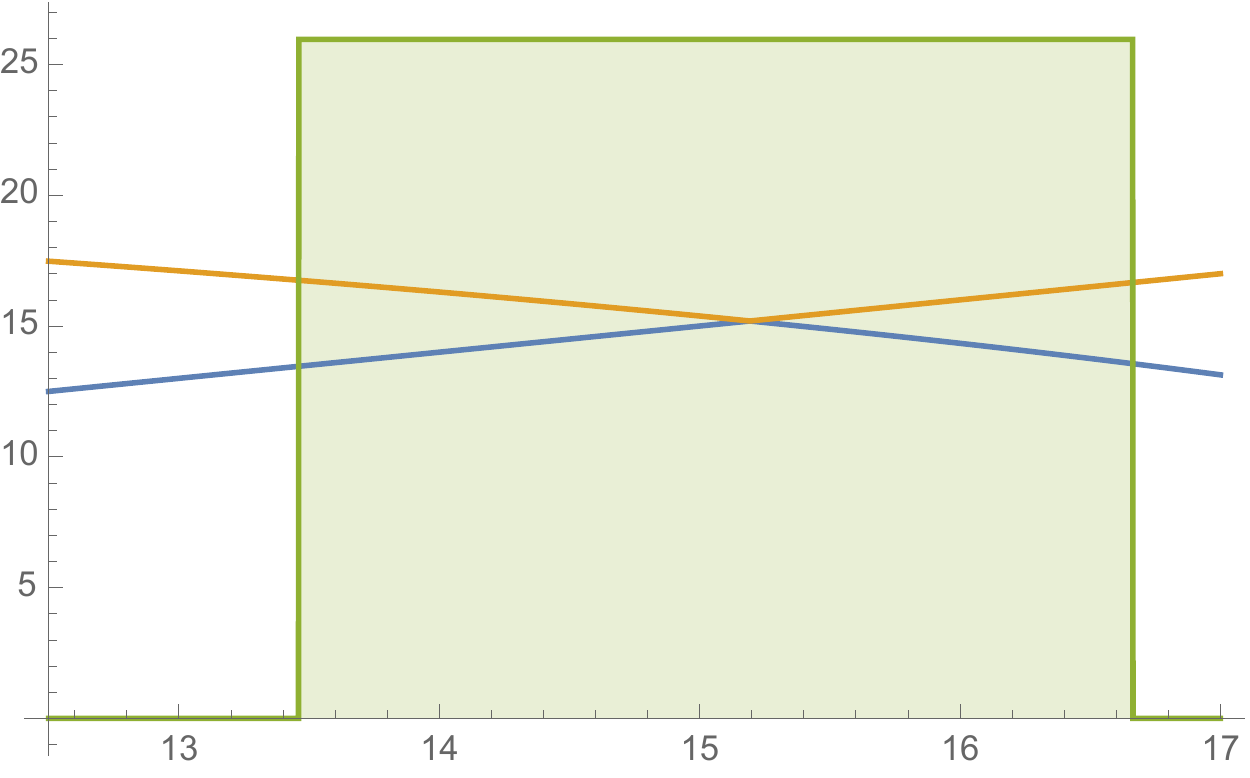}}
\put(110,65){\rotatebox{90}{ ${\cal T}_{1,2}^{(2)}$ }}
\put(290,5){$t$}
\end{picture}
\caption{ This shows the two roots (in blue and yellow) for the $t$-lengths ${\cal T}_{1,2}^{(2)}$ as parametric curves of $t$ where $A_1=A(a,a,t)$ and $A_2=A(c,c,t)$ are varying, with  $B_1=A(a,a,b)$ and $B_2=(a,a,c)$  as inputs and fixed $a=10.7,\, b=13.7$ and $c=10.9$. The green shaded region indicates the $t$-range where the generalized triangle inequalities for the yellow root is satisfied. The triangle inequalities are satisfied over the entire shown $t$-range for the blue root. To reconstruct the length parameter $t$ one has to glue part of the blue root to a part of the yellow root at the point where these roots meet.
 \label{RootJump}}
\end{figure}

Note that in order  to describe a given one-parameter family of lengths in terms of areas and 3D angles, e.g. a family where $(a,b,c)$ are fixed, whereas $t$ provides the parameter, might require to change roots at the point, where the roots meet. Fig. \ref{RootJump} shows an example.

In this work we want to concentrate on our main aim, which is to test whether implementing second class constraints weakly can lead to a satisfactory semiclassical dynamics.  We will therefore avoid configurations where such a change in roots   along the $t$-parameter family of configurations appears and will leave the study of (quantum) effects resulting from such root ambiguities for future work.

\section{Remarks on the implementation of the numerical computation}\label{speed}

\begin{figure}[ht!]
\begin{picture}(400,140)
\put(50,10){\includegraphics[scale=0.26]{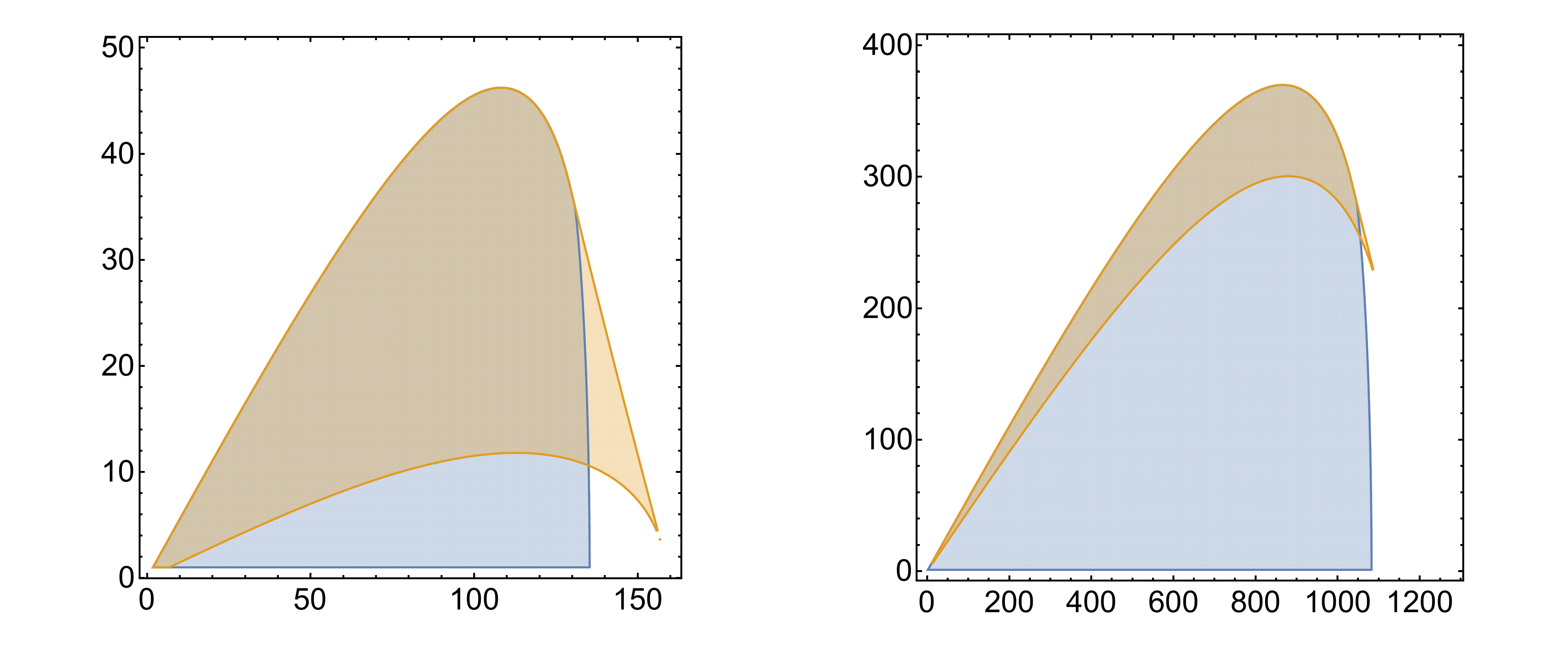}}
\put(175,5) {\footnotesize $A_1$}
\put(330,5) {\footnotesize $A_1$}
\put(59,120){\rotatebox{90}{\footnotesize $A_2$ }}
\put(210,120){\rotatebox{90}{\footnotesize $A_2$ }}
\end{picture}
\caption{ This figure shows the $(A_1,A_2)$ plane of the configuration space for the  $\numlam=10$ (left panel) and the $\numlam=80$ (right panel) scalings, for the small curvature example in section \ref{S-Curv}. The blue shaded region indicates the range for the bulk areas $A_1$ and $A_2$ of simplices of type I where the generalized triangle inequalities are satisfied.  The orange shaded region indicates  values for which  $G^\text{I}$  is larger than $10^{-10}$. 
 \label{PlotG}}
\end{figure}
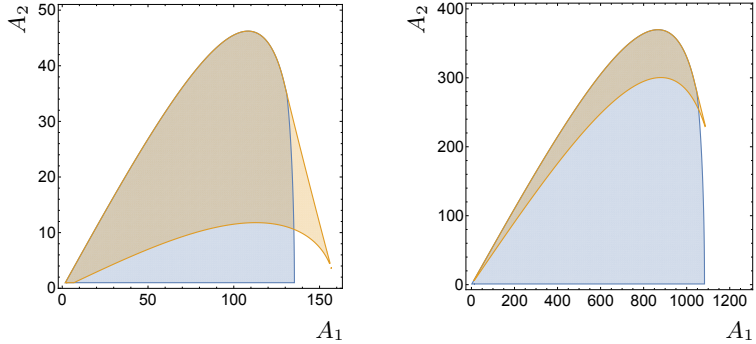

For the computation of the expectation values (\ref{ExpV1}) we have to sum over three---a priori unbounded---variables $j_{A_1},j_{A_2},$ and $j_{A_3}$.    This infinite sum can be changed into a finite one by making use of the generalized triangle inequalities. Including only area configurations for which the generalized triangle inequalities (see Fig. \ref{PlotG}) are satisfied leads to a sum over finitely many terms.  The triangle inequalities are invariant under a global rescaling, so the number of configurations allowed by the triangle inequalities scales with $\numlam^3$. If we were to work with length Regge calculus, we would only have one bulk variable, and assuming an equidistant discretization  of this variable, the number of summands would grow more slowly with $\sqrt{\numlam}$.

We can reduce the number of summands by making use of the $G$-factors, which  exponentially suppress those configurations that do not satisfy the shape matching constraints. For the computation of our examples we included only configuration for which the product of the three $G$-factors satisfies $G^\text{I}G^\text{II}G^\text{B}\geq 10^{-10}$ and for which the generalized triangle inequalities are satisfied. 

The following table shows how the number of configurations scales with $\numlam$ for the example with very small curvature: 

\begin{center}
\begin{tabular}{|c|c|c|c|c|c|}
\hline
 $\numlam$ &  1& 10 & 20 &40&80  \\ \hline
 $\,\sharp$ config.\,&\, \,117 \,\,& \,\, 73,264 \,\,&\,\,119,475 \,\, &\,\,573,878 \,\,& \,\, 1,947,237\,\,\\ \hline
\end{tabular}
\end{center}
 The number of configurations grows approximately as $\numlam^2$ for larger $\numlam$. Thus the reduction in the number of summands is particularly impactful for larger $\numlam$. This is because the widths  of the Gaussian $G$-factors scale as $\sqrt{\numlam}$, whereas the linear extent of the region allowed by the triangle inequalities grows as $\numlam$, see Fig. \ref{PlotG}.

\vspace{0.3cm}

One might be concerned about the error that is introduced by including only terms with $G^\text{I}G^\text{II}G^\text{B}\geq 10^{-10}$ into the sum. For the large curvature example with $\numlam=2$ we computed the expectation values with a summation including $G^\text{I}G^\text{II}G^\text{B}\geq 10^{-10}$ terms as well as with a summation including $G^\text{I}G^\text{II}G^\text{B}\geq 10^{-15}$ terms. The number of terms in these two cases is 194,718 and 238,215 respectively. The maximal difference between the normalized $A_1$-expectation values (in a range of $0\leq \gamma \leq 1$) was smaller than $10^{-7}$, and the differences are not visible at all for the plots of the various expectation values shown in section \ref{L-Curv}.

Computations have been performed with Mathematica on laptops. The computation of all expectation values  (for real and imaginary parts of three areas and three deficit angles as well as three area variances) and the production of the corresponding plots over $\gamma$ took---in, e.g., the large curvature case and the largest scale considered---approximately 9-12 hours of computational time.  Here, it is the production of the 18 plots showing the various expectation values as continuous functions of $\gamma$ which took the most time. The computation of the expectation values for a given value of $\gamma$  takes less than 15 minutes.
Mathematica files with example computations are available upon request from the authors.

\section*{Acknowledgements}
The authors thank Abhay Ashtekar, Eugenio Bianchi, Pietro Don\`a,  Francesco Gozzini, Laurent Freidel, Wojciech Kami\'nski,  Danielle Oriti, Aldo Riello, Carlo Rovelli, Giorgio Sarno, Lee Smolin, and Simone Speziale for discussions.
This work is  supported  by  Perimeter  Institute  for  Theoretical  Physics.   Research at Perimeter Institute is supported in part by the Government of Canada through the Department of Innovation, Science and Economic Development Canada and by the Province of Ontario through the Ministry of Colleges and Universities.

\providecommand{\href}[2]{#2}
\begingroup
\bibliographystyle{ieeetr}

\begin{thebibliography}{100}\small



\bibitem{quantumgeometry}
A.~Ashtekar and C.~J.~Isham,
``Representations of the holonomy algebras of gravity and nonAbelian gauge theories,''
Class. Quant. Grav. \textbf{9} (1992), 1433-1468
[arXiv:hep-th/9202053 [hep-th]].
A.~Ashtekar and J.~Lewandowski,
``Projective techniques and functional integration for gauge theories,''
J. Math. Phys. \textbf{36} (1995), 2170-2191
[arXiv:gr-qc/9411046 [gr-qc]].
A.~Ashtekar and J.~Lewandowski,
``Representation theory of analytic holonomy C* algebras,''
[arXiv:gr-qc/9311010 [gr-qc]].
T.~Koslowski and H.~Sahlmann,
``Loop quantum gravity vacuum with nondegenerate geometry,''
SIGMA \textbf{8} (2012), 026
[arXiv:1109.4688 [gr-qc]].

\bibitem{NewQG}
B.~Dittrich and M.~Geiller,
``A new vacuum for Loop Quantum Gravity,''
Class. Quant. Grav. \textbf{32} (2015) no.11, 112001
[arXiv:1401.6441 [gr-qc]].
B.~Bahr, B.~Dittrich and M.~Geiller,
``A new realization of quantum geometry,''
[arXiv:1506.08571 [gr-qc]].
B.~Dittrich and M.~Geiller,
``Quantum gravity kinematics from extended TQFTs,''
New J. Phys. \textbf{19} (2017) no.1, 013003
[arXiv:1604.05195 [hep-th]].
B.~Dittrich,
``(3 + 1)-dimensional topological phases and self-dual quantum geometries encoded on Heegaard surfaces,''
JHEP \textbf{05} (2017), 123
[arXiv:1701.02037 [hep-th]].

\bibitem{LQG} 
A. Ashtekar and J. Lewandowski,
``Background independent quantum gravity: a status report'',
Class. Quant. Grav. \textbf{21}  (2004), R53, [arXiv:gr-qc/0404018].
C. Rovelli,
\textit{Quantum Gravity},
(Cambridge University Press, Cambridge, 2004).
T. Thiemann,
\textit{Introduction to Modern Canonical Quantum General Relativity},
(Cambridge University Press, Cambridge, 2007).

    \bibitem{DiscreteGeom}
   C.~Rovelli and L.~Smolin,
  ``Discreteness of area and volume in quantum gravity,''
  Nucl.\ Phys.\ B {\bf 442} (1995) 593
   Erratum: [Nucl.\ Phys.\ B {\bf 456} (1995) 753]
  [gr-qc/9411005].
  %
   A.~Ashtekar and J.~Lewandowski,
  ``Quantum theory of geometry. 1: Area operators,''
  Class.\ Quant.\ Grav.\  {\bf 14} (1997) A55
  [gr-qc/9602046];
   A.~Ashtekar and J.~Lewandowski,
  ``Quantum theory of geometry. 2. Volume operators,''
  Adv.\ Theor.\ Math.\ Phys.\  {\bf 1} (1998) 388
  [gr-qc/9711031].
%


\bibitem{Immirzi-Barbero}
J.~F.~Barbero G.,
``Real Ashtekar variables for Lorentzian signature space times,''
Phys. Rev. D \textbf{51} (1995), 5507-5510
[arXiv:gr-qc/9410014 [gr-qc]].
G.~Immirzi,
``Real and complex connections for canonical gravity,''
Class. Quant. Grav. \textbf{14} (1997), L177-L181
[arXiv:gr-qc/9612030 [gr-qc]].

 
 \bibitem{ADP} 
 S.~K.~Asante, B.~Dittrich, J.~Padua Arguelles, 
 ``Effective spin foams for Lorentzian quantum gravity", [arXiv:2104.00485[gr-qc]]


 \bibitem{BHCounting}
    J.~D.~Bekenstein and V.~F.~Mukhanov,
  ``Spectroscopy of the quantum black hole,''
  Phys.\ Lett.\ B {\bf 360} (1995) 7
  [gr-qc/9505012].
 %
   A.~Ashtekar, J.~Baez, A.~Corichi and K.~Krasnov,
  ``Quantum geometry and black hole entropy,''
  Phys.\ Rev.\ Lett.\  {\bf 80} (1998) 904
  [gr-qc/9710007].
 %
   J.~D.~Bekenstein,
  ``Statistics of black hole radiance and the horizon area spectrum,''
  Phys.\ Rev.\ D {\bf 91} (2015) no.12,  124052
  [arXiv:1505.03253 [gr-qc]].
 J.~F.~Barbero G. and A.~Perez,
  ``Quantum Geometry and Black Holes,''
  arXiv:1501.02963 [gr-qc].

\bibitem{EffSFM} S.~K.~Asante, B.~Dittrich and H.~M.~Haggard,
``Effective Spin Foam Models for Four-Dimensional Quantum Gravity,'' to appear in Phys.\ Rev.\ Lett.\, 
[arXiv:2004.07013 [gr-qc]].

\bibitem{MV}
  A.~Mikovic and M.~Vojinovic,
  ``Poincare 2-group and quantum gravity,''
  Class.\ Quant.\ Grav.\  {\bf 29} (2012) 165003
  [arXiv:1110.4694 [gr-qc]];
   M.~Vojinovic,
  ``Causal Dynamical Triangulations in the Spincube Model of Quantum Gravity,''
  Phys.\ Rev.\ D {\bf 94} (2016) no.2,  024058
  [arXiv:1506.06839 [gr-qc]].
  


\bibitem{SchraderEtAl}
J.~Cheeger, W.~Muller and R.~Schrader,
``On the Curvature of Piecewise Flat Spaces,''
Commun. Math. Phys. \textbf{92} (1984), 405.


\bibitem{WilliamsEtAl} 
M.~Rocek and R.~M.~Williams,
``QUANTUM REGGE CALCULUS,''
Phys. Lett. B \textbf{104} (1981), 31.
M.~Rocek and R.~M.~Williams,
``The Quantization of Regge Calculus,''
Z. Phys. C \textbf{21} (1984), 371.

   \bibitem{AreaRegge}
 J.~W.~Barrett, M.~Rocek and R.~M.~Williams,
  ``A Note on area variables in Regge calculus,''
  Class.\ Quant.\ Grav.\  {\bf 16} (1999) 1373
  [gr-qc/9710056].
    J.~Makela,
  ``Variation of area variables in Regge calculus,''
  Class.\ Quant.\ Grav.\  {\bf 17} (2000) 4991
  [gr-qc/9801022];
   J.~Makela and R.~M.~Williams,
  ``Constraints on area variables in Regge calculus,''
  Class.\ Quant.\ Grav.\  {\bf 18} (2001) L43
  [gr-qc/0011006].
  
  \bibitem{ADHAreaR}
S.~K.~Asante, B.~Dittrich and H.~M.~Haggard,
  ``The Degrees of Freedom of Area Regge Calculus: Dynamics, Non-metricity, and Broken Diffeomorphisms,''
  Class.\ Quant.\ Grav.\  {\bf 35} (2018) no.13,  135009
  [arXiv:1802.09551]

 

\bibitem{Regge}
 T.~Regge,
  ``General Relativity Without Coordinates,''
Nuovo Cim.  {\bf 19} (1961) 558.
  
  
  
    \bibitem{AreaAngle}
 B.~Dittrich and S.~Speziale,
  ``Area-angle variables for general relativity,''
  New J.\ Phys.\  {\bf 10} (2008) 083006
  [arXiv:0802.0864 [gr-qc]].
 B.~Bahr and B.~Dittrich,
``Regge calculus from a new angle,''
New J. Phys. \textbf{12} (2010), 033010
[arXiv:0907.4325 [gr-qc]].

\bibitem{DittrichRyan}
  B.~Dittrich and J.~P.~Ryan,
  ``Phase space descriptions for simplicial 4d geometries,''
  Class.\ Quant.\ Grav.\  {\bf 28} (2011) 065006
  [arXiv:0807.2806 [gr-qc]];
  B.~Dittrich and J.~P.~Ryan,
  ``Simplicity in simplicial phase space,''
  Phys.\ Rev.\ D {\bf 82} (2010) 064026
  [arXiv:1006.4295 [gr-qc]];
B.~Dittrich and J.~P.~Ryan,
  ``On the role of the Barbero-Immirzi parameter in discrete quantum gravity,''
  Class.\ Quant.\ Grav.\  {\bf 30} (2013) 095015
  [arXiv:1209.4892 [gr-qc]].



  \bibitem{Perez}
  A.~Perez,
  ``The Spin Foam Approach to Quantum Gravity,''
  Living Rev.\ Rel.\  {\bf 16} (2013) 3
  [arXiv:1205.2019].
  
\bibitem{Plebanski} 
J.~F.~Plebanski,
  ``On the separation of Einsteinian substructures,''
  J.\ Math.\ Phys.\  {\bf 18} (1977), 2511 .

\bibitem{Corichi}
A.~Ashtekar, A.~Corichi and J.~A.~Zapata,
``Quantum theory of geometry III: Noncommutativity of Riemannian structures,''
Class. Quant. Grav. \textbf{15} (1998), 2955-2972
[arXiv:gr-qc/9806041 [gr-qc]].

\bibitem{FluxRepOfLQG}
A.~Baratin, B.~Dittrich, D.~Oriti and J.~Tambornino,
``Non-commutative flux representation for loop quantum gravity,''
Class. Quant. Grav. \textbf{28} (2011), 175011
doi:10.1088/0264-9381/28/17/175011
[arXiv:1004.3450 [hep-th]].

\bibitem{DittrichGeillerFlux}
B.~Dittrich and M.~Geiller,
``Flux formulation of loop quantum gravity: Classical framework,''
Class. Quant. Grav. \textbf{32} (2015) no.13, 135016
[arXiv:1412.3752 [gr-qc]].




\bibitem{PerezCattaneo}
L.~Freidel and A.~Perez,
``Quantum gravity at the corner,''
Universe \textbf{4} (2018) no.10, 107
[arXiv:1507.02573 [gr-qc]].
A.~S.~Cattaneo and A.~Perez,
``A note on the Poisson bracket of 2d smeared fluxes in loop quantum gravity,''
Class. Quant. Grav. \textbf{34} (2017) no.10, 107001
[arXiv:1611.08394 [gr-qc]].


\bibitem{CornerSimpl} L.~Freidel, M.~Geiller and D.~Pranzetti,
``Edge modes of gravity - III: Corner simplicity constraints,''
[arXiv:2007.12635 [hep-th]].



  
   \bibitem{NSFM}
    J.~Engle, R.~Pereira and C.~Rovelli,
  ``The Loop-quantum-gravity vertex-amplitude,''
  Phys.\ Rev.\ Lett.\  {\bf 99} (2007) 161301
  [arXiv:0705.2388 [gr-qc]];
  L.~Freidel and K.~Krasnov,
  ``A New Spin Foam Model for 4d Gravity,''
  Class.\ Quant.\ Grav.\  {\bf 25} (2008) 125018
  [arXiv:0708.1595 [gr-qc]];
       E.~R.~Livine and S.~Speziale,
  ``Consistently Solving the Simplicity Constraints for Spinfoam Quantum Gravity,''
  EPL {\bf 81} (2008) no.5,  50004
  [arXiv:0708.1915 [gr-qc]];
    J.~Engle, E.~Livine, R.~Pereira and C.~Rovelli,
  ``LQG vertex with finite Immirzi parameter,''
  Nucl.\ Phys.\ B {\bf 799} (2008) 136
  [arXiv:0711.0146 [gr-qc]];
   M.~Dupuis and E.~R.~Livine,
  ``Holomorphic Simplicity Constraints for 4d Spinfoam Models,''
  Class.\ Quant.\ Grav.\  {\bf 28} (2011) 215022
  [arXiv:1104.3683 [gr-qc]];
  A.~Baratin and D.~Oriti,
  ``Group field theory and simplicial gravity path integrals: A model for Holst-Plebanski gravity,''
  Phys.\ Rev.\ D {\bf 85} (2012) 044003
  [arXiv:1111.5842 [hep-th]].

 
  \bibitem{SFLimit}
   J.~W.~Barrett and R.~M.~Williams,
  ``The Asymptotics of an amplitude for the four simplex,''
  Adv.\ Theor.\ Math.\ Phys.\  {\bf 3} (1999) 209
  doi:10.4310/ATMP.1999.v3.n2.a1
  [gr-qc/9809032];
   J.~W.~Barrett and C.~M.~Steele,
  ``Asymptotics of relativistic spin networks,''
  Class.\ Quant.\ Grav.\  {\bf 20} (2003) 1341
  [gr-qc/0209023];
    J.~W.~Barrett, R.~J.~Dowdall, W.~J.~Fairbairn, H.~Gomes and F.~Hellmann,
  ``Asymptotic analysis of the EPRL four-simplex amplitude,''
  J.\ Math.\ Phys.\  {\bf 50} (2009) 112504
  doi:10.1063/1.3244218
  [arXiv:0902.1170 [gr-qc]];
   J.~W.~Barrett, R.~J.~Dowdall, W.~J.~Fairbairn, F.~Hellmann and R.~Pereira,
  ``Lorentzian spin foam amplitudes: Graphical calculus and asymptotics,''
  Class.\ Quant.\ Grav.\  {\bf 27} (2010) 165009
  doi:10.1088/0264-9381/27/16/165009
  [arXiv:0907.2440 [gr-qc]];
    M.~X.~Han and M.~Zhang,
  ``Asymptotics of Spinfoam Amplitude on Simplicial Manifold: Euclidean Theory,''
  Class.\ Quant.\ Grav.\  {\bf 29} (2012) 165004
  [arXiv:1109.0500 [gr-qc]].
 
 \bibitem{FlatProblem}
   F.~Conrady and L.~Freidel,
  ``On the semiclassical limit of 4d spin foam models,''
  Phys.\ Rev.\ D {\bf 78} (2008) 104023
  [arXiv:0809.2280 [gr-qc]].
  V.~Bonzom,
  ``Spin foam models for quantum gravity from lattice path integrals,''
  Phys.\ Rev.\ D {\bf 80} (2009) 064028
  [arXiv:0905.1501 [gr-qc]].
   F.~Hellmann and W.~Kaminski,
  ``Holonomy spin foam models: Asymptotic geometry of the partition function,''
  JHEP {\bf 1310} (2013) 165
  [arXiv:1307.1679 [gr-qc]].
     J.~R.~Oliveira,
  ``EPRL/FK Asymptotics and the Flatness Problem,''
  Class.\ Quant.\ Grav.\  {\bf 35} (2018) no.9,  095003
  [arXiv:1704.04817 [gr-qc]].
P.~Don\`a, F.~Gozzini, G.~Sarno,
``Numerical analysis of spin foam dynamics and the flatness problem,''
[arXiv:2004.12911].


   \bibitem{ILQGS}
 E. Bianchi, J. Engle, S. Speziale, ILQGS seminar (March 3rd 2020): Panel on the status of the vertex,
  \href{http://relativity.phys.lsu.edu/ilqgs/bianchienglespeziale030320.pdf}{\tt Slides}
  

 
  \bibitem{CE}
      E.~Magliaro and C.~Perini,
  ``Regge gravity from spinfoams,''
  Int.\ J.\ Mod.\ Phys.\ D {\bf 22} (2013) 1
  [arXiv:1105.0216 [gr-qc]];
  E.~Magliaro and C.~Perini,
  ``Emergence of gravity from spinfoams,''
  EPL {\bf 95} (2011) no.3,  30007
  [arXiv:1108.2258 [gr-qc]].
  

  
  \bibitem{MHan}
   M.~Han,
  ``Semiclassical Analysis of Spinfoam Model with a Small Barbero-Immirzi Parameter,''
  Phys.\ Rev.\ D {\bf 88} (2013) 044051
  [arXiv:1304.5628 [gr-qc]].
  

   \bibitem{DonaNum}
    S.~Speziale,
  ``Boosting Wigner's $nj$-symbols,''
  J.\ Math.\ Phys.\  {\bf 58} (2017) no.3,  032501
  [arXiv:1609.01632 [gr-qc]];
   P.~Don\`a and G.~Sarno,
  ``Numerical methods for EPRL spin foam transition amplitudes and Lorentzian recoupling theory,''
  Gen.\ Rel.\ Grav.\  {\bf 50} (2018) 127
  [arXiv:1807.03066 [gr-qc]];
  P.~Don\`a, M.~Fanizza, G.~Sarno and S.~Speziale,
  ``Numerical study of the Lorentzian Engle-Pereira-Rovelli-Livine spin foam amplitude,''
  Phys.\ Rev.\ D {\bf 100} (2019) no.10,  106003;
   P.~Don\`a, F.~Gozzini and G.~Sarno,
  ``Searching for classical geometries in spin foam amplitudes: a numerical method,''
  arXiv:1909.07832 [gr-qc].

  \bibitem{HigherGauge}
  F.~Girelli, H.~Pfeiffer and E.~M.~Popescu,
  ``Topological Higher Gauge Theory - from BF to BFCG theory,''
  J.\ Math.\ Phys.\  {\bf 49} (2008) 032503
  [arXiv:0708.3051 [hep-th]].
   A.~Baratin and L.~Freidel,
  ``Hidden Quantum Gravity in 4-D Feynman diagrams: Emergence of spin foams,''
  Class.\ Quant.\ Grav.\  {\bf 24} (2007) 2027
  [hep-th/0611042].
   A.~Baratin and L.~Freidel,
  ``A 2-categorical state sum model,''
  J.\ Math.\ Phys.\  {\bf 56} (2015) no.1,  011705
  [arXiv:1409.3526 [math.QA]].
     S.~K.~Asante, B.~Dittrich, F.~Girelli, A.~Riello and P.~Tsimiklis,
  ``Quantum geometry from higher gauge theory,''
  arXiv:1908.05970 [gr-qc].
%




  \bibitem{CG} 
    B.~Dittrich,
  ``The continuum limit of loop quantum gravity - a framework for solving the theory,''
  in {\it Loop Quantum Gravity - The First 30 Years}, ed. by A. Ashtekar And J. Pullin, World Scientific Publishing 2017, pp.  153-179,
  [arXiv:1409.1450 [gr-qc]].
  B.~Bahr,
``On background-independent renormalization of spin foam models,''
Class. Quant. Grav. \textbf{34} (2017) no.7, 075001
[arXiv:1407.7746 [gr-qc]].
  B.~Dittrich, F.~C.~Eckert and M.~Martin-Benito,
``Coarse graining methods for spin net and spin foam models,''
New J. Phys. \textbf{14} (2012), 035008
[arXiv:1109.4927 [gr-qc]].
  B.~Dittrich, M.~Martin-Benito and S.~Steinhaus,
``Quantum group spin nets: refinement limit and relation to spin foams,''
Phys. Rev. D \textbf{90} (2014), 024058
[arXiv:1312.0905 [gr-qc]].
   B.~Dittrich, S.~Mizera and S.~Steinhaus,
  ``Decorated tensor network renormalization for lattice gauge theories and spin foam models,''
  New J.\ Phys.\  {\bf 18} (2016) no.5,  053009
  [arXiv:1409.2407 [gr-qc]];
  B.~Bahr and S.~Steinhaus,
``Investigation of the Spinfoam Path integral with Quantum Cuboid Intertwiners,''
Phys. Rev. D \textbf{93} (2016) no.10, 104029
[arXiv:1508.07961 [gr-qc]].
  B.~Dittrich, E.~Schnetter, C.~J.~Seth and S.~Steinhaus,
  ``Coarse graining flow of spin foam intertwiners,''
  Phys.\ Rev.\ D {\bf 94} (2016) no.12,  124050
  [arXiv:1609.02429 [gr-qc]];
   C.~Delcamp and B.~Dittrich,
  ``Towards a phase diagram for spin foams,''
  Class.\ Quant.\ Grav.\  {\bf 34} (2017) no.22,  225006
  [arXiv:1612.04506 [gr-qc]];
   B.~Bahr, G.~Rabuffo and S.~Steinhaus,
  ``Renormalization of symmetry restricted spin foam models with curvature in the asymptotic regime,''
  Phys.\ Rev.\ D {\bf 98} (2018) no.10,  106026
  [arXiv:1804.00023 [gr-qc]].
  
 

\bibitem{Picard} 
M.~V. Fedoryuk, ``The saddle-point method,"  {\em Izdat. ``Nauka,'' Moscow,
  MR 58:22580} (1977);
D.~Kaminski, ``Exponentially improved stationary phase approximations for
  double integrals,"  Methods Appl. Anal. 1, 44 -- 56. MR 95i:41056
  (1994);
F.~Pham, ``Vanishing homologies and the n variable saddlepoint method,"
   Proc. Symp. Pure Math {\bf 2} (1983), no.~40 319--333;
M.~V. Berry and C.~J. Howls, ``Hyperasymptotics for integrals with saddles,"
   Proceedings of the Royal Society of London A: Mathematical, Physical
  and Engineering Sciences {\bf 434} (1991), no.~1892 657--675;
A.~Behtash, G.~V.~Dunne, T.~Schaefer, T.~Sulejmanpasic, M.~Unsal, 
``Toward Picard-Lefschetz Theory of Path Integrals, Complex Saddles and Resurgence,"
Annals of Mathematical Sciences and Applications {\bf 2} (2017) [arXiv:1510.03435 [hep-th]].

\bibitem{BianchiHaggard1} 
    E.~Bianchi and H.~M.~Haggard, 
    ``Discreteness of the volume of space from Bohr-Sommerfeld quantization,"
    Phys. Rev. Lett. {\bf 107} (2011) 011301, [arXiv:1102.5439 [gr-qc]];
    E.~Bianchi and H.~M.~Haggard, 
    ``Bohr-Sommerfeld Quantization of Space,"    
   Phys. Rev. D {\bf 86} (2012) 124010, [arXiv:1208.2228 [gr-qc]].
   
\bibitem{BianchiLength}
E.~Bianchi,
``The length operator in Loop Quantum Gravity,"
Nuc.\ Phys.\ B {\bf 807} (2009)  591.  


\bibitem{DFS07} B.~Dittrich, L.~Freidel and S.~Speziale,
``Linearized dynamics from the 4-simplex Regge action,''
Phys. Rev. D \textbf{76} (2007), 104020
[arXiv:0707.4513 [gr-qc]].

\bibitem{Schlafli:1858}
L.~Schl\"afli, 
``On the multiple integral $\int^{n} dx dy \dots dz$ whose limits are $p=a_1 x+b_1 y+\dots + h_1 z>0, p_2>0, \dots, p_n>0,$ and $x^2+y^2+\cdots+z^2<1$",
{ Quart. J. Pure Appl. Math.} {\bf 2} (1858) 269.

\bibitem{HaggardSch:2015}
H.~M.~Haggard, A.~Hedeman, E.~Kur, and R.~G.~Littlejohn,
``Symplectic and semiclassical aspects of the Schl\"afli identity,"
{ \href{http://dx.doi.org/10.1088/1751-8113/48/10/105203}{J. Phys. A: Math. Theor.}} {\bf 48} (2015) 105203.




     \bibitem{Twisted}
   L.~Freidel and S.~Speziale,
  ``Twisted geometries: A geometric parametrisation of SU(2) phase space,''
  Phys.\ Rev.\ D {\bf 82} (2010) 084040
  [arXiv:1001.2748 [gr-qc]].
  
 
 \bibitem{Barbieri}
  A.~Barbieri,
  ``Quantum tetrahedra and simplicial spin networks,''
  Nucl.\ Phys.\ B {\bf 518} (1998) 714
  [gr-qc/9707010].
    J.~C.~Baez and J.~W.~Barrett,
  ``The Quantum tetrahedron in three-dimensions and four-dimensions,''
  Adv.\ Theor.\ Math.\ Phys.\  {\bf 3} (1999) 815
  [gr-qc/9903060].
  
  

  \bibitem{Coherent}
  E.~R.~Livine and S.~Speziale,
  ``A New spinfoam vertex for quantum gravity,''
  Phys.\ Rev.\ D {\bf 76} (2007) 084028
  [arXiv:0705.0674 [gr-qc]].
   V.~Bonzom and E.~R.~Livine,
  ``Generating Functions for Coherent Intertwiners,''
  Class.\ Quant.\ Grav.\  {\bf 30} (2013) 055018
  [arXiv:1205.5677 [gr-qc]].
   L.~Freidel and J.~Hnybida,
  ``A Discrete and Coherent Basis of Intertwiners,''
  Class.\ Quant.\ Grav.\  {\bf 31} (2014) 015019
  [arXiv:1305.3326 [math-ph]].


  \bibitem{PImeasure}
    B.~Bahr, B.~Dittrich and S.~Steinhaus,
  ``Perfect discretization of reparametrization invariant path integrals,''
  Phys.\ Rev.\ D {\bf 83} (2011) 105026
  [arXiv:1101.4775 [gr-qc]];
    B.~Dittrich and S.~Steinhaus,
  ``Path integral measure and triangulation independence in discrete gravity,''
  Phys.\ Rev.\ D {\bf 85} (2012) 044032
  [arXiv:1110.6866 [gr-qc]];
   B.~Dittrich, W.~Kaminski and S.~Steinhaus,
  ``Discretization independence implies non-locality in 4D discrete quantum gravity,''
  Class.\ Quant.\ Grav.\  {\bf 31} (2014) no.24,  245009
  [arXiv:1404.5288 [gr-qc]];
     B.~Bahr and S.~Steinhaus,
  ``Numerical evidence for a phase transition in 4d spin foam quantum gravity,''
  Phys.\ Rev.\ Lett.\  {\bf 117} (2016) no.14,  141302
  


 \bibitem{Engle}
   J.~Engle,
  ``Proposed proper Engle-Pereira-Rovelli-Livine vertex amplitude,''
  Phys.\ Rev.\ D {\bf 87} (2013) no.8,  084048
  [arXiv:1111.2865 [gr-qc]];
  J.~Engle,
  ``A spin-foam vertex amplitude with the correct semiclassical limit,''
  Phys.\ Lett.\ B {\bf 724} (2013) 333
  [arXiv:1201.2187 [gr-qc]].
 



 
\bibitem{3DHol}
  B.~Dittrich, C.~Goeller, E.~Livine and A.~Riello,
  ``Quasi-local holographic dualities in non-perturbative 3d quantum gravity I -- Convergence of multiple approaches and examples of Ponzano-Regge statistical duals,''
  Nucl.\ Phys.\ B {\bf 938} (2019) 807
  [arXiv:1710.04202 [hep-th]];
   B.~Dittrich, C.~Goeller, E.~R.~Livine and A.~Riello,
  ``Quasi-local holographic dualities in non-perturbative 3d quantum gravity II -- From coherent quantum boundaries to BMS$_3$ characters,''
  Nucl.\ Phys.\ B {\bf 938} (2019) 878
  [arXiv:1710.04237 [hep-th]];
   B.~Dittrich, C.~Goeller, E.~R.~Livine and A.~Riello,
  ``Quasi-local holographic dualities in non-perturbative 3d quantum gravity,''
  Class.\ Quant.\ Grav.\  {\bf 35} (2018) no.13,  13LT01
  [arXiv:1803.02759 [hep-th]];
    C.~Goeller, E.~R.~Livine and A.~Riello,
  ``Non-Perturbative 3D Quantum Gravity: Quantum Boundary States and Exact Partition Function,''
  Gen.\ Rel.\ Grav.\  {\bf 52} (2020) no.3,  24
  [arXiv:1912.01968 [hep-th]].
%

 
\bibitem{Ditt05} 
B.~Dittrich,
``Partial and complete observables for Hamiltonian constrained systems,''
Gen. Rel. Grav. \textbf{39} (2007), 1891-1927
[arXiv:gr-qc/0411013 [gr-qc]].
B.~Dittrich,
``Partial and complete observables for canonical general relativity,''
Class. Quant. Grav. \textbf{23} (2006), 6155-6184
[arXiv:gr-qc/0507106 [gr-qc]].

\bibitem{BahrDittrich09a} B.~Dittrich,
``Diffeomorphism symmetry in quantum gravity models,''
Adv. Sci. Lett. \textbf{2}, 151
[arXiv:0810.3594 [gr-qc]].
B.~Bahr and B.~Dittrich,
``(Broken) Gauge Symmetries and Constraints in Regge Calculus,''
Class. Quant. Grav. \textbf{26} (2009), 225011
[arXiv:0905.1670 [gr-qc]].


\bibitem{AshtekarBaez:1998} A.~Ashtekar,  J.~Baez, A.~Corichi, and K.~Krasnov,
``Quantum Geometry and Black Hole Entropy,"
Phys. Rev. Lett. \textbf{80} (1998), 904
[arXiv:gr-qc/9710007].

\bibitem{DomagalaLewandowski:2004}
M.~Domagala, J.~Lewandowski, 
``Black hole entropy from Quantum Geometry,"
Class.Quant.Grav. \textbf{21} (2004), 5233
[arXiv:gr-qc/0407051].

\bibitem{PigozzoBacelarCarneiro:2021}
C.~Pigozzo, F.~S.~Bacelar, S.~Carneiro,
``On the value of the Immirzi parameter and the horizon entropy,"
Class. Quant. Grav. \textbf{38} (2021), 045001
[arXiv:2001.03440 [gr-qc]].

\bibitem{RT} S.~Ryu and T.~Takayanagi,
``Holographic derivation of entanglement entropy from AdS/CFT,''
Phys. Rev. Lett. \textbf{96} (2006), 181602
[arXiv:hep-th/0603001 [hep-th]].
E.~Bianchi and R.~C.~Myers,
``On the Architecture of Spacetime Geometry,''
Class. Quant. Grav. \textbf{31} (2014), 214002
[arXiv:1212.5183 [hep-th]].

\bibitem{QGTNW} B.~Swingle,
``Entanglement Renormalization and Holography,''
Phys. Rev. D \textbf{86} (2012), 065007
[arXiv:0905.1317 [cond-mat.str-el]].

\bibitem{Improved}
   B.~Bahr and B.~Dittrich,
  ``Improved and Perfect Actions in Discrete Gravity,''
  Phys.\ Rev.\ D {\bf 80} (2009) 124030
  [arXiv:0907.4323 [gr-qc]];
  B.~Bahr, B.~Dittrich and S.~He,
``Coarse graining free theories with gauge symmetries: the linearized case,''
New J. Phys. \textbf{13} (2011), 045009
[arXiv:1011.3667 [gr-qc]].
  B.~Dittrich,
``How to construct diffeomorphism symmetry on the lattice,''
PoS \textbf{QGQGS2011} (2011), 012
[arXiv:1201.3840 [gr-qc]].


 
  
  
\bibitem{DittrichHoehn1}
B.~Dittrich and P.~A.~H\"ohn,
  ``From covariant to canonical formulations of discrete gravity,''
  Class.\ Quant.\ Grav. {\bf 27} (2010) 155001
  [arXiv:0912.1817 [gr-qc]];
   V.~Bonzom and B.~Dittrich,
  ``Dirac's discrete hypersurface deformation algebras,''
   Class.\ Quant.\ Grav.  {\bf 30} (2013) 205013
  [arXiv:1304.5983 [gr-qc]].
  

  
    \bibitem{Krasnov}
   K.~Krasnov,
  ``Gravity as BF theory plus potential,''
  Int.\ J.\ Mod.\ Phys.\ A {\bf 24} (2009) 2776
  [arXiv:0907.4064 [gr-qc]];
    K.~Krasnov,
  ``Effective metric Lagrangians from an underlying theory with two propagating degrees of freedom,''
  Phys.\ Rev.\ D {\bf 81} (2010) 084026
  [arXiv:0911.4903 [hep-th]].
 
 
\bibitem{NewRegge}
B.~Bahr and B.~Dittrich,
  ``Regge calculus from a new angle,''
  New J.\ Phys.\  {\bf 12} (2010) 033010
  [arXiv:0907.4325 [gr-qc]].

\bibitem{TuraevViro:1992}
V. Turaev and O. Viro,
``State sum invariants of 3 manifolds and quantum 6j symbols'',
Topology \textbf{31} 865 (1992).

\bibitem{Crane:94}
  L.~Crane, L.~H.~Kauffman and D.~N.~Yetter,
  ``State sum invariants of four manifolds. 1.,''
  hep-th/9409167.

\bibitem{Major:96} 
S. Major and L. Smolin,
``Quantum deformation of quantum gravity'',
Nucl. Phys. B \textbf{473} (1996) 267, arXiv:gr-qc/9512020.
 
 \bibitem{Barrett:03}
   J.~W.~Barrett,
  ``Geometrical measurements in three-dimensional quantum gravity,''
  Int.\ J.\ Mod.\ Phys.\ A {\bf 18S2} (2003) 97
  [gr-qc/0203018].
  
  \bibitem{Dupuis:14}
  M. Dupuis and F. Girelli,
``Observables in Loop Quantum Gravity with a cosmological constant'',
Phys. Rev. D \textbf{90} (2014), [arXiv:1311.6841 [gr-qc]].

\bibitem{Bonzom:14} 
V.~Bonzom, M.~Dupuis and F.~Girelli,
  ``Towards the Turaev-Viro amplitudes from a Hamiltonian constraint,''
  Phys.\ Rev.\ D {\bf 90} (2014) no.10,  104038
  [arXiv:1403.7121 [gr-qc]].
 
 \bibitem{Haggard:15}  H.~M.~Haggard, M.~Han, W.~Kaminski and A.~Riello,
``SL(2,C) Chern--Simons Theory, a non-Planar Graph Operator, and 4D Loop Quantum Gravity with a Cosmological Constant: Semiclassical Geometry'',
Nucl. Phys. B \textbf{900} (2015) 1, [arXiv:1412.7546 [hep-th]].

 \bibitem{Haggard:16}
H.~M.~Haggard, M.~Han, and A.~Riello,
``Encoding Curved Tetrahedra in Face Holonomies: a Phase Space of Shapes from Group-Valued Moment Maps''
Annales Henri Poincar\'e {\bf 17} (2016) 2001, [arXiv:1506.03053 [math-ph]].

 \bibitem{Haggard:162}  H.~M.~Haggard, M.~Han, W.~Kaminski and A.~Riello,
``Four-dimensional Quantum Gravity with a Cosmological Constant from Three-dimensional Holomorphic Blocks'',
Phys Lett. B {\bf 752} (2016) 258, [arXiv:1509.00458 [hep-th]]

\bibitem{Haggard:18}
  H.~M.~Haggard, M.~Han, W.~Kaminski and A.~Riello,
  ``SL(2,C) Chern-Simons Theory, Flat Connections, and Four-dimensional Quantum Geometry,''
  arXiv:1512.07690 [hep-th].
  
  \bibitem{Dittrich:17}
B.~Dittrich and M.~Geiller,
  ``Quantum gravity kinematics from extended TQFTs,''
  New J.\ Phys.\  {\bf 19} (2017) no.1,  013003
  [arXiv:1604.05195 [hep-th]].
  
    \bibitem{Dittrich:172}
   B.~Dittrich,
  ``(3+1)-dimensional topological phases and self-dual quantum geometries encoded on Heegaard surfaces,''
  JHEP {\bf 1705} (2017) 123
  [arXiv:1701.02037 [hep-th]].

  \bibitem{HaggardLittlejohn} 
  H.~M.~Haggard and R.~G.~Littlejohn,
  ``Asymptotics of the Wigner $9j$ symbol,"
    Class.\ Quant.\ Grav.\ {\bf 27} (2010) 135010 [arXiv:0912.5384 [gr-qc]].

\bibitem{Dekster-Wilker} 
B. V. Dekster and J. B. Wilker,
``Edge lengths guaranteed to form a simplex,"
Arch.\ Math.\ {\bf 49} (1987) 351.



  
  


   
 
 
 
 


















 
     
 
 
 

 
  


  
 



 
    

  
  
  


   
 
  
 

  


  
  
  
  
  
  
     







  
\end{thebibliography}
 \endgroup

\end{document}